\documentclass[preprint,aps,amsmath,nofootinbib]{revtex4-1}
\usepackage{graphicx,array,dcolumn}
\usepackage{calc,tabularx, epsfig,mathrsfs}
\usepackage{hyperref}

\usepackage{amsmath,verbatim,enumerate}
\usepackage{amssymb}
\usepackage{wasysym}
\usepackage{xcolor}
\usepackage{subfigure}
\usepackage{tcolorbox}
\usepackage{multirow}
\usepackage{xspace}
\usepackage{slashed}
\usepackage{mathtools}
\usepackage{float}
\usepackage{cancel}
\allowdisplaybreaks[1]
\DeclareMathOperator\arctanh{arctanh}
\DeclareMathOperator\arcsinh{arcsinh}
\newlength{\figurewidth}
\newcommand{\beq}{\begin{equation}}
\newcommand{\eeq}{\end{equation}}
\newcommand{\bea}{\begin{eqnarray}}
\newcommand{\eea}{\end{eqnarray}}
\newcommand{\ba}{\begin{array}}
\newcommand{\ea}{\end{array}}

\newcommand{\mn}{{\mu\nu}}

\newcommand{\pt}{\partial}

\newcommand{\Bl}{\biggl}
\newcommand{\Br}{\biggr}
\newcommand{\bl}{\bigl}
\newcommand{\br}{\bigr}
\newcommand{\al}{\alpha}
\newcommand{\bt}{\beta}

\newcommand{\ep}{\epsilon}
\newcommand{\ta}{\theta}
\newcommand{\lam}{\lambda}
\newcommand{\Lam}{\Lambda}
\newcommand{\G}{\Gamma}

\newcommand{\de}{\delta}
\newcommand{\D}{\Delta}
\newcommand{\OM}{\Omega}

\newcommand{\sg}{\sigma}

\makeatother
\begin{document}
%
%%%%%%%%%%%%%%%%%%%%%%%%%%%%%%%%%%%%%%%%%%%%%%%
\title{
Resolving Degeneracies in Complex $\mathbb{R}\times S^3$ and $\theta$-KSW
}
\setlength{\figurewidth}{\columnwidth}
%%%%%%%%%%%%%%%%%%%%%%%%%%%%%%%%%%%%%%%%%%%%%%%
%
\author{Manishankar Ailiga}
\email{manishankara@iisc.ac.in}

\author{Shubhashis Mallik} 
\email{shubhashism@iisc.ac.in}

\author{Gaurav Narain}
\email{gnarain@iisc.ac.in}

\affiliation{
Center for High Energy Physics, Indian Institute of Science,
C V Raman Road, Bangalore 560012, India.
\vspace{10mm}
}

%%%%%%%%%%%%%%%%%%%%%%%%%%%%%%%%%%%%%%%%%%%%%%%
\vspace{20mm}

%%%%%%%%%%%%%%%%%%%%%%%%%%%%%%%%%%%%%%%%%%%%%%%
\begin{abstract}
Lorentzian gravitational path integral for the Gauss-Bonnet gravity in $4D$ is studied in the mini-superspace ansatz for metric. The gauge-fixed path-integral for Robin boundary choice is computed exactly using {\it Airy}-functions, where the dominant contribution comes from No-boundary geometries. The lapse integral is further analysed using saddle-point methods to compare with exact results. Picard-Lefschetz methods are utilized to find the {\it relevant} complex saddles and deformed contour of integration, thereby using WKB methods to compute the integral along the deformed contour in the saddle-point approximation. However, their successful application is possible only when system is devoid of degeneracies, which in present case appear in two types: {\bf type-1} where the flow-lines starting from neighbouring saddles overlap leading to ambiguities in deciding the {\it relevance} of saddles, {\bf type-2} where saddles merge for specific choices of boundary parameters leading to failure of WKB. Overcoming degeneracies using artificial {\it defects} introduces ambiguities due to the choice of {\it defects} involved. Corrections from quantum fluctuations of scale-factor overcome degeneracies only partially (lifts {\bf type-2} completely with partial resolution of {\bf type-1}), with the residual lifted fluently by complex deformation of $(G\hbar)$. {\it Anti-linear} symmetry present in various forms in the lapse action is the reason behind all the {\bf type-1} degeneracies. Any form of {\it defect} or {\it deformation} breaking anti-linearity resolves {\bf type-1} degeneracies, indicating complex deformation of $(G\hbar)$ as an ideal choice. Compatibility with the KSW criterion is analyzed after symmetry breaking. Complex deformation of $(G\hbar)$ modifies the KSW criterion, imposing a strong constraint on the deformation if No-boundary geometries are required to be always KSW-allowed.
\end{abstract}
\vspace{5mm}

%%%%%%%%%%%%%%%%%%%%%%%%%%%%%%%%%%%%%%%%%%%%%%%

\maketitle

\tableofcontents

%%%%%%%%%%%%%%%%%%%%%%%%%%%%%%%%%%%%%%%%%%%%%%%

%%%%%%%%%%%%%%%%%%%%%%%%%%%%%%%%%%%%%%%%%%%%%%%
\section{Introduction}
\label{intro}
%%%%%%%%%%%%%%%%%%%%%%%%%%%%%%%%%%%%%%%%%%%%%%%

Symmetries are known to lead to degeneracies in the system, with resolution often coming via symmetry breaking. There are several prime examples, some of which are: spontaneous symmetry breaking via the Higgs mechanism, splitting of energy levels in atoms via adding a symmetry-breaking term, and symmetry breaking via radiative corrections. In some cases external input (for example magnetic field or electric field) is needed to break the symmetry, while in some cases the quantum corrections themselves are sufficient to resolve the degeneracies of the system, a well-known example of this is Lamb-shift and symmetry breaking via radiative corrections. 

In the case of gravity, situations involving degeneracies need to be suitably treated to arrive at unambiguous results. Being a diffeomorphism invariant theory, the gravitational path-integral needs to be suitably gauge-fixed to prevent over-counting of gauge-orbits which otherwise might lead to degenerate configurations. For example, even at classical level, the linearized gravitational field equations need to be gauge-fixed to have well-defined Green's function. At the quantum level, the systematic framework of quantum field theory offers a well-defined platform where the problem can be systematically setup in a Lorentz covariant manner. The gravitational path-integral with metric as the degree of freedom, on a manifold with boundaries, can be written as 
\beq
\label{eq:PI_LI_g}
G[g_\mn] = \int_{{\cal M} + \pt {\cal M}} {\cal D} g_\mn e^{iS[g_\mn]/\hbar} \, ,
\eeq
where $g_\mn$ is the metric and $S[g_\mn]$ is the corresponding action. The path integral is defined on a manifold ${\cal M}$ with boundary $\pt{\cal M}$. This quantity represents summation over all possible metrics (modulo diffeomorphism) respecting specific boundary conditions, with each geometry coming along with a `weight-factor' $e^{iS/\hbar}$. 

However, one often has to deal with infinities while computing transition amplitudes, which need to be properly addressed to define a path-integral. A careful regularization with a follow up of renormalization of parameters systematically achieves this. For gauge theories including gravity, the corresponding path-integral needs a systematic treatment preventing over-counting of gauge-orbits, which is done via Faddeev-Popov procedure \cite{Faddeev:1967fc}. An integration contour needs to be further specified for the relevant convergence of the Lorentzian path-integral, which in flat spacetime QFTs is achieved via Wick's rotation (going from real-time to imaginary time). 

In the case when gravity is dynamical these successes are hard to replicate, as the standard application of QFT methods to Einstein-Hilbert gravity leads to non-renormalizability \cite{tHooft:1974toh, Deser:1974nb, Deser:1974cz, Goroff:1985sz, Goroff:1985th, vandeVen:1991gw} if higher-derivative terms are not included, which comes along their own set of complications like compromised unitarity \cite{Stelle:1976gc}. However, these can be carefully handled if certain conditions get satisfied \cite{Salam:1978fd, Narain:2011gs, Narain:2016sgk, Buccio:2024hys}. In this paper we will work with a minimalistic extension of Einstein-Hilbert gravity given by the action (see \cite{Narain:2021bff, Narain:2022msz} for earlier works investigating the path integral of such gravitational theories)
\bea
\label{eq:act}
S[g_\mn] = \frac{1}{16\pi G} \int {\rm d}^Dx \sqrt{-g}
\biggl[
-2\Lam + R + \al
\biggl( R_{\mu\nu\rho\sg} R^{\mu\nu\rho\sg} - 4 R_\mn R^\mn + R^2 \biggr)
\biggr] \, . 
\eea
This is the Gauss-Bonnet gravity action where $G$ is Newton's gravitational constant, $\Lam$ is the cosmological constant term, $\al$ is the Gauss-Bonnet (GB) coupling and $D$ is spacetime dimensionality. The mass dimensions of various couplings are: $[G] = M^{2-D}$, $[\Lam] = M^2$ and $[\al] = M^{-2}$. In gravitational systems, the standard flat spacetime Wick's rotation doesn't work. Moreover, working with the Euclidean gravitational path-integral brings up the conformal factor problem, where the path integral over the conformal factor is unbounded from below \cite{Gibbons:1978ac}. This issue motivates one to work directly with the Lorentzian path-integral. 

The gravitational action in eq. (\ref{eq:act}) is the second term in the expansion of Lovelock gravity action \cite{Lovelock:1971yv, Lovelock:1972vz, Lanczos:1938sf}, and is a subclass of higher-derivative gravity. Such theories also arise in the low-energy effective action of the heterotic String theory with $\al>0$ \cite{Zwiebach:1985uq, Gross:1986mw, Metsaev:1987zx}. When $D=4$, the dynamical evolution of the metric becomes independent of $\alpha$. Our goal in this paper is to study the gravitational path-integral given in eq. (\ref{eq:PI_LI_g}) with the action of theory given in eq. (\ref{eq:act}), for the boundary choices motivated by previous studies \cite{Feldbrugge:2017kzv, DiTucci:2018fdg, Lehners:2018eeo, Feldbrugge:2017fcc, Feldbrugge:2017mbc, DiTucci:2019bui, Narain:2021bff, Lehners:2021jmv, DiTucci:2020weq, Narain:2022msz, DiTucci:2019dji, Ailiga:2023wzl, Ailiga:2024mmt}.

In the path-integral over all possible metrics, we consider a subclass of metrics respecting spatial homogeneity and isotropy. In $D$ spacetime dimensions and polar coordinates $\{t_p, r, \ta, \cdots \}$ these can be written in the following form 
\beq
\label{eq:frwmet}
{\rm d}s^2 
=- N_p^2(t_p) {\rm d} t_p^2 
+ a^2(t_p) \rho_{ij} {\rm d}x^i \, {\rm d}x^j
= - N_p^2(t_p) {\rm d} t_p^2 
+ a^2(t_p) \left[
\frac{{\rm d}r^2}{1-kr^2} + r^2 {\rm d} \OM_{D-2}^2
\right] \, .
\eeq
It has two unknown time-dependent functions: lapse $N_p(t_p)$ and scale-factor $a(t_p)$, $k=(0, \pm 1)$ is the curvature, and ${\rm d}\OM_{D-2}$ is the metric for the unit sphere in $D-2$ spatial dimensions. Where $t_p $ can be rescaled to $0\leq t_p\leq 1$ without losing generality and the geometries described by eq. \ref{eq:frwmet} have topology $\mathbb{R}\times S^{D-1}$. This is called the \textit{mini-superspace} approximation, where the full path-integral is reduced to a path-integral over the class of metrics stated in eq. (\ref{eq:frwmet}). The full diffeomorphism invariance reduces to the time reparametrization invariance of the coordinate $t_p$.

The full path-integral with metric as the field variable reduces to the following for the class of metric ansatz stated in eq. (\ref{eq:frwmet})
\beq
\label{eq:Gform_sch}
G[{\rm Bd}_f, {\rm Bd}_i]
= \int_{\cal C} {\rm d}N_p
\int_{{\rm Bd}_i}^{{\rm Bd}_f} 
{\cal D} a(t_p) \,\,
e^{i S_{\rm grav}[a, N_p]} \, ,
\eeq
where ${\cal C}$ refers to the contour of integration, and Batalin-Fradkin-Vilkovisky (BFV) formalism \cite{Batalin:1977pb, Feldbrugge:2017kzv} is used to perform the integration over the fermionic ghosts and momenta corresponding to each field leading to the residual path-integral stated in eq. (\ref{eq:Gform_sch}). Gauge-fixing the time-reparameterization invariance via proper-time gauge $N^\prime_p=0$ reduces the path-integral over $N_p(t)$ to an ordinary integral over constant lapse $N_p$ (for more elaborate discussion on BFV quantization process and ghosts, see \cite{Teitelboim:1981ua, Teitelboim:1983fk, Halliwell:1988wc}). The residual quantity gives the transition amplitude for the Universe to evolve from one boundary configuration to another in proper time $N_p$ \cite{Teitelboim:1983fh}.

Our purpose in this paper is to study the transition amplitude given in Eq. (\ref{eq:Gform_sch}) for various boundary choices. For the kind of gravitational actions considered in Eq. (\ref{eq:act}), the path integral over the scale factor is exactly doable. The ordinary integration over the lapse needs to be done carefully in the complex plane. This integral, depending on the boundary choices is sometimes doable exactly via usage of special functions. However, in many cases it can be computed approximately following WKB methods and Picard-Lefschetz techniques \cite{Witten:2010cx, Witten:2010zr, Basar:2013eka, Tanizaki:2014xba}. Picard-Lefschetz method helps in finding a convergent integration contour to which the original integration contour ($\mathcal{C}$) can be smoothly deformed, thereby generalizing the notion of Wick's rotation to curved spacetime, while WKB gives an approximate answer to the integration done along the deformed contour. However, these two methods work unambiguously only when the gravitational system is devoid of any degeneracies, the presence of which leads to the failure of either of them. Hence, one needs to resolve these degeneracies systematically before proceeding further. 

In the gravitational system studied in this paper, we come across two types of degeneracies: one where the flow-lines emanating from neighbouring saddles overlap ({\bf type-1}) and the other where, for certain boundary choices, the saddles merge ({\bf type-2}). Presence of former leads to breakdown of Picard-Lefschetz methods as it is not possible to decide the {\it relevance} of saddles unambiguously which is crucial for finding the deformed integration contour. Presence of later leads to breakdown of WKB. In this paper we investigate these degenerate situations and find ways of resolving them. This also raises one curiosity, and makes one wonder about the cause of such degeneracies and whether there could be any symmetry responsible for them. Hoping that these symmetries when broken lead to resolution of degeneracies systematically. In this paper we study this deeply and find systematic methods of lifting these degeneracies via symmetry breaking.   

It is worth asking how this procedure fit with the Kontsevich-Segal-Witten (KSW)-allowability criterion, which is a diagnostic tool to identify ``allowable'' complex spacetime metrics on which sensible QFTs can be defined. This directly follows from the requirements of convergence of the path-integral \cite{Kontsevich:2021dmb, Witten:2021nzp}. The complex plane gets divided into KSW allowed and disallowed regions, with the saddles of the theory scattered around. It is, however, interesting to note that the No-boundary saddles and standard Lorentzian geometry lie on the boundary of KSW-allowed/disallowed regions. It is crucial to know, how the situation changes once the symmetries leading to degeneracies gets broken. This lead to three scenarious: (1) where the saddles gets modified without affecting the KSW criterion, (2) where the saddles are not affected but the KSW criterion is modified, and (3) where both the saddles and KSW criterion are affected. This paper investigates the first two possibilities and systematically arrives at the relevant scenario based on the situation. 

This paper is organized as follows: In Sec. \ref{rbc_free}, we discuss the Lorentzian Robin path integral in quantum mechanics. Results from these are used in the follow-up section \ref{minisup} which deals with the path integral of the mini-superspace model of Gauss-Bonnet gravity with Robin boundary conditions. In Sec. \ref{trans}, we compute the transition amplitude exactly by performing the lapse $N_c$-integration using Airy-functions. In section \ref{lab:SadRel} we do the saddle analysis to compare and contrast the exact transition amplitude with the one obtained in WKB approximation and Picard-Lefschetz methods. This also helps in finding the deformed integration contour along which the integrand is well-behaved. In this process, we come across degenerate situations whose resolution is mandatory for the implementation of Picard-Lefschetz. Resolving them via artificial {\it defect} is studied in \ref{lab:SadRel}. In Sec. \ref{sec:qc_as_def}, we exploit corrections coming from quantum fluctuations of the scale-factor to resolve the degeneracies. It is seen that unlike artificial {\it defects} they don't resolve all the degeneracies. The residual ones are seen to be lifted via complex deformation of $G\hbar$ and covered in section \ref{complex_Ghbar}. Symmetries responsible for such degenerate situations are explored in section \ref{sec:sym_type1}, where it has been shown how their breaking helps in lifting the degeneracies. Section \ref{sec:KS} analyzes the KSW criterion and its compatibility with the degeneracy resolution and symmetries. In section \ref{Mod_KSW_bound}, we introduce a $\theta$-modified KSW bound and study the allowability of saddles under this modification. Finally, summary of our findings and conclusions are presented in section \ref{sec:conc}.

%%%%%%%%%%%%%%%%%%%%%%%%%%%%%%%%%%%%%%%%%%%%%%%
\section{$1D$-Robin path-integral}
\label{rbc_free}
%%%%%%%%%%%%%%%%%%%%%%%%%%%%%%%%%%%%%%%%%%%%%%%

Let's consider the path integral of a one-dimensional quantum mechanical system whose initial and final boundary conditions are specified \cite{Ailiga:2023wzl}. We will consider a particle moving in a linear potential, which is of interest to our studies in quantum cosmology. Our purpose will be to compute the path integral
\beq
\label{1partPI}
\bar{G}[{\rm Bd_f}, {\rm Bd_i}] =
\int_{\rm Bd_i}^{\rm Bd_f} {\cal D} q(t) \,\, e^{i S_{\rm tot}[q]/\hbar} \, ,
\eeq
where $S_{\rm tot}[q]$ is the action for the one particle system 
\beq
\label{act1part}
S_{\rm tot}[q] = S[q] + S_{\rm bd} = 
\int_0^1 {\rm d}t \left[ \frac{m}{2} \dot{q}^2 - V(q) \right]
+ S_{\rm bd} \, ,
\eeq
$m$ is a $t$-independent parameter, `dot' denotes $t$-derivative, $V(q)$ is the potential, $S_{\rm bd}$ is the surface term added to have a consistent variational problem. Following the time-slicing method, the path integral can be computed from first principles. However, extra care is needed at end points to account for the nontrivial boundary condition. Due to the ease of incorporating boundary choices, the path integral is studied in the {\it Hamiltonian} language. The Hamiltonian of the system is given by
\beq
\label{eq:Ham}
H(p,q) = p \dot{q} - L (q, \dot{q}) 
= \frac{p^2}{2m} + V(q) \, ,
\eeq
where $p = m \dot{q}$. The transition from initial state $|\rm Bd_i\rangle$ to the final one $|\rm Bd_f\rangle$ in the Hamiltonian picture is given by,
\beq
\label{eq:transH}
\bigl\langle {\rm Bd_f}, t=1 \big| {\rm Bd_i}, t=0 \bigr\rangle
= \bigl\langle {\rm Bd_f} \big| e^{-i H(\hat{p}, \hat{q})/\hbar} \big| {\rm Bd_i} \bigr\rangle \, .
\eeq
$H(\hat{p},\hat{q})$ is the Hamiltonian operator corresponding to the classical Hamiltonian in Eq. (\ref {eq:Ham}).
The operators $\hat{p}$ and $\hat{q}$ obey the commutation relation $\big[\hat{q},\hat{p} \big] = i \hbar$. Consider the case when Robin boundary condition (RBC) is specified at the initial time $t=0$ and Dirichlet condition (DBC) is specified at $t=1$, then the path integral becomes 
\beq
\label{eq:Grbc-def}
\bar{G}_{\rm RBC} (q_f, t=1; p_{i}+\bt \,q_{i}, t=0)
= \langle q_f \big| e^{-i H(\hat{p}, \hat{q})/\hbar} \big| p_{i}+\bt \,q_{i} \rangle \, .
\eeq
where the boundary configurations are 
\beq
\label{eq:rbc}
{\rm Bd_f}: q ({t=1})= q_{f}  \quad \text{and} \quad {\rm Bd_i}: p_{i}+\bt \,q_{i} = P_i \, ,
\eeq
with $q_{i}\equiv q(t=0)$ and $p_{i}\equiv p(t=0)$. A detailed evaluation of $G_{\rm RBC}$ has been presented in \cite{Ailiga:2023wzl}, here for completeness we will state the relevant steps and key result which will be needed in later parts of paper. 

We proceed by making a change of variables, and define new canonical variables $(P,Q)$ as 
\beq
\label{eq:newP}
P(t) = p(t) + \bt q(t) 
\quad {\rm and} \quad
Q(t) = q(t) \, ,
\eeq
which preserves the commutation relation thereby leading to a trivial Jacobian of transformation. The Hamiltonian however acquires a new form $H(\hat{p},\hat{q})=H(\hat{P} - \bt \hat{Q}, \hat{Q}) \equiv H_\bt(\hat{P}, \hat{Q})$ 
(where subscript `$\bt$' symbolizes $\bt$-dependence of the Hamiltonian). Performing the Fourier transform of the initial state as  
\beq
\label{PiF}
\big| P_i \rangle =
\int_{-\infty}^{\infty} {\rm d} Q_0 \, \, e^{i P_i Q_0/\hbar} \,\, \big| Q_0 \rangle \, ,
\eeq
the Robin transition amplitude becomes the following 
\beq
\label{Grbc_four}
\bar{G}_{\rm RBC} (q_f, t=1; p_{i}+\bt \,q_{i}, t=0)
= \int_{-\infty}^{\infty} {\rm d} Q_0 \, \, e^{i P_i Q_0/\hbar}
\langle Q_f \big\lvert e^{-i H_\bt(\hat{P},\hat{Q})/\hbar}\big\rvert Q_0 \rangle \, .
\eeq
where $\langle Q_f \big\lvert e^{-i H_\bt(\hat{P},\hat{Q})/\hbar}\big\rvert Q_0 \rangle$ is a transition amplitude with Dirichlet boundary conditions at the two endpoints for the transformed Hamiltonian $H_\bt(P,Q)$. As shown in \cite{Ailiga:2023wzl}, the integral in eq. (\ref{Grbc_four}) converges for arbitrary complex $P_i$ but requires $Re(-i\beta)<0$ for real $m$. We will consider only real $m$ as it is relevant in the context of ordinary quantum mechanics (where $m$ plays the role of mass) and as well as later in the minisuperspace cosmology (where it will play the role of ``lapse"). In the case of free particle 
$H^{\rm free}_\bt (\hat{P},\hat{Q}) = (\hat{P}-\bt \hat{Q})^2/2m $, the Robin path integral is given by \cite{Ailiga:2023wzl}
\beq
\label{Grbc_free_exp}
\bar{G}^{\rm free}_{\rm RBC} (Q_f, t=1; P_i, t=0)
= \sqrt{\frac{m}{m-\bt}}
\exp\biggl[
\frac{i}{2\hbar}
\left\{
\bt Q_f^2
- \frac{\bt m Q_f^2 + P_i^2 - 2 m P_i Q_f}{m-\bt} 
\right\}
\biggr]
\, ,
\eeq
which in limit $\bt \to0$ gives the expression for the 
Neumann boundary condition. In the case when interactions are present, the Hamiltonian gets amended by potential term $V(Q)$: $H_\bt (\hat{P},\hat{Q}) = (\hat{P}-\bt \hat{Q})^2/2m + V(\hat{Q})$. In this case one can follow the steps given in \cite{Ailiga:2023wzl} to find that for linear potential $V(Q) = \lam Q$ one has
\bea
\label{Grbc_V_exp}
\bar{G}_{\rm RBC} (Q_f, t=1; P_i, t=0)
=&& \bar{G}^{\rm free}_{\rm RBC} (Q_f, t=1; P_i, t=0)
\notag \\
&& \times
\exp\left[
\frac{i\lam \left\{
\bt q_f +(P_i - 2 m q_f)
\right\}}{2(m-\bt)\hbar} 
- \frac{i\lam^2 (4m-\bt)}{24m(m-\bt)\hbar}
\right] \, .
\eea
In the limit $\bt\to0$ Eq. (\ref{Grbc_V_exp}) reduces to the 
NBC expression for the linear potential case. The Robin path integral for interacting theory becomes a product of the Robin path integral for free theory and a coupling-dependent part. In the following, we will make use of this relation in the study of the gravitational path integral in mini-superspace approximation.

%%%%%%%%%%%%%%%%%%%%%%%%%%%%%%%%%%%%%%%%%%%%%%%
\section{Robin Mini-superspace}
\label{minisup}
%%%%%%%%%%%%%%%%%%%%%%%%%%%%%%%%%%%%%%%%%%%%%%%

The conformally flat FLRW metric given in Eq. (\ref{eq:frwmet}) has vanishing Weyl-tensor ($C_{\mu\nu\rho\sg} =0$). The non-zero 
entries of the Riemann tensor are 
\cite{Deruelle:1989fj,Tangherlini:1963bw,Tangherlini:1986bw}
\bea
\label{eq:riemann}
R_{0i0j} &=& - \left(\frac{a^{\prime\prime}}{a} - \frac{a^\prime N_p^\prime}{a N_p} \right) \rho_{ij} \, , 
\notag \\
R_{ijkl} &=& \left(\frac{k}{a^2} + \frac{a^{\prime2}}{N_p^2 a^2} \right)
\left(\rho_{ik} \rho_{jl} - \rho_{il} \rho_{jk} \right) \, .
\eea
Here $({}^\prime)$ denotes the derivative with respect to $t_p$. On plugging the FLRW metric of Eq. (\ref{eq:frwmet}) in the gravitational action stated in Eq. (\ref{eq:act}) and utilizing the relations mentioned in Eq. (\ref{eq:riemann}), action for scale-factor $a(t_p)$ and lapse $N_p(t_p)$ in $D$-dimensions becomes 
\bea
\label{eq:midSact}
&&
S[a,N_{p}] = \frac{V_{D-1}}{16 \pi G} \int {\rm d}t_p
\biggl[
\frac{a^{D-3}}{N_p^2} \biggl\{
(D-1)(D-2) k N_p^3 - 2 \Lam a^2 N_p^3 - 2 (D-1) a a^\prime N_p^\prime
\notag \\
&&
+ (D-1)(D-2) a^{\prime2} N_p + 2 (D-1) N_p a a^{\prime\prime}
\biggr\}
+ (D-1)(D-2)(D-3) \al\biggl\{
\frac{a^{D-5}(D-4)}{N_p^3} 
\notag\\
&&
\times (kN_p^2 + a^{\prime2})^2 
+ 4 a^{D-4}\frac{{\rm d}}{{\rm d}t_p} 
\left(
\frac{k a^\prime}{N_p} + \frac{a^{\prime 3}}{3N_p^3}
\right)
\biggr\}
\biggr] \, .
\eea
Here, $V_{D-1}$ is the volume of the $D-1$ dimensional spatial hyper-surface which is given by,
\beq
\label{eq:volDm1}
V_{D-1} = \frac{\G(1/2)}{\G(D/2)} \left(\frac{\pi}{k}\right)^{(D-1)/2} \, .
\eeq
In $D=4$, terms coming from the Gauss-Bonnet sector become total-derivative, thereby giving the $4D$ mini-superspace gravitational action as 
\beq
\label{eq:mini_sup_d4}
S[a,N_{p}] = \frac{V_{3}}{16 \pi G} \int {\rm d}t_p
\biggl[
6k a N_p - 2 \Lam a^3 N_p - \frac{6 a^2 a^\prime N_p^\prime}{N_p}
+ \frac{6 a a^{\prime 2}}{N_p} + \frac{6 a^{\prime\prime} a^2}{N_p}
+ 24 \al \frac{{\rm d}}{{\rm d}t_p} 
\left(
\frac{k a^\prime}{N_p} + \frac{a^{\prime 3}}{3N_p^3}
\right)
\biggr] \, .
\eeq
The mini-superspace action can be cast into a more appealing form by rescaling of lapse $N_p$ and scale factor $N_p(t_p) {\rm d} t_p = (N(t)/a(t)) {\rm d} t$ and $q(t) = a^2(t)$. This transformation re-expresses the original FLRW metric in Eq. (\ref{eq:frwmet}) as
\beq
\label{eq:frwmet_changed}
{\rm d}s^2 = - \frac{N^2}{q(t)} {\rm d} t^2 
+ q(t) \left[
\frac{{\rm d}r^2}{1-kr^2} + r^2 {\rm d} \OM_{D-2}^2
\right] \, .
\eeq
The mini-superspace action in the new variables acquires the following simple form ($D=4$)
\bea
\label{eq:Sact_frw_simp}
S[q,N] = \frac{V_3}{16 \pi G} \int_0^1 {\rm d}t \biggl[
(6 k - 2\Lam q) N + \frac{3 \dot{q}^2}{2N}
+ 3q \frac{{\rm d}}{{\rm d} t} \left(\frac{\dot{q}}{N} \right)
+ 24 \al \frac{{\rm d}}{{\rm d} t} \left(
\frac{k\dot{q}}{2N} + \frac{\dot{q}^3}{24 N^3} 
\right)
\biggr] \, .
\eea
Here $(\dot{})$ represent time $t$ derivative. One can do integration by parts to recast the action in a more recognizable form along with some surface terms
\bea
\label{eq:Sact_frw_simp_inp}
S[q,N] &&
=\frac{V_3}{16 \pi G} \int_0^1 {\rm d}t \biggl[
(6 k - 2\Lam q) N - \frac{3 \dot{q}^2}{2N} \biggr] 
+ \frac{V_3}{16 \pi G} \biggl[
\frac{3q_f \dot{q}_f}{N} - \frac{3q_i \dot{q}_i}{N}
\notag \\
&&
+ 24 \al \left(
\frac{k\dot{q_f}}{2N} + \frac{\dot{q_f}^3}{24 N^3} 
- 
\frac{k\dot{q_i}}{2N} - \frac{\dot{q_i}^3}{24 N^3} 
\right)
\biggr]
\, .
\eea
The collection of surface terms consists of two sets: one coming from EH-part of gravitational action while the other is from Gauss-Bonnet sector. The bulk action resembles an action of one-particle system in a linear potential. We will be using this form of mini-superspace action throughout the paper and in the computation of the path-integral. 

To setup the variational problem consistently, we write 
\beq
\label{eq:qfluc}
q(t) = \bar{q}(t) + \ep \de q(t) \, ,
\eeq
where $\bar{q}(t)$ is some background and $\de q(t)$ is the fluctuation around this, the parameter $\ep$ is placed to keep track of the of the order of fluctuation terms. In the ADM gauge $\dot{N}=0$ ($N(t) = N_c$ (constant)), when the decomposition in Eq. (\ref{eq:qfluc}) is plugged in the action Eq. (\ref{eq:Sact_frw_simp}), and subsequently expanded in powers to $\ep$, then at first order in $\ep$ we get the following set of terms 
\beq
\label{eq:Sexp_qvar}
\de S = \frac{\ep}{2} \int_{0}^{1} {\rm d}t \biggl[
\left(-2 \Lam N_c + \frac{3 \ddot{\bar{q}}}{N_c} \right) \de q
+ \frac{3}{N_c} \frac{{\rm d}}{{\rm d} t} \left(\bar{q} \de \dot{q} \right)
+ 24 \al \frac{{\rm d}}{{\rm d} t} \left\{ 
\left(\frac{k}{2N_c} + \frac{\dot{\bar{q}}^2}{8N_c^3} \right) \de \dot{q} \right\}
\biggr] \, .
\eeq
The two total time-derivative pieces contribute at the boundaries. For variational consistency these boundary terms must either vanish due to a specific choice of boundary conditions or need to be appropriately canceled by supplementing the action with suitable surface terms. The terms proportional to $\de q$ on the other hand will lead to the equation of motion
\beq
\label{eq:dyn_q_eq}
\ddot{\bar{q}} = \frac{2}{3} \Lam N_c^2 \, .
\eeq
The general solution of it involving two arbitrary constants $c_{1,2}$ is given by
\beq
\label{eq:qsol_gen}
\bar{q}(t) = \frac{\Lam N_c^2}{3} t^2 + c_1 t + c_2 \, .
\eeq
The constants are fixed once the boundary conditions are specified. We will consider the case when Robin boundary condition is imposed at $t=0$ and Dirichlet boundary condition is imposed at $t=1$. Robin condition we take is the linear combination of the scale factor $q_i$ and the corresponding conjugate momentum $\pi_i=\pt {\cal L}/\pt \dot{q} = - 3\dot{q}/2N_c$ at $t=0$. The final scale factor $q_f$ is fixed at $t=1$. This means 
\beq
\label{eq:robin_cond}
\pi_i + \bt q_i = P_i= {\rm fixed} \hspace{3mm} \&  \hspace{3mm}
q_f = {\rm fixed} 
\hspace{5mm} 
\Rightarrow 
\hspace{5mm}
\de P_i = 0  \hspace{3mm} \&  \hspace{3mm}
\de q_f = 0 \, .
\eeq
These boundary configurations are chosen as they lead to a stable universe where metric perturbations are well-behaved and suppressed \cite{DiTucci:2019bui, Narain:2021bff, DiTucci:2020weq, Lehners:2021jmv, Ailiga:2024nkz}. In this case the set of boundary terms generated during the variation of action is given by the following 
\beq
\label{eq:Sbd_mom_RBC}
\biggl. S_{\rm bdy} \biggr|_{\rm RBC} = - \ep\biggl[
\bar{q}_f \de \pi_f + \bt \bar{q}_i \de q_i 
+4 \al \left\{ 
k\de \pi_f + \frac{\bar{\pi}_f^2\de \pi_f}{9} 
+ \bt\left(k\de q_i + \frac{(\bar{P}_i- \bt \bar{q}_i)^2\de q_i}{9} \right)\right\}
\biggr] \, .
\eeq
These need to be suitably canceled for variational consistency by supplementing the original action with appropriate surface terms whose variation in turn will precisely cancel these. The surface term that suitably does the job is given by
\bea
\label{eq:Sact_surf_RBC}
\biggl. S_{\rm surface} \biggr|_{\rm RBC} 
&=& q_f \pi_f 
+ 4 \al \left(k\pi_f +\frac{\pi_f^3}{27} \right) 
+ \frac{\bt}{2} (q_f^2 + q_i^2)
\notag \\
&&
+ 4 \al \bt \biggl[
k q_i + \frac{q_i}{9} \left(
P_i^2 - \bt P_i q_i + \frac{\bt^2 q_i^2}{3} 
\right)
\biggr]\, .
\eea
In the case when $\bt=0$ (Neumann boundary case), this precisely gives the surface terms for the Neumann boundary problem. The surface term $\bt q_f^2/2$ in principle is not needed as $q_f$ is fixed at $t=1$, hence its variation vanishes. However, it is added for later convenience in the computation of Robin path integral of the gravitational system. The surface term $\bt q_i^2/2$ in Eq. (\ref {eq:Sact_surf_RBC}) is coming from the Einstein-Hilbert gravity part. The term proportional to $\al \bt$ corresponds to surface terms that need to be supplemented for the Gauss-Bonnet gravity part \cite{Ailiga:2023wzl}. This set of surface terms when added to the action in Eq. (\ref{eq:Sact_frw_simp_inp}), leads to (setting $V_3 = 8\pi G$) 
\beq
\label{eq:Sact_frw_simp_RBC}
S_{\rm tot}[q,N_c]
=\frac{1}{2} \int_0^1 {\rm d}t \biggl[
(6 k - 2\Lam q) N_c - \frac{3 \dot{q}^2}{2N_c} \biggr] 
+ q_i P_i + \frac{\bt}{2} (q_f^2 - q_i^2)
+ 4 \al \left(k P_i +\frac{P_i^3}{27} \right)
\, ,
\eeq
which is variationally consistent for the choice of boundary condition mentioned in Eq. (\ref{eq:robin_cond}).
It should be specified that these are canonical boundary terms and don't always have a covariant analogue. In the case of Einstein-Hilbert gravity, the covariant boundary action for the Robin boundary condition has been recently computed in \cite{Deruelle:2009zk, Krishnan:2017bte, Ailiga:2024nkz, Brizuela:2023vmb}. The covariant boundary term for the Gauss-Bonnet sector with Robin boundary condition is not known yet. In this paper, we won't need the covariant form of the boundary action.

The constants $c_{1,2}$ appearing in the solution to the equation of motion given in Eq. (\ref{eq:qsol_gen}) can be determined, leading to 
\beq
\label{eq:qsol_RBC}
\bar{q}(t) = \frac{\Lam N_c^2}{3} t^2 
+ \frac{P_i}{\bt}
+ \left(1 + \frac{3}{2 \bt N_c} \right)^{-1}
 \left(t + \frac{3}{2 \bt N_c} \right)
 \left(q_f - \frac{P_i}{\bt} - \frac{\Lam N_c^2}{3} \right) \, .
\eeq
The on-shell action for the Robin case can also be determined by substituting the solution of the equation of motion and $\bar{q}_i = \bar{q}(t=0)$ into the action given in Eq. (\ref{eq:Sact_frw_simp_RBC}). This is given by
\bea
\label{eq:stot_onsh_rbc}
\bar{S}_{\rm tot}^{\rm on-shell}[ N_c] &=& \frac{1}{18(3+ 2 N_c \bt)} \biggl[
\bt \Lam^2 N_c^4 + 6 \Lam^2 N_c^3 +
N_c^2 \{108 \bt  k-18 \Lam  (P_i+\bt q_f)\}
\notag \\
&&
+18 N_c \left\{9
k+P_i^2+ q_f \left(\bt^2 q_f-3 \Lambda \right)\right\}
+ 54 P_i q_f
\biggr] 
+ 4 \al \left(k P_i +\frac{P_i^3}{27} \right) .
\eea
This is the action for the lapse $N_c$ for the case of Robin boundary condition at $t=0$ and Dirichlet boundary condition at $t=1$.

%======================================

%%%%%%%%%%%%%%%%%%%%%%%%%%%%%%%%%%%%%%%%%%%%%%%
\section{Transition Amplitude}
\label{trans}
%%%%%%%%%%%%%%%%%%%%%%%%%%%%%%%%%%%%%%%%%%%%%%%

After having computed the on-shell action for the Robin mini-superspace, we move forward to study the path integral, which is the transition amplitude from one $3$-hypersurface to another. The quantity of our interest in the minisuperspace approximation is as follows
\bea
\label{eq:TA_exp}
G[{\rm{Bd_f},\rm{Bd_i}}] = \int_{\cal C} {\rm{d}}N_c \int_{{\rm Bd}_i}^{{\rm Bd}_f} \, \mathcal{D}q(t) \, \exp{\Bl(\frac{i}{\hbar} S_{\rm tot}[q,N_c]\Br)}\, ,
\eea
where as before ${\rm Bd}_i$ and  ${\rm Bd}_f$ refers to initial and final boundary configurations respectively, and $S_{\rm tot}$ refers to the Robin mini-superspace action mentioned in eq. (\ref{eq:Sact_frw_simp_RBC}). We first focus our interest in computing the $q$-path integral mentioned in Eq. (\ref{eq:TA_exp}) for the relevant boundary choices. To achieve this we compare the RBC action with the quantum mechanical RBC problem discussed in section \ref{rbc_free}. This shows that the following substitution needs to be made 
\begin{gather}
\label{eq:subsNBC}
m \to -\frac{3}{2 N_c} \, , 
\hspace{1cm}
V(q) = \lam q \to \Lam N_c q 
\hspace{3mm} \Rightarrow \hspace{3mm} \lam \to \Lam N_c \, ,
\hspace{1cm}
p_i \to \pi_i \, .
\end{gather}
This eventually leads to 
\bea
\label{eq:RBC_qint}
\int_{{\rm Bd}_i}^{{\rm Bd}_f} 
&&
{\cal D} q(t) \,\, 
\exp \left(\frac{i}{\hbar} S_{\rm tot}[q, N_c] \right)
= \exp\left[\frac{i}{\hbar} \left\{3k N_c  
+ 4 \al \left(k P_i + \frac{P_i^3}{27}\right) \right\}\right]
\notag \\
&& \times \int_{{\rm Bd}_i}^{{\rm Bd}_f} {\cal D} q(t)
\exp\left[
\frac{i}{2\hbar} \int_0^1 {\rm d}t \biggl\{
- 2\Lam q N_c - \frac{3 \dot{q}^2}{2N_c} \biggr\}
+ \frac{i}{\hbar}q_i P_i + \frac{\bt}{2\hbar}(q_f^2 - q_i^2)
\right] 
\notag \\
&&
= \exp\left[\frac{i}{\hbar} \left\{3k N_c  
+ 4 \al \left(k P_i + \frac{P_i^3}{27}\right) \right\}\right] 
\bar{G}_{\rm RBC}[q_f, t=1; P_i, t=0]
\, ,
\eea
where
\bea
\label{eq:gbar_rbc_ms_qint}
\bar{G}_{\rm RBC}[q_f, t=1; P_i, t=0]
= && {\cal N} \int_{{\rm Bd}_i}^{{\rm Bd}_f} {\cal D} q(t)
\exp\biggl[
\frac{i}{2\hbar} \int_0^1 {\rm d}t \biggl\{
- 2\Lam q N_c - \frac{3 \dot{q}^2}{2N_c} \biggr\}
\notag \\
&&
+ \frac{i}{\hbar}q_i P_i + \frac{\bt (q_f^2-q_i^2)}{2\hbar} 
\biggr] \,.
\eea
The numerical constant ${\cal N}$ has been introduced as an adjustable prefactor to be fixed appropriately to adhere to the requirements of classicality, where the amplitude is expected to be classical with the expansion of the Universe in the WKB sense \cite{Feldbrugge:2017kzv, DiTucci:2019bui}. By making use of Eq. (\ref{Grbc_free_exp}), (\ref{Grbc_V_exp}) and (\ref{eq:subsNBC}), one immediately obtains 
\bea
\label{eq:gbar_RBC_ms}
&&
\int_{{\rm Bd}_i}^{{\rm Bd}_f} {\cal D} q(t) 
\exp \left(\frac{i}{\hbar} S_{\rm tot}[q, N_c] \right)
= \frac{1}{\sqrt{1+2\beta N_c/3}} \exp\Bl(\frac{i}{\hbar} \bar{S}_{\rm tot}^{\rm on-shell}[ N_c] \Br) \, ,
\eea
where $\bar{S}_{\rm tot}^{\rm on-shell}[ N_c]$ is given  in Eq. (\ref{eq:stot_onsh_rbc}).
To obtain the transition amplitude, lapse $N_c$-integration needs to be done with the integrand given by RHS of Eq. (\ref{eq:gbar_RBC_ms}). The transition amplitude is given by
\beq
\label{eq:TA_lps_int}
G[{\rm{Bd_f},\rm{Bd_i}}] = \int_{\cal C}  \, \frac{{\rm{d}}N_c}{\sqrt{1+2\beta N_c/3}} \exp\Bl(\frac{i}{\hbar} \bar{S}_{\rm tot}^{\rm on-shell}[ N_c] \Br)\, ,
\eeq
where ${\cal C}$ is the original integration contour, which we take to be along the real line from $-\infty$ to $+\infty$. This integration needs to be performed by analytically continuing $N_c$ into the complex plane. Note the integrand is singular at $N_c = - 3/2\beta$. To avoid the singularity appearing on the real integration contour, we avoid taking the $\beta$-parameter to be real.

The lapse $N_c$-integration can be done exactly following a set of transformations. We first do a rescaling of the lapse-$N_c$ along with a constant shift
\beq
\label{eq:lap_redef}
 N_c = \frac{3}{2\beta}\left(\bar{N}_c-1\right) \implies dN_c \rightarrow \frac{3}{2\beta}d\bar{N}_c\, .
\eeq
This transformation converts our total on-shell action mentioned in eq. (\ref{eq:stot_onsh_rbc}) to following 
\bea
\label{eq:on_shell_Act_nb}
\mathcal{\bar{S}}^{\rm on-shell}_{\rm tot}[\bar{N}_c] = && \frac{3\Lambda^2}{32\beta^{3}}\bar{N}_c^{3} - \Bl(\frac{9\Lambda^2}{16\beta^{3}} + \frac{3\Lambda P_i}{4\beta^{2}} + \frac{3\Lambda q_{f} - 18 k}{4\beta}\Br)\bar{N}_c + 4\al \biggl(k P_i + \frac{P_i^3}{27} \biggr)
\notag \\
&&
-\frac{1}{2\bar{N}_c\beta}\left(\frac{3\Lambda}{4\beta} + 
P_i - \beta q_{f}\right)^{2} + \left(\frac{3\Lambda^{2}}{4\beta^{3}} + \frac{3\Lambda P_i}{2\beta^{2}} + \frac{P_i^2 - 9k}{2\beta}\right) \, .
\eea
In the language of $\bar{N}_c$, the transition amplitude acquires the following form
\beq
\label{eq:TA_lps_int_N_bar}
G[{\rm{Bd_f},\rm{Bd_i}}] = \int_{\bar{\bf C}}  \,{\rm{d}}\bar{N}_c\, \frac{3}{2\beta \sqrt{\bar{N}_c}} \exp\Bl(\frac{i}{\hbar} \mathcal{\bar{S}}^{\rm on-shell}_{\rm tot}[\bar{N}_c] \Br)\, ,
\eeq
where the contour $\bar{\bf C}$  runs from $1-\infty e^{i\omega}$ to $1+\infty e^{i\omega}$, with $\omega$ being phase corresponding to complexified $\beta=|\beta|e^{i\omega}$. The above integral can be solved analytically with the use of Hubbard-Stratonovich (HS) transformation. Using it we can write
\bea
\label{eq:HS_trans}
\frac{1}{\sqrt{\bar{N}_c}} 
&&
\exp\Bl[- \frac{i}{2\hbar\bar{N}_c\beta}
\Bl(\frac{3\Lambda}{4\beta} + P_i - \beta q_{f}
\Br)^2\Br] 
\notag \\
&&
= \sqrt{\frac{\beta\epsilon}{2\pi i\hbar}} \int_{-\infty}^{\infty}
{\rm d}\xi 
\exp\Bl[\frac{i}{\hbar}\Bl\{\epsilon\frac{\bar{N}_c\beta}{2}\xi^{2} 
+ \sqrt{\epsilon}\Bl(\frac{3\Lambda}{4\beta} + P_i - \beta q_{f} \Br)\xi\Br\}
\Br] \, ,
\eea
where $Re(i\epsilon\bar{N}_c\beta)<0$ is required to ensure the convergence. Without loss of any generality, we take $\epsilon=e^{i\omega_{\ep}}$. To ensure convergence of the above integral at asymptotic values of $\bar{N}_c$, one needs to satisfy the following conditions: 
\bea
\label{eq:cond_conv}
&& 
\cos(\omega+\pi/2+\omega_\ep)-\infty \, \cos(2\omega +\pi/2+\omega_\ep)<0 \, ,
\notag \\
&&
\cos(\omega+\pi/2+\omega_\ep)+\infty \, \cos(2\omega +\pi/2+\omega_\ep)<0 \, ,
\eea
for a given pair of $(\omega, \omega_\ep)$. These conditions are satisfied if and only if the following requirement is met
\begin{equation}
\label{cond}
\cos(2\omega +\pi/2+\omega_\ep)=0 \, ,
\hspace{5mm}
\cos(\omega+\pi/2+\omega_\ep)< 0.
\end{equation} 
The above set of equations has many solutions in the $\omega-\omega_\ep$ plane. Some possibilities are: $(-\pi/2,\pi)$, $(\pi/2, 0)$. It should be highlighted that choice of $\omega$ and $\omega_\ep$ is not arbitrary. For a given $\omega$, $\omega_\ep$ is fixed in such a way to ensure convergence. 

With the use of the Hubbard-Stratonovich transformation, the transition amplitude becomes the following
\bea
\label{eq:TA_exp_2}
G[{\rm{Bd_f},\rm{Bd_i}}]  = && \frac{3\sqrt{\epsilon}}{ 2 \sqrt{2 i \pi\hbar\beta}} \exp\left[\frac{i}{\hbar}\left(\frac{3\Lambda^{2}}{4\beta^{3}} + \frac{3\Lambda P_i}{2\beta^{2}} + \frac{P_i^2 - 9k}{2\beta}\right)\right] \exp{\left[ 
 4 \al \left(k P_i + \frac{P_i^3}{27}\right) \right]}  \notag \\
&&\int_{\bar{\bf C}} {\rm d}\bar{N}_c {\rm d}\xi \exp\Bigg[\frac{i}{\hbar}\Bigg\{\frac{3\Lambda^2}{32\beta^{3}}\bar{N}_c^{3} - \Bl(\frac{9\Lambda^2}{16\beta^{3}} + \frac{3\Lambda P_i}{4\beta^{2}} + \frac{3\Lambda q_{f} - 18 k}{4\beta}\Br)\bar{N}_c\notag \\
&& + \epsilon \frac{\bar{N}_c\beta}{2}\xi^{2} + \sqrt{\epsilon}\left(\frac{3\Lambda}{4\beta} + P_i - \beta q_{f} \right)\xi\Bigg\}\Bigg]\, .
\eea
Due to the appearance of the mixing term $\bar{N}_c\beta\xi^{2}/2$, the above integrations over $\bar{N}_c$ and $\xi$ are not trivial. However, they can be decoupled and can be done with the following transformation
\bea
\label{eq:var_trans}
&&
\bar{N}_c  =  2\beta\left(\frac{\hbar}{9\Lambda^2}\right)^{1/3} \left(\eta_{1} + \eta_{2}\right) \, ,
\hspace{5mm}
\xi =\frac{(3\Lambda\hbar)^{1/3}}{2 \beta \sqrt{\epsilon}}\left(-\eta_{1} +\eta_{2}\right) \, ,
\notag \\
&&
{\rm d}\bar{N} {\rm d}\xi = \frac{2\left(\hbar^2/3\Lambda\right)^{1/3}}{\sqrt{\epsilon}} {\rm d}\eta_{1} {\rm d}\eta_{2} \, .
\eea
In terms of new integration variables, decoupling is achieved thereby allowing the transition amplitude to be expressed in the following manner
\bea
\label{eq:TA_new_var}
G[{\rm{Bd_f},\rm{Bd_i}}] = && 3\pi J  \sqrt{\frac{ 2 \pi}{i\hbar\beta}} \exp\left[\frac{i}{\hbar}\left(\frac{3\Lambda^{2}}{4\beta^{3}} + \frac{3\Lambda P_i}{2\beta^{2}} + \frac{P_i^2 - 9k}{2\beta}\right)\right] \notag \\
 && \times \exp{\left[ 
 4 \al \left(k P_i + \frac{P_i^3}{27}\right) \right]} \Psi_{1}(P_i,\beta)  \Psi_{2}(q_{f}) \, .
\eea
The special functions $\Psi_{1}(P_i,\beta)$ and $\Psi_{2}(q_{f})$ introduced are given by 
\bea
\label{airy}
&&
\Psi_{1} = \int_{\mathcal{C}_1} \frac{d\eta_{1}}{2\pi} \exp\left[i\left\{\frac{\eta_1^3}{3}+\left(\frac{9}{\hbar\Lambda}\right)^{2/3}\left(k - \frac{\Lambda P_i}{3\beta} - \frac{\Lambda^2}{4\beta^2}\right)\eta_{1}\right\}\right]
\notag \\
&&
\Psi_{2} = \int_{\mathcal{C}_2} \frac{d\eta_{2}}{2\pi} \exp\left[i\left\{\frac{\eta_2^3}{3} + \left(\frac{\sqrt{3}}{\hbar\Lambda}\right)^{2/3} \left(3k - \Lambda q_{f}\right) \eta_{2}\right\}\right] \, .
\eea
Here, $\mathcal{C}_1$ and $\mathcal{C}_2$ denote the integration contour in the complex plane of $\eta_1$ and $\eta_2$, respectively. The transformations mentioned in Eq. (\ref{eq:var_trans}) can be inverted to express $\eta_1$ and $\eta_2$ in terms of $N_c$ and $\xi$. This is given by
\beq 
\label{eq:inv_trans}
\eta_{1} = \frac{1}{(3\Lam \hbar)^{1/3}}\biggl[\frac{3\Lam}{4\beta} + \frac{\Lam N_c}{2} - \beta \sqrt{\ep} \xi\biggr]\, , 
\hspace{5mm}
\eta_{2} = \frac{1}{(3\Lam \hbar)^{1/3}}\biggl[\frac{3\Lam}{4\beta} + \frac{\Lam N_c}{2} + \beta \sqrt{\ep} \xi\biggr] \, .
\eeq
We note that if $\beta\sqrt{\epsilon}$ is real with $-\pi<\omega<0$, the complex contours are simply shifts of the real line along the positive imaginary direction. In this special case, the above integrals in Eq. (\ref{airy}) can be cast into Airy-integral with $\Psi_{1}$ entirely dependent on the initial parameters $P_i$, $\beta$, and $\Psi_{2}$ is dependent only on the final parameter $q_{f}$. On restoring the factors $V_{3}$ and $8\pi G$, we obtain the exact expression of transition amplitude
\bea
\label{eq:transAmp_rbc_exact}
G[{\rm{Bd_f},\rm{Bd_i}}] 
= &&
2\pi\sqrt{\frac{ 2 \pi \hbar}{i\beta}} \left(\frac{9}{\Lambda\hbar}\right)^{1/3}\left(\frac{8\pi G}{V_{3}}\right)^{1/6} \exp\left[\frac{iV_{3}}{8\pi G\hbar}\left\{\frac{P_i^2 - 9k}{2\beta} + \frac{3\Lambda P_i}{2\beta^2} + \frac{3\Lambda^2}{4\beta^3}\right\}\right] \notag \\
&& \times \exp{\left[ \frac{V_3}{8\pi G}\frac{4 i \al}{\hbar} \left(k P_i + \frac{P_i^3}{27}\right) \right]}
\mathcal{A}i\left[\left(\frac{9V_{3}}{8\pi G\hbar\Lambda}\right)^{2/3}\left( k -\frac{\Lambda P_i}{3\beta}-\frac{\Lambda^2}{4\beta^2}\right)\right] 
\notag \\
&& 
\times \mathcal{A}i\left[\left(\frac{\sqrt{3}V_{3}}{8\pi G\hbar\Lambda}\right)^{2/3}(3k - \Lambda q_{f})\right],
\eea
which is valid for $\beta=\lvert \beta \rvert e^{i\omega}$, $-\pi<\omega<0$. 
Note that for these values of $\beta$, the integrand in Eq. \ref{eq:TA_lps_int} is an entire function in the upper half of the complex $N_c$-plane. It allows the original integral contour to be deformable to any contour lying in the upper half-plane without encountering any singularities. The result obtained here agrees with those derived using other methods \cite{Ailiga:2023wzl,Ailiga:2024mmt}. Moreover, if one takes $\omega=-\pi/2$ along with an imaginary $P_i$, it is noticed that the above expression becomes purely real. Surprisingly, these are the only relevant physical ranges of $\beta$ and $P_i$, as obtained from saddle point analysis in the next section. In the following we will work with $\hbar=1$ conventions, where $V_3 =2\pi^2 $ is the volume of the spatial hypersurface ($3$-sphere).

%%%%%%%%%%%%%%%%%%%%%%%%%%%%%%%%%%%%%%%%%%%%%%%
\section{$N_c$-integral, Saddles and Artificial Defects}
\label{lab:SadRel}
%%%%%%%%%%%%%%%%%%%%%%%%%%%%%%%%%%%%%%%%%%%%%%%

Having obtained an exact expression for the transition amplitude, it is crucial to understand its structure and behaviour as a function of the universe size. This is achieved by studying the structure of saddle geometries contributing to the path integral and their relevance. Such studies give a semiclassical understanding of the path integral. Schematically, one can express the lapse-integration in a generic case as follows
\beq
\label{eq:sch_form}
G[{\rm{Bd_f},\rm{Bd_i}}] = \int_{\cal C} \, {\rm d} N_c
\,\, \bl[ \D(N_c) \br]^{-1/2}
\, \exp \bl[i \bar{S}_{\rm tot}^{\rm on-shell}[N_c]/\hbar \br] \, ,
\eeq
where 
\beq
\label{eq:DelNc_rbc}
\D(N_c) \Br \rvert_{\rm RBC}=
1+\frac{2\beta N_c}{3}\, ,
\hspace{5mm}
\beta=-i\frac{\Lam x}{2\sqrt{k}} \, .
\eeq
and $\bar{S}_{\rm tot}^{\rm on-shell}[N_c]$ is the on shell action for the lapse $(N_c)$ which for the case of Robin boundary condition is mentioned in Eq. (\ref{eq:stot_onsh_rbc}). The contour integral in Eq. (\ref{eq:sch_form}) is analyzed by analytically continuing $N_c$ into the complex plane, where one writes the complex exponent as $i \bar{S}_{\rm tot}^{\rm on-shell} [N_c] = h(N_c) + iH (N_c)$. The contour integral in Eq. (\ref{eq:sch_form}) is highly oscillatory if ${\cal C}$ is along the real line. However, it can be made convergent if the contour ${\cal C}$ can be deformed to a suitable contour along which the integrand is well-behaved. Picard-Lefschetz methodology helps in systematically finding the convergent contour of integration \cite{Witten:2010cx, Witten:2010zr, Basar:2013eka, Tanizaki:2014xba, Lehners:2018eeo, Narain:2021bff, Ailiga:2024mmt}. The procedure involves finding the set of saddle points and their corresponding steepest ascent/descent lines. Picard-Lefscehtz methodology then dictates the saddles which become {\it relevant} (as the steepest ascent starting from them hits the original integration contour ${\cal C}$), with the corresponding steepest descent lines (also called {\it Lefschetz thimbles}) constituting the deformed integration contour. Schematically, this can be written as 
\begin{equation}
\label{eq:calC_def_breakup}
\mathcal{C}=\sum_\sigma n_\sigma \mathcal{J}_\sigma,
\hspace{5mm} \,n_\sigma={\rm Int} (\mathbb{D}, {\cal K}_\sg),
\end{equation}
where $\rm Int (. , .)$ counts the intersection between two curves and satisfies $\rm Int (\mathcal{J}_\sigma,\mathcal{K}_{\sigma'})=\delta_{\sigma\sigma'}$. $n_\sigma$ takes values $(0,\pm 1)$, sign of $n_\sigma$ signifies the orientation, $\mathcal{J}_\sigma$ is the steepest descent while $\mathcal{K}_\sigma$ is the steepest ascent. 

Saddle geometries are computed systematically following the action $\bar{S}_{\rm tot}^{\rm on-shell}[N_c]$ mentioned in Eq. (\ref{eq:stot_onsh_rbc}). In the gravitational path-integral, each value of the lapse $N_c$ gives an appropriately weighted contribution, the dominant ones coming from the saddles. The saddles can be determined following
\beq
\label{eq:sad_eq}
{\rm d} \bar{S}_{\rm tot}^{\rm on-shell}[ N_c] / {\rm d} N_c = 0 \, ,
\eeq
where $\bar{S}_{\rm tot}^{\rm on-shell}[ N_c]$ is mentioned in eq. (\ref{eq:stot_onsh_rbc}). On solving it, we get the following set of saddle points 
\beq
\label{eq:sads_pts}
N_{\sg} = -\frac{3i\sqrt{k}}{\Lam x} 
+ \frac{3 c_1}{\Lambda} \sqrt{\frac{\Lambda q_{f}}{3}-k}
+\frac{3 c_1 c_2 \sqrt{k}}{\Lam x}
\sqrt{\frac{2 i x  P_{i}}{3\sqrt{k}} -  (1 + x^2)} \, , 
\hspace{5mm} 
c_1,\, c_2\in\{-1,1\}\,.
\eeq
Once the {\it relevance} of the saddle-points is sorted out and the deformed integration contour is figured, we proceed to compute the contour integral. This can be performed numerically, where the integral is computed along each of the {\it Lefschetz thimble} and summed together appropriately. However, in the saddle-point approximation, it can be computed analytically in some cases. To do so, one expands the lapse $N_c$-action around the saddle giving 
\beq
\label{eq:Act_on_exp_ptsig}
\bar{S}_{\rm tot}[N_c]
= \bar{S}_{\rm tot}[N_\sg]
+ \frac{1}{2} \bar{S}_{\rm tot}^{\prime\prime} [N_\sg] (\de N_c)^2
+  \frac{1}{6} \bar{S}_{\rm tot}^{\prime\prime\prime}[N_\sg](\de N_c)^3 \, + \mathcal{O}(\de N_c^4),
\eeq
where $\de N_c = N_c - N_\sg$ with $N_\sg$ being a generic saddle, and for brevity we have used $\bar{S}_{\rm tot}^{\rm on-shell}[N_c] \equiv \bar{S}_{\rm tot}[N_c]$. On writing the double-derivative of the lapse action at the saddle-point $\bar{S}_{\rm tot}^{\prime\prime}(N_\sg)$ as $r_{\sg} e^{i\rho_{\sg}}$, the change in $iH (N_c)$ is given by
\beq
\label{eq:changeH}
\de (iH) \propto i \,
\bar{S}_{\rm tot}^{\prime\prime} [N_\sg] (\de N_c)^2
\sim v_\sg^2 e^{i\left(\pi/2 + 2\ta_\sg + \rho_\sg \right)} \, ,
\eeq
where we have written $\de N_c = v_\sg e^{i\ta_\sg}$. $\ta_\sg$ is the direction of flow lines (steepest-ascent and descent) at $N_\sg$. This is given by
\beq
\label{eq:flowang}
\ta_\sg = \frac{(2m-1)\pi}{4} - \frac{\rho_\sg}{2} \, ,
\eeq
where $m \in \mathbb{Z}$. The $\ta_\sg$ corresponding to steepest descent and ascent is determined by requiring the phase in the $\de (iH)$ to satisfy $e^{i\left(\pi/2 + 2\ta_\sg + \rho_\sg \right)} = \mp1$. This implies
\beq
\label{eq:TaDesAes}
\ta_\sg^{\rm des} = m \pi + \frac{\pi}{4} - \frac{\rho_\sg}{2} \, ,
\hspace{5mm}
\ta_\sg^{\rm aes} = m \pi - \frac{\pi}{4} - \frac{\rho_\sg}{2} \, .
\eeq
In the saddle-point approximation and using Picard-Lefschetz methods, the contour integral given in Eq. (\ref{eq:TA_lps_int}) can be approximated as 
\beq
\label{eq:G_RBC_sad_appr}
G[{\rm{Bd_f},\rm{Bd_i}}] \approx \sum_{s} n_s \sqrt{2\pi\hbar}
\,\, 
e^{i\pi/4} (-1)^m 
\Bl[\Bl(1 + \frac{2N_s\beta}{3} \Br)
{\bar S}_{\rm tot}^{\prime\prime}[N_s]\Br]^{-1/2}
\exp\bl[i {\bar S}_{\rm tot}[N_s]\br] \, ,
\eeq
where $n_s$ takes value $0$ or $\pm1$ depending on the orientation, $N_s$ corresponds to the {\it relevant} saddles. The Gauss-Bonnet term in the gravitational action, being topological in nature, although doesn't play any role in determining the saddles, it however still enters ${\bar S}_{\rm tot}(N_s)$ and eventually dictates the behaviour of dominant contribution to the path-integral. In the $\hbar\to 0$ limit, the Gauss-Bonnet contribution dictates that the dominant contribution comes from configurations corresponding to $P_i = - 3i\sqrt{k}$ and $P_{i} = 3i\sqrt{k}$. For the case of positive Gauss-Bonnet coupling (as required by causality and GW data constraining black hole area-laws \cite{Chakravarti:2022zeq, Isi:2020tac}), the former is found to be the dominant configuration \cite{Ailiga:2023wzl} which also leads to the well-behaved metric fluctuations \cite{DiTucci:2019bui, Narain:2021bff, DiTucci:2020weq, Lehners:2021jmv}. For this value of $P_i$, the set of saddles reduces to two categories: one where Universe starts with zero size and other where it starts with non-zero size. We label the former as $N_\pm^{\rm nb}$, while the later as $N_\pm^{\rm \cancel{\rm nb}}$. These are given by
\beq
\label{eq:nb_sads}
N_{\pm}^{\rm nb} = \frac{3}{\Lam}\biggl(-i\sqrt{k} \pm \sqrt{\frac{\Lam q_f}{3} - k}\biggr)\quad , \quad N_{\pm}^{\cancel{\rm nb}} = \frac{3}{\Lam}\biggl(\frac{x-2}{x}i\sqrt{k} \pm \sqrt{\frac{\Lam q_f}{3} - k}\biggr)\, .
\eeq
It is seen that only for the saddles $N_{\pm}^{\rm nb}$, the regularity condition is satisfied, which also enjoys $\bt$-independence. In this case the Universe starts with zero size. The later set of saddles $N_{\pm}^{\cancel{\rm nb}}$, depend on $\bt$-parameter and lead to Universe starting from non-zero size $\bar{q}(0) = 3 (\Lam - 2 i \sqrt{k}\bt)/\bt^2$. As $\bt\to0$, these $N_{\pm}^{\cancel{\rm nb}}$ saddles are pushed to infinity. The on-shell action at all these saddle geometries is given by 
\bea
\label{eq:NB_on_act}
\bar{S}_{\rm tot}[N_{\pm}^{\rm nb}] &=& 
\mp\frac{6}{\Lam}\biggl(\frac{\Lam q_f}{3} - k\biggr)^{3/2} 
- 2 i k^{3/2} \Bl(\frac{3}{\Lam}+4\al\Br) \, , \\
\bar{S}_{\rm tot}[N_{\pm}^{\cancel{\rm nb}}] & = & \mp\frac{6}{\Lam}\biggl(\frac{\Lam q_f}{3} - k\biggr)^{3/2} 
+ \frac{6ik^{3/2} }{\Lam x^3} (x^3 - 6x^2 + 6 x -2 ) - 8i\al k^{3/2}.
\eea
Note, that in the limit $\bt\to0$, the action corresponding to $N_{\pm}^{\cancel{nb}}$ blows-up! Although this is not very pleasant, luckily the usage of Picard-Lefschetz formalism naturally discards them as they become {\it irrelevant} in the $\bt\to0$ limit. 

%%%%%%%%%%%%%%%%%%%%%%%%%%%%%%%%%%%%%%%%%%%%%%%
\subsection{Type-1 Degeneracy}
\label{subs:type1_dg}
%%%%%%%%%%%%%%%%%%%%%%%%%%%%%%%%%%%%%%%%%%%%%%%

To study the structure of {\it thimbles}, we write $i \bar{S}_{\rm tot}(N_c)= h(N_c) + i H(N_c)$ and analyze it at the saddles as $q_f$ varies. For $q_f>3k/\Lam$, $h$ and $H$ at the saddle points are given by
\bea
\label{eq:hH_qf>3k/lam_type1}
&&
h(N_{\pm}^{\rm nb})
= 2 k^{3/2} \frac{\bl(3+4\al \Lam\br)}{\Lam} \, ,
\hspace{5mm}
H(N_{\pm}^{\rm nb})
= \mp \frac{2\bigl(\Lam q_f - 3 k\bigr)^{3/2}}{\Lam\sqrt{3}}
\notag \\
&&
h(N_{\pm}^{\cancel{\rm nb}})
= -\frac{12k^{3/2}(x-1)^{3}}{\Lam x^3} 
+ \frac{2 k^{3/2} \bl(3+4\al \Lam\br)}{\Lam} \, ,
\hspace{5mm}
H(N_{\pm}^{\cancel{\rm nb}})
= \mp\frac{2\bigl(\Lam q_f - 3 k\bigr)^{3/2}}{\Lam\sqrt{3}} \, .
\eea
For $q_f \leq 3k/\Lam$, $H=0$, as the action at the saddle points becomes purely imaginary and the Morse function $(h)$ changes appropriately. By definition, the steepest descent/ascent thimbles connected to a saddle have constant $H$. As in the present case for $q_f>3k/\Lam$, we have $H(N_{\pm}^{\rm nb}) = H(N_{\pm}^{\cancel{\rm nb}})$, it therefore implies that the flowline of one saddle will overlap with the flowline of another saddle. The same will happen for $q_f \leq 3k/\Lam$. This indicates the presence of degeneracies. In order to find out which flowline (ascent or descent) of one saddle will overlap with which flowline of another saddle (descent or ascent), one needs to compute the second derivative of the action at the saddle points. In the saddle-point approximation of the transition amplitude, the second-derivative also enters via  $\bl[\bl(1 + 2N_s\beta/3 \br){\bar S}_{\rm tot}^{\prime\prime}(N_s)\br]^{-1/2}$ as mentioned in Eq. (\ref{eq:G_RBC_sad_appr}). This is given by
\bea
\label{eq:pre_fac_act_sads}
\bl(1 + 2N_{\pm}^{\rm nb}\beta/3 \br)
{\bar S}_{\rm tot}^{\prime\prime}(N_{\pm}^{\rm nb}) 
&& = \pm 2\Lam  (1-x) \sqrt{\Lam q_f - 3k}/\sqrt{3}\, , 
\notag \\
\bl(1 + 2N_{\pm}^{\cancel{\rm nb}}\beta/3 \br)
{\bar S}_{\rm tot}^{\prime\prime}(N_{\pm}^{\cancel{\rm nb}})
&& = \mp 2\Lam  (1-x) \sqrt{\Lam q_f - 3k}/\sqrt{3}\,  .
\eea
From the second-derivative of action at the saddles ${\bar S}_{\rm tot}^{\prime\prime}(N_\sg)$, one can compute the corresponding $\rho_\sg$ using ${\rm Arg}[{\bar S}_{\rm tot}^{\prime\prime}(N_\sg)] = \rho_\sg$. This in turn gives the direction of the flowline emanating from the saddle. For the case of $q_f>3k/\Lam$, the $\rho_\sg$ at the saddles is given by
\beq
\label{eq:sec_der_act_sads}
\tan \bl[\rho(N_{\pm}^{\rm nb})\br]
= \pm \frac{x \sqrt{\Lam q_f - 3 k}}{2(1 - x)\sqrt{3k}} \, , 
\hspace{5mm}
\tan \bl[\rho(N_{\pm}^{\cancel{\rm nb}}) \br]
= 
\mp \frac{x \sqrt{\Lam q_f - 3 k}}{2(1 - x)\sqrt{3k}}
\, .
\eeq
This clearly shows that the $\rho(N_{+}^{\rm nb}) =- \rho(N_{+}^{\cancel{\rm nb}})$ and $\rho(N_{-}^{\rm nb}) = -\rho(N_{-}^{\cancel{\rm nb}})$, thereby implying that steepest ascent of one will overlap with the steepest descent of another. This kind of degeneracy, where the flowline of one saddle overlaps with the flowline of another, we classify it as {\bf type-1} degeneracy.
\begin{table}[t]
\centering
\begin{tabular}{|c|c|c|c|}
\hline
$q_f$ & Defect-Saddle & $\mathfrak{h}_\ep$ & $\mathcal{H}_\ep$ \\ 
\hline
$q_f>3k/\Lam$ 
& $N_{\pm,\ep}^{\rm nb}$ 
& $2 k^{3/2} \bl(3+4\al \Lam\br)/\Lam$  
& $\mp2\bigl(\Lam q_f - 3 k\bigr)^{3/2}/(\Lam\sqrt{3})$ \\ 
& 
& $\pm \ep (3\Lam q_{f} - 9k)^{1/2}/\Lam$
& $-3\ep k^{1/2}/\Lam$ \\
& & & \\
& $N_{\pm,\ep}^{\cancel{\rm nb}}$ 
& $-12k^{3/2}(x-1)^{3}/(\Lam x^3)$ 
& $\mp2\bigl(\Lam q_f-3 k\bigr)^{3/2}/(\Lam\sqrt{3})$ \\
&
& $+2 k^{3/2} \bl(3+4\al \Lam\br)/\Lam$ 
& $+ 3\ep k^{1/2}(x-2)/(\Lam x)$\\ 
&  & &$\pm \ep (3\Lam q_{f} - 9k)^{1/2}/\Lam$ \\
\hline
$q_f<3k/\Lam$
& $N_{\pm,\ep}^{\rm nb}$ 
& $2 k^{3/2} \bl(3+4\al \Lam\br)/\Lam$  
& $\pm \ep ( 9k - 3\Lam q_{f})^{1/2}/\Lam$ \\ 
&
& $\mp 2\bigl( 3 k - \Lam q_f\bigr)^{3/2}/(\Lam\sqrt{3})$
& $- 3\ep k^{1/2}/\Lam$ \\
& & & \\
& $N_{\pm,\ep}^{\cancel{\rm nb}}$ 
& $-12k^{3/2}(x-1)^{3}/(\Lam x^3)$ 
& $ \pm \ep ( 9k - 3\Lam q_{f})^{1/2}/\Lam $ \\
&
& $ +2 k^{3/2} \bl(3+4\al \Lam\br)/\Lam $ 
& $+ 3\ep k^{1/2}(x-2)/(\Lam x)$\\ 
&
& $\mp 2\bigl( 3 k - \Lam q_f\bigr)^{3/2}/(\Lam\sqrt{3})$ & \\
\hline
\end{tabular}
\caption{Type-1 degeneracy getting resolved due the presence of artificial {\it defect} for various values of $q_f$. This is evident from the behaviour of $\mathfrak{h}_\ep$ and $\mathcal{H}_\ep$ at the corrected saddles.}
\label{tab:hH_qf_def_ep}
\end{table}
The drawback of this degeneracy is that it prevents us from unambiguously deciding the {\it relevant} saddle points, which comes from determining $n_\sigma$. Recall that in order to obtain a unique value of $n_\sigma$, one requires $\rm Int (\mathcal{J}_\sigma,\mathcal{K}_{\sigma'})=\delta_{\sigma\sigma'}$ which implies
steepest descent $\mathcal{J}_\sg$ of one saddle can never intersect with the steepest ascent $\mathcal{K}_{\sg^{\prime}}$ of another saddle. Clearly, the presence of {\bf type-1} degeneracies violates this requirement and hence prevents one from implementing the PL method. Hence, lifting this degeneracy is crucial to figure out the {\it relevant} saddles and the correct integration contour. This kind of degeneracy can be lifted by adding a small-\textit{defect} artificially in the lapse-$N_c$ action mentioned in Eq. (\ref{eq:stot_onsh_rbc}). This method of artificially lifting the degeneracy has been largely utilized in the past works, and helps in unambiguously finding the {\it relevant} saddles \cite{Feldbrugge:2017kzv}. A particular choice of such a defect will be to add a linear term $ iN_c$ to the action mentioned in Eq. (\ref{eq:stot_onsh_rbc}). The action with the defect is given by
\beq
\label{eq:Sact_arti_defect}
\bar{S}^{\ep}_{\rm tot}[N_c] = \bar{S}_{\rm tot}^{\rm on-shell}[ N_c] - i \ep N_{c}, \, \hspace{7mm} \ep \ll 1.
\eeq

The saddle point equation with the {\it defect} becomes
\beq
\label{eq:SadRBC_degen_eq}
(N_\sigma - N_{+}^{\rm nb})(N_\sigma - N_{-}^{\rm nb})(N_\sigma - N_{+}^{\cancel{\rm nb}})(N_\sigma - N_{-}^{\cancel{\rm nb}}) =  12 i \ep ( 3i \sqrt{k} + x \Lam N_\sigma )^{2}/(\Lam^4 x^2) \, .
\eeq
Solving it, we get the corrected saddles in the leading order as
\bea
\label{eq:Ep_cor_sad_RBC}
N_{\pm, \ep}^{\rm nb} && = N_{\pm}^{\rm nb} -  \ep \frac{x\sqrt{q_{f}\Lam/3k - 1} \mp (x-1)}{2\Lam(x-1)\sqrt{q_{f} \Lam/3k -1}} + \mathcal{O}(\ep^2) \, , 
\notag \\
N_{\pm, \ep}^{\cancel{\rm nb}} && = N_{\pm}^{\cancel{\rm nb}} +  \ep \frac{x\sqrt{q_{f}\Lam/3k - 1} \pm (x-1)}{2\Lam(x-1)\sqrt{q_{f} \Lam/3k -1}} + \mathcal{O}(\ep^2)\, .
\eea
At these corrected saddle points, one can compute $i\bar{S}^{\ep}_{\rm tot}[N_c]=\mathfrak{h}_\ep+i\mathcal{H}_\ep$ to first order in $\ep$, and re-examine the nature of $\mathfrak{h}_\ep$ and $\mathcal{H}_\ep$ at the corrected saddles which are given in the table \ref{tab:hH_qf_def_ep}. It is evident that due to the presence of a small {\it defect} ($\epsilon$), degeneracy that was noticed and mentioned in Eq. \ref{eq:hH_qf>3k/lam_type1} gets lifted as both $\mathfrak{h}_\ep$ and $\mathcal{H}_\ep$ are corrected.
%
%%%%%%%%%%%%%%%%%%%%%%%%%%%%%%%%%%%%%%%%%%%%%%%
\begin{figure}[h]
\centering
\includegraphics[trim={2.2cm 0 0 0.5cm},clip,
scale=0.8
]
{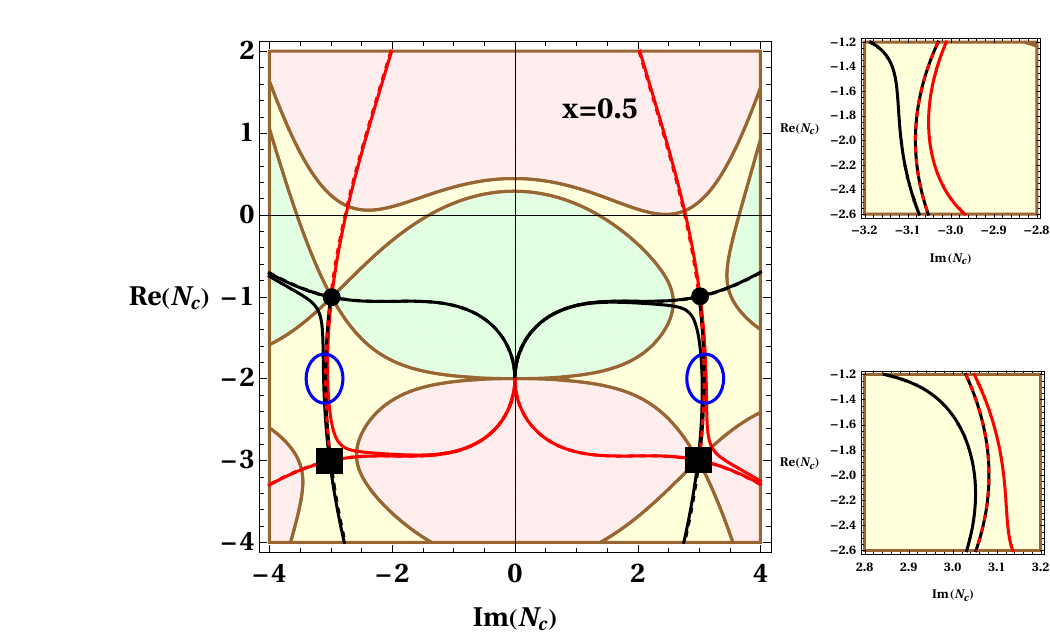}
\caption{
Plot depicting type-1 degeneracy where the flow-lines overlap and get resolved by the addition of an artificial defect. Red lines are the steepest ascent lines, while black lines are the steepest descent lines. The saddles are depicted by black dots (no-boundary) and squares (non-no-boundary). For illustration purpose, we set $P_{i} = -3i\sqrt{k}$, $\al = 2, \Lam=3, x = 1/2, k=1, q_{f} = 10$ and $\ep = 1/10$. Note that the allowed and forbidden region also gets corrected due to the {\it defect}. 
Sign of $\ep$ merely changes the orientation of the thimbles, without affecting the relevance of the saddle.}
\label{fig:PL_w_wo_LD_x0.5_cr}
\end{figure}
%%%%%%%%%%%%%%%%%%%%%%%%%%%%%%%%%%%%%%%%%%%%%%%
%
%
Once the degeneracy is lifted, one can determine the {\it relevant} saddles unambiguously. Once that is done, the {\it defect} can be safely put to zero, and the transition amplitude can be computed in the saddle point approximation with the deformed integration contour lying along the thimbles attached to that {\it relevant} saddle. 
The parameter space for the present minisuperspace can be divided into four regimes. After breaking the type-1 degeneracies, the relevant saddles become
\begin{itemize}
\item $q_f>3k/\Lam$, $0<x<1$: saddles $N_{\pm}^{\rm nb}$ are {\rm relevant}
\item $q_f<3k/\Lam$, $0<x<1$: saddles $N_{+}^{\rm nb}$ is {\rm relevant}
\item $q_f>3k/\Lam$, $x>1$: saddles $N_{\pm}^{\cancel{\rm nb}}$ are {\rm relevant}  
\item $q_f<3k/\Lam$, $x>1$: saddles $N_{+}^{\cancel{\rm nb}}$ is {\rm relevant} 
\end{itemize}
The transition amplitude in each of these cases can be computed following Picard-Lefschetz methods. In the saddle-point approximation, the transition amplitude for each case is mentioned in Table \ref{tab:Gqf_type1}. Although this proposal of lifting degeneracy by incorporating artificial defects suitably resolves the issues, it is still {\it artificial} in the sense that there is no clear guideline for the construction of such defects. It is then becomes crucial to find alternative methods of lifting degeneracies. In this paper, we will describe two such methods of resolving degenerate situations. 
\begin{table}[h]
\centering
\begin{tabular}{|c|c|c|}
% {|p{2cm}|p{6.5cm}|p{6.5cm}|}
\hline
 & $0<x<1$ & $x>1$ \\ \hline
& $N_{\pm}^{\rm nb}$ {\rm relevant} 
& $N_{\pm}^{\cancel{\rm nb}}$ {\rm relevant} \\
& & \\
$q_f>3k/\Lam$ 
& $\approx \frac{3^{1/4}\sqrt{4\pi\hbar} }{\sqrt{\Lambda\left(1-x\right)}(\Lambda q_f -3k)^{1/4}}$ 
& $\approx \frac{3^{1/4}\sqrt{4\pi\hbar} }{\sqrt{\Lambda\left(x-1\right)}(\Lambda q_f - 3k)^{1/4}}$
\\ 
& $\times \cos\bl[
2\bl(\Lam q_f - 3k\br)^{3/2}/(\hbar \Lam \sqrt{3})-\pi/4
\br] $ 
& $\times \exp\bl[-12 k^{3/2}
(x-1)^3/(\hbar \Lam x^3) \br]$ \\ 
& & $\times \cos\bl[
2\bl(\Lam q_f - 3k\br)^{3/2}/(\hbar \Lam \sqrt{3})-\pi/4
\br] $ \\ \hline
%%%%%%%
%%%%%%%
& $N_{+}^{\rm nb}$ {\rm relevant} 
& $N_{+}^{\cancel{\rm nb}}$ {\rm relevant} \\
& & \\
$q_f<3k/\Lam$ 
& $\approx \frac{3^{1/4}\sqrt{4\pi\hbar} }{\sqrt{\Lambda\left(1-x\right)}(3k-\Lambda q_f)^{1/4}}$ 
& $\approx \frac{3^{1/4}\sqrt{4\pi\hbar} }{\sqrt{\Lambda\left(x-1\right)}(3k-\Lambda q_f)^{1/4}}$ \\ 
& $\times \exp\bl[
-2 \bl(3k -\Lam q_f \br)^{3/2}/\bl(\hbar\Lam\sqrt{3}\br)
\br]$
& $\times \exp\bl[
-12 k^{3/2}
(x-1)^3/(\hbar \Lam x^3) \br]$ \\ 
& & $\times \exp\bl[
-2 \bl(3k -\Lam q_f \br)^{3/2}/\bl(\hbar\Lam\sqrt{3}\br)
\br]$ \\\hline
\end{tabular}
\caption{The transition amplitude in the Robin boundary configuration $G_{\rm RBC}[q_f]$ in the case of Type-1 degeneracy. Here we mention $G_{\rm RBC}[q_f] \times \exp\bl[+2k^{3/2}\bl(3+4\al \Lam\br)/(\Lam \hbar)\br]$ for the four possibilities in the saddle-point approximation. }
\label{tab:Gqf_type1}
\end{table}

%%%%%%%%%%%%%%%%%%%%%%%%%%%%%%%%%%%%%%%%%%%%%%%
\subsection{Type-2 Degeneracy}
\label{subs:type2_dg}
%%%%%%%%%%%%%%%%%%%%%%%%%%%%%%%%%%%%%%%%%%%%%%%

Together with the type-1 degeneracy, another kind of degeneracy appears in the study of {\it thimbles} in the complex $N_c$ plane.
It happens when saddles coalesce because of the specific choices of the boundary parameters. This corresponds to the vanishing of the argument of the Airy function in Eq. (\ref{eq:transAmp_rbc_exact}). We call it \textbf{type-2} degeneracy. In the present case, such situations arise when 
\begin{itemize}
\item the final boundary parameter $q_f = 3k/\Lam$
\item the initial boundary parameter $x=1$
\item both initial and final boundary parameters satisfy $x=1$ and  $q_f = 3k/\Lam$, respectively. 
\end{itemize}
In all these cases, the saddle point approximation breaks down, and one has to go beyond the saddle-point approximation to compute the transition amplitude in these special cases. However, when {\it defect} is present, this degeneracy gets resolved, which we will discuss now.

%%%%%%%%%%%%%%%%%%%%%%%%%%%%%%%%%%%%%%%%%%%%%%%
\subsubsection{$q_{f} = 3k/\Lam$}
\label{sss:qf=3_by_lam}
%%%%%%%%%%%%%%%%%%%%%%%%%%%%%%%%%%%%%%%%%%%%%%%

This is a degenerate situation, where the Hartle-Hawking saddles $N_{\pm}^{\rm nb}$ and the non-Hartle-Hawking saddles $N_{\pm}^{\cancel{\rm nb}}$ merge: 
\beq
\label{eq:MEG_sad_qf=3/lam}
N_{+}^{\rm nb} = N_{-}^{\rm nb}=N^{\rm nb} \, ,
\hspace{5mm}
N_{+}^{\cancel{\rm nb}} = N_{-}^{\cancel{\rm nb}}=N^{\cancel {\rm nb}} \, .
\eeq
We also observe {\bf type-1} degeneracy between $N^{\rm nb}$ and $N^{\cancel{\rm nb}}$ saddles which causes problem in determining the relevant saddles, (see table \ref{tab:hH_qf_qc_def_ep} with $\ep=0$) .
To find the relevant saddle, we consider adding the {\it defect} mentioned in Eq. (\ref{eq:Sact_arti_defect}). Under the presence of this defect, the corrected saddle point equation is given by
\bea
\label{eq:sad_qf_qc_2_deg_dft}
(\Lam^4 x^2)\bigl(N_{\sigma} - N^{\rm nb}\bigr)^2 \bigl(N_{\sigma} - N^{\cancel{\rm nb}}\bigr)^2 - 12 i \ep \bigl( x \Lam N_{\sigma} + 3 i \sqrt{k}\bigr)  = 0\, .
\eea
With the linear defect, the leading order correction to the degenerate saddles is
\bea
\label{eq:del_cor_qf_qc_dft}
N_{\ep}^{\rm nb} && = N^{\rm nb}  \pm  e^{i\pi/4} \sqrt{3\ep}/\Lam + \cdots \, , \notag \\
N_{\ep}^{\cancel{\rm nb}} &&= N^{\cancel{\rm nb}}  \pm  e^{i\pi/4} \sqrt{3\ep}/\Lam + \cdots \, .
\eea
From Eq. (\ref{eq:del_cor_qf_qc_dft}), it is easy to see that the saddle degeneracy is resolved as the background saddles split into two. Note that the correction to saddles comes at order $\sqrt{\ep}$, instead of order $\ep$, which happens in the case of type-1 degeneracies. The leading order corrections to $\mathfrak{h}_\ep$ and $\mathcal{H}_\ep$ are given in table \ref{tab:hH_qf_qc_def_ep}. The behaviour of $\mathcal{H}_\ep$ under the presence of {\it defect}, clearly shows that there is no \textbf{type-1} degeneracy between the corrected saddles.
\begin{table}[h]
\centering
\begin{tabular}{|c|c|c|}
\hline
Defect-Saddle & $\mathfrak{h}_\ep$ & $\mathcal{H}_\ep$ \\ \hline
$N_{\ep}^{\rm nb}$ 
& $2 k^{3/2} \bl(3+4\al \Lam\br)/\Lam \pm \sqrt{2}\ep^{3/2}/(\sqrt{3}\Lam)$  
& $- 3\ep k^{1/2}/\Lam \pm \sqrt{2}\ep^{3/2}/(\sqrt{3}\Lam)$ \\ 
\hline
$N_{\ep}^{\rm \cancel{nb}}$ 
& $-12k^{3/2}(x-1)^{3}/(\Lam x^3) + 6k^{3/2}/\Lam + 8\al k^{3/2}$  
& $    3\ep k^{1/2}(x-2)/(\Lam x) \pm \sqrt{2}\ep^{3/2}/(\sqrt{3}\Lam)$ \\
& $\pm \sqrt{2}\ep^{3/2}/(\sqrt{3}\Lam)$ & \\ \hline
\end{tabular}
\caption{On-shell values of $\mathfrak{h}_\ep$ and $\mathcal{H}_\ep$ for $q_{f}=3k/\Lam$ degeneracy at the corrected saddles. For non zero $\epsilon$,
\textbf{type-1} degeneracy gets resolved, as explained in the text. }
\label{tab:hH_qf_qc_def_ep}
\end{table}
From this one can infer that $N_{+/-}^{\rm nb}$ is the relevant saddle for $0<x<1$ and $N_{+/-}^{\cancel{\rm nb}}$ being relevant for $x>1$. However, in either of these cases the second derivative computed at the relevant saddles vanishes: ${\bar S}_{\rm tot}^{\prime\prime}[N_{+/-}^{\rm nb}]=0$ and ${\bar S}_{\rm tot}^{\prime\prime}[N_{+/-}^{\cancel{\rm nb}}]=0$. This signals a breakdown of the saddle-point approximation, indicating one needs to go beyond second order approximation. The transition amplitude at the third order is given by
\beq
\label{eq:G_sad_trd_odr_exp}
G_{\rm RBC}[q_{f}=3k/\Lam] \approx \sum_{s} n_{s} \frac{(6\hbar)^{1/3}\Gamma(1/3)}{\sqrt{3+2N_s\beta}} \bl[{\bar S}_{\rm tot}^{\prime\prime\prime}[N_s]\br]^{-1/3} e^{iS[N_s]/\hbar}\, .
\eeq 
The third derivative of the lapse action computed at all saddle points is given by
\beq
\label{eq:trd_der_act_sads}
{\bar S}_{\rm tot}^{\prime\prime\prime}[N_s] = 2 \Lam^2/3 \, .
\eeq
The transition amplitude in the degenerate case where $q_f=3k/\Lam$, is given by
\bea
\label{eq:G_sad_q=3/Lam_x<1}
&&
G_{\rm RBC}\bl[0<x<1\br]
\approx 
\biggl(\frac{3\hbar^2}{\Lam^4}\biggr)^{1/6} \frac{\Gamma(1/3)}{\sqrt{1-x}} \exp{\Big[\frac{2 k^{3/2}}{\hbar}\Big(\frac{3}{\Lam} + 4\al\Big)\Big]} \, ,
\notag \\
&&
G_{\rm RBC}\bl[x>1\br]
\approx
\biggl(\frac{3\hbar^2}{\Lam^4}\biggr)^{1/6} \frac{\Gamma(1/3)}{\sqrt{x-1}} 
\exp{\Big[\frac{2 k^{3/2}}{\hbar}\Big(\frac{3}{\Lam} + 4\al\Big)\Big]}
e^{-\frac{12 k^{3/2}(x-1)^3}{\hbar \Lam x^3}} \, .
\eea
The above expressions are not valid for $x=1$, which is a double degenerate case and addressed in Sec. (\ref{sss:x=1_qf3_by_Lam}).

%%%%%%%%%%%%%%%%%%%%%%%%%%%%%%%%%%%%%%%%%%%%%%%
\subsubsection{$x=1$}
\label{sss:x=1}
%%%%%%%%%%%%%%%%%%%%%%%%%%%%%%%%%%%%%%%%%%%%%%%

In the case when the initial boundary parameter $x=1$, the no-boundary saddles merge with the  non-no-boundary saddles
\beq
\label{eq:MEG_sad_x=1}
N_{+}^{\rm nb} = 
N_{+}^{\cancel{\rm nb}} \, ,
\hspace{5mm}
N_{-}^{\rm nb} = N_{-}^{\cancel{\rm nb}}.
\eeq
Similar to the previous case, the presence of {\it defect} will break this saddle degeneracy. The corrected saddle point equation with artificial {\it defect} for $x=1$ is given by
\bea
\label{eq:sad_x_1_deg_dft}
\Lam^4\bigl(N_{s} - N_{+}^{\rm nb}\bigr)^2 \bigl(N_{s} - N_{-}^{\rm nb}\bigr)^2 - 12 i \ep (\Lam N_{s} + 3 i\sqrt{k})^2 = 0\,  .
\eea
In the limit $\ep\to 0$ one gets the degeneracy back and for non-zero $\epsilon$ Eq. (\ref{eq:sad_x_1_deg_dft}) will have four non-degenerate saddle solutions.
The corrected saddles to leading order in $\ep$ are given by 
\bea
\label{eq:del_cor_x_dft}
N_{+,\ep}^{\rm nb} && = N_{+}^{\rm nb}  \pm  e^{i\pi/4} \sqrt{3\ep}/\Lam\, + \cdots , 
\notag \\
N_{-,\ep}^{\rm nb} &&= N_{-}^{\rm nb}  \pm  e^{i\pi/4} \sqrt{3\ep}/\Lam + \cdots \, .
\eea
This clearly shows that the presence of {\it defect} splits the set of two degenerate saddles into four saddles. Again, the correction to saddles comes at order $\sqrt{\ep}$, instead of order $\ep$ like in type-1 degeneracies. The $\mathfrak{h}_\ep$ and $\mathcal{H}_\ep$ at the corrected saddle points are given in table \ref{tab:hH_x_1_qf>3/Lam_def_ep}.
%%%%%%%%%%%%%%%%%%%%%%%%%%%%%%%%%%%%%%%%%%%%%%%
\begin{figure}[h]
\centering
\includegraphics[trim={2.2cm 0 0 0},clip,
scale=0.8
%\textwidth
]
{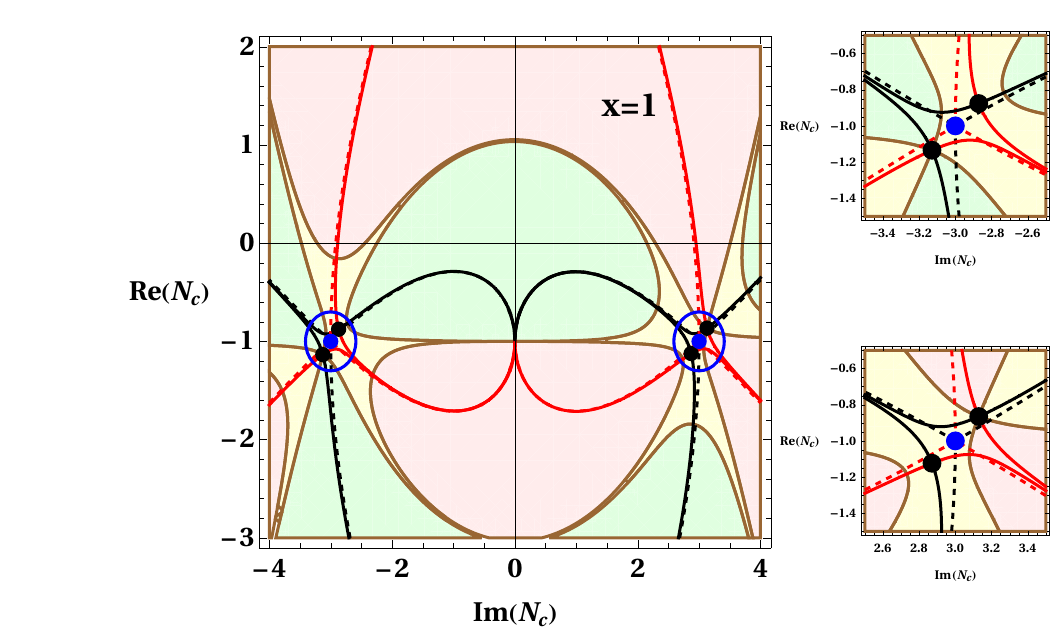}
\caption{
Plot depicting type-2 degeneracy of second kind, where not only there is overlap of flow lines, but also due to the choice of external parameter $x=1$ leading to merging of the saddles. This degeneracy gets resolved due to the inclusion of {\it defect}. Red lines are the steepest ascent lines, while black lines are the steepest descent lines. The saddles are depicted black dots and squares. We take $P_{i} = -3i\sqrt{k}$ the dominant contribution from Gauss Bonnet Sector. For illustration purpose, we set $\al = 2$, $\Lam=3$, $k=1$, $q_{f} = 10$ and $\ep = 1/10$. Note, that the allowed and forbidden region also gets corrected due to the {\it defect}. The two degenerate saddles split into two individually.
}
\label{fig:PL_w_wo_LD_x1}
\end{figure}
%%%%%%%%%%%%%%%%%%%%%%%%%%%%%%%%%%%%%%%%%%%%%%%
%
\begin{table}[h]
\centering
\begin{tabular}{|c|c|c|}
\hline
Defect-Saddle & $\mathfrak{h}_\ep$ & $\mathcal{H}_\ep$ \\ \hline
$N_{+,\ep}^{\rm nb}$ 
& $2 k^{3/2} \bl(3+4\al \Lam\br)/\Lam + \ep (3\Lam q_{f} - 9k)^{1/2}/\Lam$  
& $- 2\bigl(\Lam q_f - 3 k\bigr)^{3/2}/(\Lam\sqrt{3}) - 3\ep k^{1/2}/\Lam$ \\ 
& $\pm \sqrt{2}\ep^{3/2}/(\sqrt{3}\Lam)$ & $\pm \sqrt{2}\ep^{3/2}/(\sqrt{3}\Lam)$ \\
\hline
$N_{-,\ep}^{\rm nb}$ 
& $2 k^{3/2} \bl(3+4\al \Lam\br)/\Lam - \ep (3\Lam q_{f} - 9k)^{1/2}/\Lam$  
& $ 2\bigl(\Lam q_f - 3 k\bigr)^{3/2}/(\Lam\sqrt{3})  - 3\ep k^{1/2}/\Lam $ \\
& $\pm \sqrt{2}\ep^{3/2}/(\sqrt{3}\Lam)$ & $\pm \sqrt{2}\ep^{3/2}/(\sqrt{3}\Lam)$\\ \hline
\end{tabular}
\caption{On-shell values of $\mathfrak{h}_\ep$ and $\mathcal{H}_\ep$ for $q_f>3k/\Lam$ and $x=1$ degeneracy case at the corrected saddles.}
\label{tab:hH_x_1_qf>3/Lam_def_ep}
\end{table}
\begin{table}[h]
\centering
\begin{tabular}{|c|c|c|}
\hline
Defect-Saddle & $\mathfrak{h}_\ep$ & $\mathcal{H}_\ep$ \\ \hline
$N_{+,\ep}^{\rm nb}$ 
& $2 k^{3/2} \bl(3+4\al \Lam\br)/\Lam  - 2\bigl( 3 k - \Lam q_f \bigr)^{3/2}/(\Lam\sqrt{3}) $  
& $ \ep ( 9k - 3\Lam q_{f} )^{1/2}/\Lam - 3\ep k^{1/2}/\Lam $ \\ 
& $\pm \sqrt{2}\ep^{3/2}/(\sqrt{3}\Lam)$ & $\pm \sqrt{2}\ep^{3/2}/(\sqrt{3}\Lam)$ \\
\hline
$N_{-,\ep}^{\rm nb}$ 
& $2 k^{3/2} \bl(3+4\al \Lam\br)/\Lam + 2\bigl(  3 k - \Lam q_f\bigr)^{3/2}/(\Lam\sqrt{3})$  
& $ - \ep (  9k - 3\Lam q_{f})^{1/2}/\Lam  - 3\ep k^{1/2}/\Lam $ \\
& $\pm \sqrt{2}\ep^{3/2}/(\sqrt{3}\Lam)$ & $\pm \sqrt{2}\ep^{3/2}/(\sqrt{3}\Lam)$\\ \hline
\end{tabular}
\caption{On-shell values of $\mathfrak{h}_\ep$ and $\mathcal{H}_\ep$ for $q_f<3k/\Lam$ and $x=1$ degeneracy case at the corrected saddles.}
\label{tab:hH_x_1_qf<3/Lam_def_ep}
\end{table}
From table \ref{tab:hH_x_1_qf>3/Lam_def_ep}, it is evident that there is no \textbf{type-1} degeneracy between any corrected two saddles for arbitrary value of $q_{f} \neq 3k/\Lam$. From this one can now decide the relevance of the saddles for various values of $q_f$. For $q_f>3k/\Lam$, both saddles $N_{\pm}^{\rm nb}$ are relevant. While for $q_f<3k/\Lam$, only the saddle $N_{+}^{\rm nb}$ is relevant. In this case also, the second derivatives of the lapse action computed at the saddles vanish: ${\bar S}_{\rm tot}^{\prime\prime}(N_{\pm}^{\rm nb})=0$. However, unlike the earlier case, this degeneracy arises irrespective of the size of the universe, and the saddle point analysis breaks down for all values of $q_f$. The transition amplitude is computed going to third order in both $q_f<3k/\Lam$ and $q_f>3k/\Lam$ regimes. These are given by
\bea
\label{eq:G_sad_q>3/Lam_x=1}
&&
G_{\rm RBC}[q_{f}>3k/\Lambda] 
\approx 
2 \biggl(\frac{3\hbar}{\Lam^2}\biggr)^{1/3}
\frac{3^{1/12}k^{1/4}\Gamma(1/3) }{(\Lambda q_f - 3k)^{1/4}} 
e^{\frac{2k^{3/2}}{\hbar }\left(\frac{3}{\Lambda}+4\alpha\right)}   
\cos\left[\frac{2\left(3k-\Lambda q_{f}\right)^{3/2}}{\hbar \Lambda \sqrt{3}}
-\frac{\pi}{4}\right] \, ,
\notag \\
&&
G_{\rm RBC}[q_{f}<3k/\Lambda] 
\approx 
\biggl(\frac{3\hbar}{\Lam^2}\biggr)^{1/3}
\frac{3^{1/12}k^{1/4} 
\Gamma(1/3)}{( 3k - q_f \Lam)^{1/4}} 
e^{\frac{2k^{3/2}}{\hbar }\left(\frac{3}{\Lambda}+4\alpha\right)} 
\exp\left[-\frac{2\left(3k-\Lambda q_{f}\right)^{3/2}}{\hbar \Lambda\sqrt{3}}\right]\, .
\eea
%

%%%%%%%%%%%%%%%%%%%%%%%%%%%%%%%%%%%%%%%%%%%%%%%
\subsubsection{$x=1$ and $q_f=3k/\Lam$}
\label{sss:x=1_qf3_by_Lam}
%%%%%%%%%%%%%%%%%%%%%%%%%%%%%%%%%%%%%%%%%%%%%%%

This is a doubly degenerate situation in which both the initial and final boundary parameters acquire special values that are $x=1$ and $q_f=3k/\Lambda$, respectively. In this special case, the $\bar{S}_{\rm tot}^{\rm on-shell}[N_c]$ becomes a cubic polynomial. It has two coalescing saddles leading to the degeneracy. This degenerate situation can be resolved by supplementing the theory with a {\it defect} as before. The corrected saddle point equation with the {\it defect} is given by
\beq
\label{eq:Sad_DD_deg_def1}
(\Lam N_{s} + 3i\sqrt{k})^2 - 12 i \ep = 0 \, .
\eeq
In the limit $\ep\to0$ limit, we get back the doubly-degenerate complex saddle $N_{\sigma} = -3i\sqrt{k}/\Lam$. With non-zero $\ep$, Eq. (\ref{eq:Sad_DD_deg_def1}) can be solved exactly, with the $\ep$-dependent corrected saddles given by
\beq
\label{eq:Sad_DD_deg_def}
N_{\pm,\ep} = - 3i\sqrt{k}/\Lam \pm e^{i\pi/4} 2\sqrt{3\ep}/\Lam \, .
\eeq
Note that the correction is of the order $\sqrt{\ep}$, 
similar to the previous degenerate cases and the background saddle splits into two. In this special doubly-degenerate situation, once the {\it defect} is incorporated, $\mathfrak{h}_\ep$ and $\mathcal{H}_\ep$ at the corrected saddles $N_{\pm,\ep}$ are given by
\bea
\label{eq:dd_LD_hH}
&&
\mathfrak{h}_\ep(N_{\pm,\ep}) = 2 k^{3/2} \bl(3+4\al \Lam\br)/\Lam \pm 2\sqrt{6}\ep^{3/2}/(3\Lam)
\, ,
\notag \\
&&
\mathcal{H}_\ep(N_{\pm,\ep}) = - 3\ep k^{1/2}/\Lam \pm 2\sqrt{6}\ep^{3/2}/(3\Lam) \, .
\eea
In this case, one needs additional care in the computation of the transition amplitude equation, given in Eqs. (\ref{eq:sch_form}), and (\ref{eq:G_RBC_sad_appr}). This is because, at the special values of the boundary parameters $x=1$ and $q_f=3k/\Lambda$, the fluctuation determinant vanishes at the background saddle point $N_\sigma=-3i\sqrt{k}/\Lam$; $\D(N_\sigma)=0$, eventually introducing an infinity in the integrand as can be seen from Eqs. (\ref{eq:sch_form}), and (\ref{eq:G_RBC_sad_appr}). To deal with this situation carefully, and obtain a correct answer matching the appropriate limit of the exact expression computed in Eq. (\ref{eq:transAmp_rbc_exact}), we make a change of variable and go from the complex $N_c$-plane to the complex $u$-plane, and do the computation of the transition amplitude for this special situation directly in the $u$-plane. This is achieved via the transformation (for generic $x$ and $q_f$)
\begin{equation}
\label{eq:uTR_pl}
{\rm d}u= \bl[ \D(N_c) \br]^{-1/2} {\rm d}N_c \, ,
\end{equation}
where $\D(N_c)$ is mentioned in Eq. (\ref{eq:DelNc_rbc}). 
The lapse action in terms of the $u$-variable becomes 
\begin{equation}
\label{eq:action_u_label}
\bar{S}_{\rm tot}^{\rm on-shell}[\bar{q},u]=A u^6+Bu^2+C+\frac{D}{u^2} \, ,
\end{equation}
where,
\bea
\label{eq:ABCD_upl}
&&
A=\frac{i \Lambda ^5 x^3}{62208 k^{3/2}},
\hspace{5mm}
B=\frac{i \Lambda  \left(k (-6x^2+6x -3)+\Lambda  q_f x^2\right)}{24 \sqrt{k} x}
\notag \\
&&
C=-\frac{6 i k^{3/2} (3 x^2-3x+1)}{\Lambda  x^3}-8 i \alpha  k^{3/2}\, ,
\hspace{5mm}
D=-\frac{9 i \sqrt{k} \left(k (3-6 x)+\Lambda  q_f x^2\right)^2}{\Lambda ^3 x^5} \, .
\eea
The benefit of this field redefinition is that the prefactor $\bl[ \D(N_c) \br]^{-1/2}$ appearing in Eq. (\ref{eq:sch_form}) disappears (similar redefinition procedure can be applied generically and has been first implemented in \cite{Honda:2024aro,Honda:2024hdr}). It is evident from Eq. (\ref{eq:action_u_label}) that in the transformed $u$-variable, the number of semiclassical saddles increases, and hence, it will lead to the degeneracies of higher order. In our present case, when $x=1$ and $q_f=3k/\Lambda$, one needs to go to the sixth order in the perturbation expansion around the corresponding saddle as $B=0$, $D=0$. The saddle at $N_c=-3 i \sqrt{k}/\Lambda $ will now be at $u=0$. Performing the integration (with $x=1$) in the $u$-plane along the suitable contour, the transition amplitude can be expressed as follows 
\begin{equation}
\label{eq:du_inte}
G_{\rm RBC} (q_f=3k/\Lam, x=1)
=
\int_{\mathcal{C}_u} du \exp\left[-{\frac{ \Lambda ^5 u^6}{62208k^{3/2}\hbar}}+\frac{2k^{3/2}}{\hbar}\left(\frac{3}{\Lambda}+4\alpha\right)\right] \, ,
\end{equation}
where $\mathcal{C}_u$ is the equivalent contour in $u$-plane corresponding to the real contour ($\mathcal{C}$) in the $N_c$ plane. This integral can be rewritten as
\bea
\label{eq:Grbc_dd_degen}
G_{\rm RBC} (q_f=3k/\Lam, x=1)
&&
=\left(e^{i\pi/3}-e^{2i\pi/3}\right)
\int_{0}^\infty du \exp\left[-{\frac{ \Lambda ^5 u^6}{62208k^{3/2}\hbar}}+\frac{2k^{3/2}}{\hbar}\left(\frac{3}{\Lambda}+4\alpha\right)\right] \, ,
\notag \\
&&
= \frac{2^{1/3}\Gamma \left(1/6\right)k^{1/4} \hbar^{1/6}}{\sqrt[6]{3} \Lam^{5/6}}\exp\left[{\frac{2k^{3/2}}{\hbar}\left(\frac{3}{\Lambda}+4\alpha\right)} \right] \, ,
\eea
which matches exactly with the Eq. (\ref{eq:transAmp_rbc_exact}) when $q_f=3/\Lam$ and $x=1$. It is worth checking how the same will respond once {\it defect} is incorporated and the above steps are repeated in the presence of the {\it defect}. The modified action with {\it defect} becomes in the $u$-plane
\begin{equation}
\label{eq:action_u_label_def}
S^{\ep}[\bar{q}, u]=\bar{S}_{\rm tot}^{\rm on-shell}[\bar{q},u]-\epsilon\left(\frac{\Lambda  x}{12\sqrt{k}}  u^2 +\frac{3\sqrt{k}}{\Lambda  x}\right),
\end{equation}
where $\bar{S}_{\rm tot}^{\rm on-shell}[\bar{q},u]$ is mentioned in Eq. (\ref{eq:action_u_label}). As before, when $q_f=3/\Lam$, $x=1$; we have $B=0$ and $D=0$. The contour integral in the $u$-plane for the RBC case with the {\it defect} is given by the following 
\bea
\label{eq:du_inte_ep}
G_{\rm RBC}^\ep \Bl[q_f=\frac{3k}{\Lam}, x=1\Br]=
\int_{\mathcal{C}_u} {\rm d}u
\exp\Bl[
&&
-\frac{ \Lambda ^5 u^6}{62208k^{3/2}\hbar}
+\frac{2k^{3/2}}{\hbar}\left(\frac{3}{\Lambda}+4\alpha\right)
\notag \\
&&
-i \ep 
\Bl\{\frac{\Lambda u^2}{12\sqrt{k}\hbar}
+\frac{3\sqrt{k}}{\Lam \hbar} \Br\}\Br] \, .
\eea
In the $\hbar\to0$ limit, the dominant contribution comes from the saddle configurations. The saddles are given by 
\begin{equation}
\label{eq:saddup}
u=0\, , 
\,\, 
u=\exp[i\pi(7+4n)/8] 12^{3/4}(k\epsilon)^{1/4}/\Lam\, ,
\hspace{5mm} 
n=0, 1, 2, 3\, .
\end{equation}
However, among these saddles, the relevant one according to the Picard-Lefschetz methodology occurs only for $n=3$ and is given by $u_{\rm rel}=(12i)^{3/4}(k\epsilon)^{1/4}/\Lam $. Once the relevant saddle has been identified, the integral in Eq.(\ref{eq:du_inte_ep}) can be evaluated using the saddle point approximation. Evaluating it, we get
\begin{equation}
\label{eq:Grbc_dd_degenlindef}
G_{\rm RBC}^\ep \Bl[q_f=\frac{3k}{\Lam}, x=1\Br]
\sim \sqrt{\frac{3\pi \hbar\sqrt{k} i}{\Lambda  \epsilon }}\exp\left[\frac{2k^{3/2}}{\hbar}\left(\frac{3}{\Lambda}+4\alpha\right)+{\frac{(2+2 i) \sqrt{6} \epsilon ^{3/2}-9 i \epsilon\sqrt{k}}{3 \hbar \Lambda }}\right] \, .
\end{equation}
This diverges in the $\epsilon\rightarrow 0$ limit, as the saddle-point approximation breaks down when the {\it defect} is switched off. One has to go beyond the standard WKB (to sixth order) in order to compute the integral. It should be emphasized that in the presence of {\it defect}, the computation of transition amplitude for the doubly-degenerate case can also be performed directly in the complex $N_c$-plane. This is because in the presence of {\it defect}, the prefactor $\bl[ \D(N_c) \br]^{-1/2}$ appearing in eq. (\ref{eq:sch_form}) no longer diverges, thereby allowing the computation in the WKB approximation. In this case , one obtains the same result for $G_{\rm RBC}^\ep \bl[q_f=3k/\Lam, x=1\br]$ as in Eq. (\ref{eq:Grbc_dd_degenlindef}).

%%%%%%%%%%%%%%%%%%%%%%%%%%%%%%%%%%%%%%%%%%%%%%%
\section{Quantum corrections as {\it defects}}
\label{sec:qc_as_def}
%%%%%%%%%%%%%%%%%%%%%%%%%%%%%%%%%%%%%%%%%%%%%%%

In the previous section, we noticed the emergence of two types of degeneracies in the saddle analysis of the contour integral mentioned in Eq. (\ref{eq:sch_form}). While doing the PL analysis, it is seen that for generic values of boundary parameters, there are four saddle points. In all cases, the steepest ascent flow line from one saddle overlaps with the steepest descent flow line from another saddle. This, as discussed previously, is \textbf{type-1} degeneracy and can be lifted by introducing an artificial {\it defect}. \textbf{Type-2} degeneracy arises when saddles coalesce due to the special values of boundary parameters. Again, this degeneracy is also lifted in the presence of artificial {\it defect}. One can't help wondering if the resolution of degeneracies can be achieved without introducing artificial {\it defects}. In this section, we try to answer this question by studying the effects of quantum corrections on the type-1 and type-2 degeneracies. 

Consider the contour integration stated in Eq. (\ref{eq:sch_form}), where the factor $\bl[\D(N_c)\br]^{-1/2}$ arises after integrating over the quantum fluctuations, which could be either quantum fluctuations of the scale-factor $q(t)$ and/or fluctuations of the metric perturbations, if they are also involved. In the present case, only the former is present. This, when exponentiated, leads to the quantum corrected lapse action where the effects from one-loop determinant are incorporated in $\D(N_c)$. The quantum corrected lapse action then is given by 
\beq
\label{eq:qmAact}
{\cal A}(N_c) = \bar{S}_{\rm tot}^{\rm on-shell}[N_c]
+ \frac{i\hbar}{2} \log \D(N_c) 
\equiv {\cal A}_0(N_c) + \hbar {\cal A}_1(N_c) 
\, ,
\eeq
where $\bar{S}_{\rm tot}^{\rm on-shell}[N_c]$ and $\D(N_c)$ are mentioned in Eq. (\ref{eq:stot_onsh_rbc}) and Eq. (\ref{eq:uTR_pl}) respectively. Such corrections from the one-loop determinant have also been considered in the Picard-Lefschetz framework previously in \cite{Nishimura:2014rxa, Nishimura:2014kla, Nishimura:2015loa, Tanizaki:2015pua, Tanizaki:2015gcv, Kanazawa:2014qma}. For later purposes, we define $i\mathcal{A}(N_c)=\mathfrak{h}(N_c)+i\mathcal{H}(N_c)$. Treating quantum correction as a {\it defect}, one can compute the corrections the saddles receive at leading order in $\hbar$. If saddle point is denoted by $N_\sg$, which receives $\hbar$-corrections, then it can be written in the form
\beq
\label{eq:Nsg_1lp_form}
N_\sg = N_\sg^{(q)} + \hbar 
{\cal N}_\sg^{(q)} + \cdots \, ,
\eeq
where $N_\sg^{(q)}$ is the saddle point for the zeroth-order lapse action ${\cal A}_0(N_c)$, and $
{\cal N}_\sg^{(q)}$ is the correction due to quantum effects. Saddles by definition satisfy ${\cal A}^\prime(N_\sg)=0$. Plugging eq. (\ref{eq:Nsg_1lp_form}) in ${\cal A}^\prime(N_\sg)=0$, and expanding to first order in $\hbar$, we get 
\beq
\label{eq:ANc_sad_exp_1lp}
{\cal A}_0^\prime(N_\sg^{(q)})
+ \hbar {\cal N}_\sg^{(q)} {\cal A}_0^{\prime\prime}(N_\sg^{(q)}) 
+ \hbar {\cal A}_1^\prime(N_\sg^{(q)}) = 0 \, .
\eeq
At zeroth-order by definition we have ${\cal A}_0^\prime(N_\sg^{(q)})=0$. This gives 
\beq
\label{eq:Hbar_sadcor_genform}
{\cal N}_\sg^{(q)} 
= - {\cal A}_1^\prime(N_\sg^{(q)})/{\cal A}_0^{\prime\prime}(N_\sg^{(q)}) \, .
\eeq
For the case when Robin boundary condition is applied on the initial hyper-surface and Dirichlet boundary condition at the final hyper-surface, we get
\begin{equation}
\label{eq:calNs_q_hbarCR}
\mathcal{N}_\sigma^{(q)}
=-(i\beta/3)\bl(1+2\beta N_\sigma^{(q)}/3\br)^{-1}/\bl[\bar{S}_{\rm tot}(N_\sigma^{(q)}) \br]^{\prime\prime}\, .
\end{equation}
This is the correction received due to the logarithmic term, provided the denominator is non-vanishing. Cases where the denominator can vanish are discussed in sec. \ref{subs:type2_dg_qc}.

The transition amplitude can be computed by expanding ${\cal A}(N_c)$ around the saddle point $N_\sg$. This gives (keeping terms up to $\mathcal{O}(\hbar)$)
\bea
\label{eq:ANc_exp_sad_form}
{\cal A}(N_c) 
&& = {\cal A}(N_\sg) + \frac{1}{2}(N_c - N_\sg)^2 {\cal A}^{\prime\prime}(N_\sg) + \cdots
\notag \\
&&
= {\cal A}_0(N_\sg^{(q)})+ \hbar {\cal A}_1(N_\sg^{(q)}) 
+ \hbar {\cal A}_0^\prime(N_\sg^{(q)}) 
\bl(
{\cal N}_\sg^{(q)}\br)
\notag \\
&&
+ \frac{1}{2}(N_c - N_\sg)^2 \bl[
{\cal A}_0^{\prime\prime} (N_\sg^{(q)})
+ \hbar {\cal A}_0^{\prime\prime\prime} (N_\sg^{(q)})\bl(
{\cal N}_\sg^{(q)} \br)
+ \hbar {\cal A}_1^{\prime\prime} (N_\sg^{(q)})
+ \cdots
\br] + \cdots 
\notag \\
&& 
= {\cal A}_0(N_\sg^{(q)})
+ \hbar {\cal A}_1(N_\sg^{(q)}) 
+ \frac{1}{2} (N_c - N_\sg)^2 \bl[
{\cal A}_0^{\prime\prime} (N_\sg^{(q)})
\notag \\
&&
+ \hbar {\cal A}_0^{\prime\prime\prime} (N_\sg^{(q)})\bl(
{\cal N}_\sg^{(q)} \br)
+ \hbar {\cal A}_1^{\prime\prime} (N_\sg^{(q)})
+ \cdots
\br] + \cdots \, ,
\eea
where some terms vanish as ${\cal A}^\prime(N_\sg)=0$ and ${\cal A}_0^\prime(N_\sg^{(q)})=0$. In the saddle-point approximation and retaining terms relevant at order $\hbar$, we will get (up to some irrelevant numerical factor)
\beq
\label{eq:LDordI}
\int_{{\cal J}_\sg} {\rm d}N_c\,
e^{i \mathcal{A}(N_c)/\hbar} 
= 
\frac{e^{i\phi_\sg}}{\sqrt{\bl\vert {\cal A}_0^{\prime\prime} (N_\sg^{(q)}) \br\rvert}}
\times 
\exp\Bl[ 
\frac{i}{\hbar} {\cal A}_0(N_\sg^{(q)})
+ i {\cal A}_1(N_\sg^{(q)}) + \cdots
\Br] \, ,
\eeq
Here, $\phi_\sigma$ is the direction of the flow lines (steepest descent) at the corresponding saddle point ($N_\sigma$). It is
\begin{equation}
\label{eq:theta_sg_exp}
    \phi_\sigma =\frac{(2k-1)\pi}{4}-\frac{\delta_{\sigma}}{2},\hspace{5mm}k\in \mathbb{Z}_{odd},
\end{equation} 
where $\delta_{\sg}$ is defined via $\mathcal{A}''(N_\sg)=|\mathcal{A}''(N_\sg)|e^{i\delta_\sg}$. In principle, there are corrections to $\phi_\sg$ and to the denominator in eq. (\ref{eq:LDordI}). However, they will play a role at higher orders in $\hbar$.

%%%%%%%%%%%%%%%%%%%%%%%%%%%%%%%%%%%%%%%%%%%%%%%
\subsection{Type-1 Degeneracy}
\label{subs:type1_dg_qc}
%%%%%%%%%%%%%%%%%%%%%%%%%%%%%%%%%%%%%%%%%%%%%%%

The presence of quantum fluctuations modifies the saddle points and their associated structures. To investigate whether this resolves the \textbf{type-1} degeneracy, one must analyze the steepest descent/ascent flow lines emanating from the quantum-corrected saddles, which we now proceed to discuss.

%%%%%%%%%%%%%%%%%%%%%%%%%%%%%%%%%%%%%%%%%%%%%%%
\subsubsection{$q_{f} > 3k/\Lam$}
\label{sss:qf>3_by_lam_qm}
%%%%%%%%%%%%%%%%%%%%%%%%%%%%%%%%%%%%%%%%%%%%%%%

%
\begin{table}[t]
\centering
\begin{tabular}{|c|c|c|c|}
\hline
$x$ & Corrected saddle & $\mathfrak{h}$ & $\mathcal{H}$ \\ 
\hline
$x<1$ 
& $N_{\pm}^{\rm nb} + \hbar {\cal N}_{\pm}^{\rm nb}$ 
& $2 k^{3/2} \bl(3+4\al \Lam\br)/\Lam$  
& $\mp 2\bigl(\Lam q_f - 3 k\bigr)^{3/2}/(\Lam\sqrt{3}) $ \\ 
& & $- \hbar \log{r}/2$ & $\pm \hbar \Phi(x,q_f)/2$ \\
& & & \\
& $N_{\pm}^{\rm \cancel{\rm nb}} + \hbar {\cal N}_{\pm}^{\rm \cancel{\rm nb}}$ 
& $-12k^{3/2}(x-1)^{3}/(\Lam x^3)  $  
& $ \mp2\bigl(\Lam q_f - 3 k\bigr)^{3/2}/(\Lam\sqrt{3}) $ \\
& & $+ 2 k^{3/2} \bl(3+4\al \Lam\br)/\Lam - \hbar \log{r}/2$ 
& $ \pm \hbar \Phi(x,q_f)/2- \hbar \pi/2$ \\ 
\hline
$x>1$
& $N_{\pm}^{\rm nb} + \hbar {\cal N}_{\pm}^{\rm nb}$ 
& $2 k^{3/2} \bl(3+4\al \Lam\br)/\Lam $  
& $\mp2\bigl(\Lam q_f - 3 k\bigr)^{3/2}/(\Lam\sqrt{3}) $ \\
&
& $- \hbar \log{r}/2$ 
& $\pm\hbar \Phi(x,q_f)/2- \hbar \pi/2$\\
& & & \\
& $N_{\pm}^{\rm \cancel{\rm nb}} + \hbar {\cal N}_{\pm}^{\rm \cancel{\rm nb}}$ 
& $-12k^{3/2}(x-1)^{3}/(\Lam x^3) $  
& $ \mp2\bigl(\Lam q_f - 3 k\bigr)^{3/2}/(\Lam\sqrt{3})  $ \\
&
& $ 2 k^{3/2} \bl(3+4\al \Lam\br)/\Lam - \hbar \log{r}/2$ & $\pm\hbar \Phi(x,q_f)/2$ \\ 
\hline
\end{tabular}
\caption{ On-shell values of $\mathfrak{h}$ and $\mathcal{H}$ at the corrected saddles for various values of $x$ and $q_f>3k/\Lam$. Clearly type-1 degeneracy between $N_{\pm}^{\rm nb}$ and $N_{\pm}^{\cancel{\rm nb}}$ is lifted upon taking quantum correction. $r$ and $\Phi(x, q_f)$ are mentioned in Eq. (\ref{eq:def_r_phi}).}
\label{tab:hH_QC_qf>3/Lam_x_ep}
\end{table}
For $q_f>3k/\Lam$, the corrections to the saddles can be computed using the expression mentioned in Eq. (\ref{eq:calNs_q_hbarCR}). The correction to the $N_\pm^{\rm nb}$ and $N_\pm^{\cancel{\rm nb}}$ saddles are given by
\bea
\label{eq:QC_cor_sads}
&&
\mathcal{N}_{\pm}^{\rm nb} 
= \mp x \bigl[4(1-x)\sqrt{3\Lam q_f k - 9k^2}\bigr]^{-1}\, ,
\notag \\
&&
\mathcal{N}_{\pm}^{\cancel{\rm nb}} 
= \pm x \bigl[4(1-x)\sqrt{3\Lam q_f k - 9k^2}\bigr]^{-1}\, ,
\eea
respectively.
At the corrected saddles, one can compute the on-shell action to first order in $\hbar$. These are mentioned in the table \ref{tab:hH_QC_qf>3/Lam_x_ep} for $0<x<1$ and $x>1$, 
where we have expressed the one-loop contribution as $i{\cal A}_1(N_c)  = i\Phi -\log{r}$ with
\bea
\label{eq:def_r_phi}
r = \bigl[ 1 - 2x + \Lam q_f x^2/(3k)\bigr]^{1/2} \quad \& \quad \Phi(x,q_f) = \arctan{\biggl(\frac{x\sqrt{\Lam q_f/3k -1}}{|1-x|}\biggr)}\, .
\eea
Examining $\mathcal{H}$ from the table \ref{tab:hH_QC_qf>3/Lam_x_ep}, it is evident that for $q_{f} > 3k/\Lambda$, introducing quantum corrections as defects resolves all the type-1 degeneracies, thereby ensuring that the flowlines don't overlap with each other. This in turn helps in determining unambiguously the {\it relevant} saddle using the Picard-Lefschetz methodology. Moreover, in non-gravitational cases, particularly in finite density QCD and fermion sign problem, it is a standard practice to incorporate the quantum corrections into the effective action when applying Picard-Lefschetz theory \cite{Nishimura:2014rxa, Nishimura:2014kla, Nishimura:2015loa, Tanizaki:2015pua, Tanizaki:2015gcv,Kanazawa:2014qma}. Once the relevant saddles are determined, Eq. (\ref{eq:LDordI}) can be used to compute the transition amplitude.

%%%%%%%%%%%%%%%%%%%%%%%%%%%%%%%%%%%%%%%%%%%%%%%
\subsubsection{$q_{f} < 3k/\Lam$}
\label{sss:qf<3_by_lam_qm}
%%%%%%%%%%%%%%%%%%%%%%%%%%%%%%%%%%%%%%%%%%%%%%%

%
\begin{table}[h]
\centering
\begin{tabular}{|c|c|c|c|}
\hline
$x$ & Corrected saddle & $\mathfrak{h}$ & $\mathcal{H}$ \\ 
\hline
$x < (1 + \mu)^{-1}$
& $N_{\pm}^{\rm nb} + \hbar {\cal N}_{\pm}^{\rm nb}$ 
& $2 k^{3/2} \bl(3+4\al \Lam\br)/\Lam \mp 2\bigl(3 k - \Lam q_f\bigr)^{3/2}/(\Lam\sqrt{3})$  
& $ 0 $ \\
& & $- \hbar \log|1 - x(1 \mp \mu)|/2 $&\\ 
& & & \\
& $N_{\pm}^{\rm \cancel{\rm nb}} + \hbar {\cal N}_{\pm}^{\rm \cancel{\rm nb}}$ 
& $-12k^{3/2}(x-1)^{3}/(\Lam x^3) + 6k^{3/2}/\Lam+ 8\al k^{3/2}  $  
& $ - \hbar \pi/2$  \\
& & $\mp 2\bigl(3 k - \Lam q_f\bigr)^{3/2}/(\Lam\sqrt{3}) - \hbar \log|1 - x(1 \pm \mu)|/2 $ &  \\ 
\hline
$ (1 + \mu)^{-1} < x $
& $N_{+}^{\rm nb} + \hbar {\cal N}_{+}^{\rm nb}$ 
& $2 k^{3/2} \bl(3+4\al \Lam\br)/\Lam - 2\bigl(3 k - \Lam q_f\bigr)^{3/2}/(\Lam\sqrt{3})$  
& $ 0 $ \\
$< (1 - \mu)^{-1}$ & 
& $- \hbar \log|1 - x(1 - \mu)|/2 $&\\ 
& & & \\
& $N_{-}^{\rm nb} + \hbar {\cal N}_{-}^{\rm nb}$ 
& $2 k^{3/2} \bl(3+4\al \Lam\br)/\Lam + 2\bigl(3 k - \Lam q_f\bigr)^{3/2}/(\Lam\sqrt{3})$  
& $ -\hbar\pi/2 $ \\
& & $- \hbar \log|1 - x(1 + \mu)|/2 $&\\ 
& & & \\
& $N_{+}^{\rm \cancel{\rm nb}} + \hbar {\cal N}_{+}^{\rm \cancel{\rm nb}}$ 
& $-12k^{3/2}(x-1)^{3}/(\Lam x^3) + 6k^{3/2}/\Lam+ 8\al k^{3/2}  $  
& $ 0 $  \\
& & $- 2\bigl(3 k - \Lam q_f\bigr)^{3/2}/(\Lam\sqrt{3}) - \hbar \log|1 - x(1 + \mu)|/2 $ &  \\ 
& & & \\
& $N_{-}^{\rm \cancel{\rm nb}} + \hbar {\cal N}_{-}^{\rm \cancel{\rm nb}}$ 
& $-12k^{3/2}(x-1)^{3}/(\Lam x^3) + 6k^{3/2}/\Lam+ 8\al k^{3/2}  $  
& $ - \hbar \pi/2$  \\
& & $+ 2\bigl(3 k - \Lam q_f\bigr)^{3/2}/(\Lam\sqrt{3}) - \hbar \log|1 - x(1 -\mu)|/2 $ &  \\ 
\hline
$ (1 - \mu)^{-1} < x$
& $N_{\pm}^{\rm nb} + \hbar {\cal N}_{\pm}^{\rm nb}$ 
& $2 k^{3/2} \bl(3+4\al \Lam\br)/\Lam \mp 2\bigl(3 k - \Lam q_f\bigr)^{3/2}/(\Lam\sqrt{3})$  
& $ - \hbar \pi/2 $ \\
& & $- \hbar \log|1 - x(1 \mp \mu)|/2 $&\\ 
& & & \\
& $N_{\pm}^{\rm \cancel{\rm nb}} + \hbar {\cal N}_{\pm}^{\rm \cancel{\rm nb}}$ 
& $-12k^{3/2}(x-1)^{3}/(\Lam x^3) + 6k^{3/2}/\Lam+ 8\al k^{3/2}  $  
& $ 0 $  \\
& & $\mp 2\bigl(3 k - \Lam q_f\bigr)^{3/2}/(\Lam\sqrt{3}) - \hbar \log|1 - x(1 \pm \mu)|/2 $ &  \\ 
\hline
\end{tabular}
\caption{On-shell values of $\mathfrak{h}$ and $\mathcal{H}$ at the corrected saddles for $q_f<3k/\Lam$ for various values of $x$. Observe that quantum correction can't resolve all the type-1 degeneracies. Some residual degeneracies still exist.}
\label{tab:hH_QC_qf<3/Lam_x_ep}
\end{table}
When $q_{f} < 3k/\Lambda$, all the (background) saddles lie along the imaginary axis. The phase of the on-shell action vanishes ($\mathcal{H}=0$) at all saddles, in contrast to the $q_f>3k/\Lam$ case. Thus, in the absence of any {\it defect} (either {\it artificial} or quantum effects), it is expected that flow-lines will overlap and generate type-1 degeneracy. However in the presence of quantum effects some of these degeneracies are resolved. While for $q_f>3k/\Lam$ quantum corrections completely resolve the degeneracy, for $q_f<3k/\Lam$ it resolves only partially. This becomes clearer when $\mathfrak{h}$ and $\mathcal{H}$ are analyzed for $q_f<3k/\Lam$ for various values of $x$, to leading order in $\hbar$. Defining a new parameter $\mu = \sqrt{1-\Lam q_{f}/3k}$, the range of $x$ splits into three categories: $0 < x < (1 + \mu)^{-1}$, $(1 + \mu)^{-1} < x < (1 - \mu)^{-1}$ and $(1 - \mu)^{-1}< x$. For each of these range, the value of $\mathfrak{h}$ and $\mathcal{H}$ at the corrected saddles are mentioned in the table \ref{tab:hH_QC_qf<3/Lam_x_ep} \footnote{We have verified numerically that the phase factors listed for $q_{f} < 3k/\Lambda$ remain the same to all orders in $\hbar$.}. 
From the results mentioned in tables \ref{tab:hH_QC_qf<3/Lam_x_ep}, it is evident that the presence of quantum corrections doesn't lift the degeneracies completely. Two immediate questions arise: why can't it resolve all the degeneracies, and second, what resolves the residual degeneracies? Is it possible to come up with a proposal which systematically resolves type-1 degeneracies for all possible situations? Quantum correction is something which can't be avoided, but is unfortunately not enough to lift all the degeneracies alone. We address the first question of ``why" in Sec. \ref{sec:sym_type1}. The second question is addressed in Sec. \ref{complex_Ghbar}, where it is seen that rotation of Newton's constant $G$ by a complex phase helps in overcoming the residual degeneracies.

%%%%%%%%%%%%%%%%%%%%%%%%%%%%%%%%%%%%%%%%%%%%%%%
\subsection{Type-2 Degeneracy}
\label{subs:type2_dg_qc}
%%%%%%%%%%%%%%%%%%%%%%%%%%%%%%%%%%%%%%%%%%%%%%%
%
Type-1 is not the only degeneracy that is present in this system. As seen before, the type-2 degeneracy also occurs for special values of the boundary parameters where saddles coalesce. There are cases when the flow-line from these coalescing saddles overlaps and generates type-1 degeneracy. Hence, one needs to break the degeneracy to determine the relevant saddles. 
As quantum correction is naturally present in the system, one wonders if it can resolve it. 

As seen in subsection \ref{subs:type2_dg}, there are three possibilities that emerges in the case of type-2 degeneracy: 
\begin{itemize}
\item Final boundary parameter $q_f = 3k/\Lam$
\item Initial boundary parameter $x=1$
\item Both initial and final boundary parameter satisfying $x=1$ and $q_f = 3k/\Lam$ respectively.
\end{itemize}
In all these cases, in the absence of any kind of {\it defect}, the saddle-point approximation breaks down. One can compute the path integral by going to the third order once the {\it relevant} saddle is determined. Such a degenerate situation in the theory emerges only semiclassically. If the theory itself gets modified, then the same boundary parameters need not give rise to degeneracies. In particular, when the quantum corrections are taken, these degeneracies are absent, as will be seen below.  For the above choice of parameters, one finds that $\bar{S}_{tot}^{\prime\prime}(N_{\sg}^{(q)})$ or $\mathcal{A}_{0}^{\prime\prime}(N_{\sg}^{(q)})$ vanishes with $\mathcal{A}_{0}^{\prime\prime\prime}(N_{\sg}^{q}) \neq 0$ , making Eqs. (\ref{eq:Nsg_1lp_form}), (\ref{eq:ANc_sad_exp_1lp}) and (\ref{eq:Hbar_sadcor_genform}) inappropriate for the situation. For these degenerate situations, the perturbative expansion of saddle point reads as
\beq
\label{eq:deg_pert_sad_exp}
N_{\sg} = N_{\sg}^{(q)} + \hbar^{1/2} \mathcal{N}_{\sg} ^{(q)} + ...\, .
\eeq
Plugging eq. (\ref{eq:deg_pert_sad_exp}) in the saddle point equation $\mathcal{A}_{\sg}^{\prime}(N_{\sg}) = 0$, and expanding in $\hbar$, we get
\beq
\label{eq:ANc_deg_sad_exp_1lp}
{\cal A}_0^\prime(N_\sg^{(q)})
+ \hbar [{\cal N}_\sg^{(q)}]^2 {\cal A}_0^{\prime\prime\prime}(N_\sg^{(q)})/2
+ \hbar {\cal A}_1^\prime(N_\sg^{(q)}) = 0 \, .
\eeq
At the zeroth-order we get $\mathcal{A}_{0}^{\prime}(N_{\sg}^{(q)}) = 0$, which by satisfied by definition. This gives the leading order correction to the saddle point for the \textbf{type-2} degeneracy cases as:
\beq
\label{eq:deg_sad_1lo_exp}
\mathcal{N}_{\sg}^{(q)} = \pm i [ 2\mathcal{A}_{1}^{\prime}(N_{\sg}^{(q)})/\mathcal{A}_{0}^{\prime\prime\prime}(N_{\sg}^{(q)}) ]^{1/2} \, .
\eeq
 From Eq. \ref{eq:deg_sad_1lo_exp} it is clear that in the presence of quantum corrections the \textbf{type-2} degeneracies get resolved, where each degenerate saddle splits into two. However, to see if there are any \textbf{type-1} degeneracies present between these corrected saddles, one needs to analyze $\mathfrak{h}$ and $\mathcal{H}$, which we discuss now.
%%%%%%%%%%%%%%%%%%%%%%%%%%%%%%%%%%%%%%%%%%%%%%%
\subsubsection{$q_{f} = 3k/\Lam$}
\label{sss:qf=3_by_lam_qm}
%%%%%%%%%%%%%%%%%%%%%%%%%%%%%%%%%%%%%%%%%%%%%%%

%%%%%%%%%%%%%%%%%%%%%%%%%%%%%%%%%%%%%%%%%%%%%%%
\begin{figure}[h]
\centering
\includegraphics[trim={2.2cm 0 0 0},clip,
scale=0.85
%\textwidth
]
{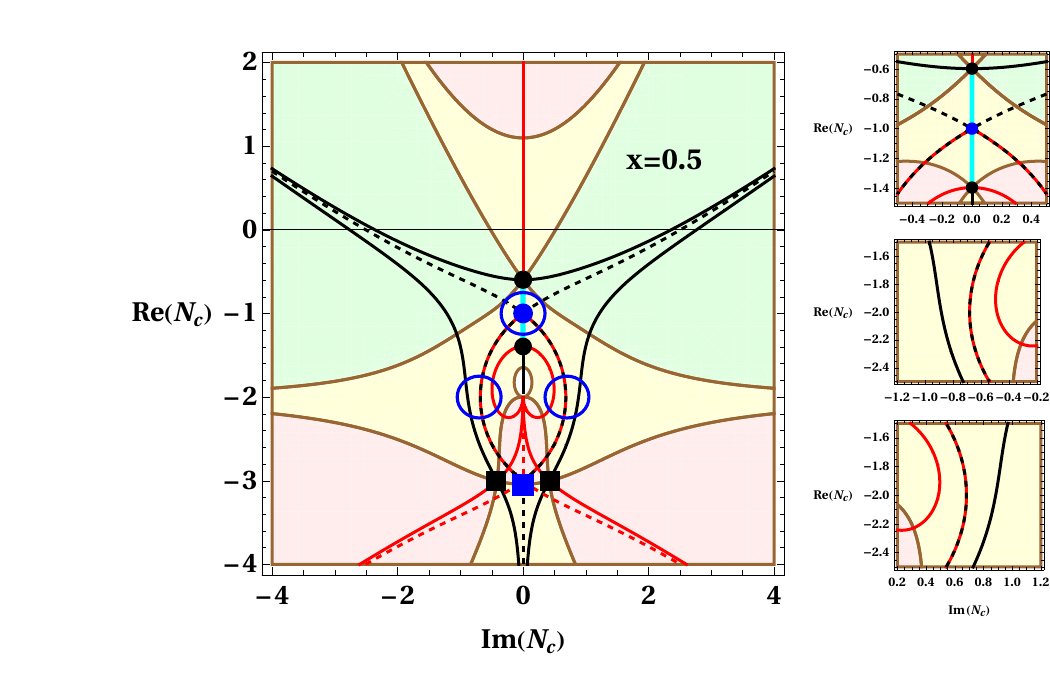}
\caption{
Plot depicting type-2 degeneracy of first kind, where not only there is overlap of flow lines, but also due to the choice of external parameter $q_f=3k/\Lam$ leading to merging of the saddles. The plot depicts the effects of quantum corrections on the degenerate situation. Red lines are the steepest ascent lines, while black lines are the steepest descent lines. The background saddles are represented by blue dot and square. The quantum corrected saddles are depicted black dots and squares. For illustration purpose, we set $P_{i} = -3i\sqrt{k}, \al = 2$, $\Lam=3$, $x = 1/2$, $\hbar = 1$ and $k=1$. Note, that the allowed and forbidden region also gets corrected due to the quantum effects too. The two degenerate saddles split: the upper blue dot saddle into two black dots, and the lower blue square splits into two black squares. The residual degeneracy (type-1) is shown by the vertical line in cyan.}
\label{fig:PL_wwo_QM_qfqc_x0.5}
\end{figure}
%%%%%%%%%%%%%%%%%%%%%%%%%%%%%%%%%%%%%%%%%%%%%%%

The case when $q_f=3k/\Lam$ (and $x \neq 1$) the set of no-boundary saddles merge among themselves and the same happens for non-no-boundary saddles (see Eq. (\ref{eq:MEG_sad_qf=3/lam})). Besides this, there is also type-1 degeneracy present between these two degenerate saddles, see Fig. \ref{fig:PL_wwo_QM_qfqc_x0.5}. In the previous section, we noticed that artificial {\it defect} neatly resolves both the degeneracies. It is worth asking if the quantum correction present in the theory itself can offer a {\it cure} to this degenerate situation? When contributions from quantum effects are incorporated, the corrected saddle point equation for $q_{f}=3k/\Lam$ is given by
\bea
\label{eq:sad_qf_qc_2_deg_QC_dft}
x \Lam^2 \bigl(N_{\sigma} - N^{\rm nb}\bigr)^2 \bigl(N_{\sigma} - N^{\cancel{\rm nb}}\bigr)^2 +  \hbar \bigl( 6 i \Lam x N_s - 18 \sqrt{k} \bigr)  = 0\, .
\eea
\begin{table}[h]
\centering
\begin{tabular}{|c|c|c|c|}
\hline
$x$ & Corrected saddle & $\mathfrak{h}$ & $\mathcal{H}$ \\ 
\hline
$0<x<1$ 
& $N^{\rm nb} + \sqrt{\hbar} \, {\cal N}_{\pm}^{\rm nb}$ 
& $2 k^{3/2} \bl(3+4\al \Lam\br)/\Lam - \hbar \log{(1-x)}/2$  
& $ 0 $ \\
& & $ \pm 6\Lam^2 \hbar^{3/2} (x/(18\Lam\sqrt{k}(1-x)))^{3/2} $ & \\
& & & \\
& $N^{\rm \cancel{\rm nb}} + \sqrt{\hbar} \,  {\cal N}_{\pm}^{\cancel{\rm nb}}$ 
& $-12k^{3/2}(x-1)^{3}/(\Lam x^3) + 6k^{3/2}/\Lam$  
& $-\hbar \pi/2 \pm  6\Lam^2 \hbar^{3/2}$ \\
& & $+ 8\al k^{3/2} -\hbar \log(1-x)/2 $ 
& $\times(x/(18\Lam\sqrt{k}(1-x)))^{3/2}$ \\ 
\hline
$x>1$ & $N^{\rm nb} + \sqrt{\hbar} \,  {\cal N}_{\pm}^{\rm nb}$ 
& $2 k^{3/2} \bl(3+4\al \Lam\br)/\Lam - \hbar \log{(x-1)}/2$  
& $  -\hbar \pi/2 \pm 6\Lam^2 \hbar^{3/2}$ \\
& & & $\times (x/(18\Lam\sqrt{k}(x-1)))^{3/2}$ \\
& & & \\
 & $N^{\rm \cancel{\rm nb}} + \sqrt{\hbar} \,  {\cal N}_{\pm}^{ \cancel{\rm nb}}$ 
& $-12k^{3/2}(x-1)^{3}/(\Lam x^3) + 6k^{3/2}/\Lam  $  
& $ 0 $   \\ 
& & $ + 8\al k^{3/2} -\hbar \log{(x-1)}/2$ & \\ 
& & $\pm 6\Lam^2 \hbar^{3/2} (x/(18\Lam\sqrt{k}(x-1)))^{3/2}$ & \\
\hline
\end{tabular}
\caption{First order correction to $\mathfrak{h}$ and $\mathcal{H}$ due to the quantum effects in the case of type-2 degeneracy with $q_f=3k/\Lam$ for various $x>0$. Although \textbf{type-2} degeneracies are resolved, \textbf{type-1}  degeneracy is still present for $\mathcal{H}=0$.}
\label{tab:hH_QC_qf=3/Lam_x}
\end{table}
Although the above equation can be solved exactly, it will be sufficient to analyze it in the leading order in $\hbar$. The first-order corrections to the saddles are given by
\beq
\label{eq:quantum_cor_x_deg}
\mathcal{N}^{\rm nb}_\pm = \pm \Bigl(\frac{x}{2\sqrt{k}\Lam(x-1)}\Bigr)^{1/2} \, ,
\hspace{5mm}
\mathcal{N}^{\cancel{\rm nb}}_\pm = \pm \Bigl(\frac{x}{2\sqrt{k}\Lam(1-x)}\Bigr)^{1/2}\, .
\eeq
This clearly shows that the inclusion of quantum contributions breaks the coalescence of the saddles, where two degenerate saddles split into four non-degenerate saddles.

It is now worthwhile to investigate whether the presence of quantum correction also lifts the overlapping of the flow-lines (type-1 degeneracy)? The answer is partly yes; quantum correction lifts only some of the degeneracies; it can't lift all. Some residual type-1 degeneracy between the corrected saddle is still present for $\mathcal{H}=0$, as can be seen from table \ref{tab:hH_QC_qf=3/Lam_x}. However, there is no degeneracy left between the corrected  $N^{\cancel{\rm nb}}$ saddles for $0 < x <1$ as $\mathcal{H}$'s are different (see table \ref{tab:hH_QC_qf=3/Lam_x})\footnote{We have verified analytically to the subleading order and numerically to all order in $\hbar$.}. Similarly, for $x>1$: the flow-lines between corrected $N^{\rm nb}$ saddles no longer overlap as $\mathcal{H}$'s are different. There persists type-1 degeneracy between the corrected $N^{{\rm nb}}$ and $N^{\cancel{\rm nb}}$ saddles for $0<x<1$ and $x>1$ respectively. The figure \ref{fig:PL_wwo_QM_qfqc_x0.5} gives a pictorial representation of the same for the case of $x=1/2$.

%%%%%%%%%%%%%%%%%%%%%%%%%%%%%%%%%%%%%%%%%%%%%%%
\subsubsection{$x=1$}
\label{sss:x=1_qm}
%%%%%%%%%%%%%%%%%%%%%%%%%%%%%%%%%%%%%%%%%%%%%%%

\begin{table}[h]
\centering
\begin{tabular}{|c|c|c|c|}
\hline
$q_f$ & Corrected saddle & $\mathfrak{h}$ & $\mathcal{H}$ \\ \hline
& $N_{+}^{\rm nb} + \hbar^{1/2} \mathcal{N}_{\pm}^{\rm nb}$ 
& $2 k^{3/2} \bl(3+4\al \Lam\br)/\Lam $  
& $ - 2\bigl(\Lam q_f - 3 k\bigr)^{3/2}/(\Lam\sqrt{3})$ \\
&
& $ - \hbar \log{r}/2 \pm \sqrt{\Lam} \hbar^{3/2} $ 
& $ - \hbar \pi/4 \mp \sqrt{\Lam} \hbar^{3/2} $\\
&
& $\times (\Lam q_f/3 - k)^{-3/4}/9$ & $\times (\Lam q_f/3 - k)^{-3/4}/9 $\\
$q_f>3k/\Lam$ & & & \\
&
$N_{-}^{\rm nb} + \hbar^{1/2} \mathcal{N}_{\pm}^{\rm nb}$ 
& $2 k^{3/2} \bl(3+4\al \Lam\br)/\Lam $  
& $  2\bigl(\Lam q_f - 3 k\bigr)^{3/2}/(\Lam\sqrt{3})$ \\
&
& $- \hbar \log{r}/2\pm \sqrt{\Lam} \hbar^{3/2} $ 
& $- \hbar \pi/4\pm \sqrt{\Lam} \hbar^{3/2} $  \\ 
&
& $\times (\Lam q_f/3 - k)^{-3/4}/9 $ & $\times (\Lam q_f/3 - k)^{-3/4}/9 $\\
\hline
& $N_{+}^{\rm nb} + \hbar^{1/2} \mathcal{N}_{\pm}^{\rm nb}$ 
& $2 k^{3/2} \bl(3+4\al \Lam\br)/\Lam $  
& $  0 $ \\
&
& $- 2\bigl( 3 k - \Lam q_f \bigr)^{3/2}/(\Lam\sqrt{3})$
& \\
&
& $- \hbar \log{r}/2\pm \sqrt{2\Lam} \hbar^{3/2}  $ & \\
&
& $\times (k - \Lam q_f/3 )^{-3/4}/9$ & \\
$q_f<3k/\Lam$& & & \\
& $N_{-}^{\rm nb} + \hbar^{1/2} \mathcal{N}_{\pm}^{\rm nb}$ 
& $2 k^{3/2} \bl(3+4\al \Lam\br)/\Lam $  
& $ -\hbar \pi/2 \pm \sqrt{2\Lam} \hbar^{3/2}$ \\
&
& $+ 2( 3 k - \Lam q_f \bigr)^{3/2}/(\Lam\sqrt{3})$ 
& $\times (k - \Lam q_f/3 )^{-3/4}/9$ \\ 
&
& $- \hbar \log{r}/2 $ & \\
\hline
\end{tabular}
\caption{First order correction to $\mathfrak{h}$ and $\mathcal{H}$ due to quantum effects in the case of type-2 degeneracy with $x=1$ and for various values of $q_f$. Notice that \textbf{type-1} degeneracy is still present whenever $\mathcal{H}=0$.}
\label{tab:hH_QC_qfx=1_ep}
\end{table}

In the case when $x=1$ (but $q_f \neq 3k/\Lam$), we notice that the non-no-boundary saddles merge with the no-boundary saddles, when quantum corrections are not incorporated (see Eq. (\ref{eq:MEG_sad_x=1})). Similar to the previous case, it is expected that the quantum fluctuations will lift the degeneracies. Here, we investigate it explicitly. When quantum correction is taken into account, the saddle point equation for $x=1$ reads
\bea
\label{eq:sadQ_x_1_deg}
\Lam^3\bigl(N_{s} - N_{+}^{\rm nb}\bigr)^2 \bigl(N_{s} - N_{-}^{\rm nb}\bigr)^2 + \hbar \bigl(6i\Lam N_{s} - 18 \sqrt{k}\bigr) = 0\, .
\eea
From the above equation, it is clearly evident that in the semiclassical limit ($\hbar\to0$), one encounters a degenerate situation where only two saddles are present $N_{+}^{\rm nb}$ and $N_{-}^{\rm nb}$. When $\hbar$ effects are incorporated, they receive correction. After incorporating the quantum corrections the set of new saddles can be written as $N_{\pm}^{\rm nb, \pm} = N^{\rm nb}_\pm + \sqrt{\hbar} {\cal N}^{\rm nb, \pm}_{\pm}$, where ${\cal N}^{\rm nb, \pm}_{\pm}$ is given by
\bea
\label{eq:quantum_cor_x_deg1}
\mathcal{N}^{\rm nb, \pm}_{+} && = \pm e^{-i\pi/4}\bigl(\Lam q_{f}/3 -k\bigr)^{-1/4}/\sqrt{2\Lam}\, , \notag \\
\mathcal{N}^{\rm nb, \pm}_{-} && = \pm e^{i\pi/4}\bigl(\Lam q_{f}/3 -k\bigr)^{-1/4}/\sqrt{2\Lam}\, .
\eea
This clearly shows that the type-2 degeneracy is lifted. Furthermore, examining $\mathcal{H}$ from table \ref{tab:hH_QC_qfx=1_ep}, we see that \textbf{type-1} degeneracy is not completely resolved. It is present when $\mathcal{H}=0$, which happens for $q_f<3k/\Lam$. For other cases, for example when $N_-^{\rm nb}$ splits into two saddles, there are no further degeneracies as $\mathcal{H}$ is different at the corrected saddles.

%%%%%%%%%%%%%%%%%%%%%%%%%%%%%%%%%%%%%%%%%%%%%%%
\subsubsection{$x=1$ and $q_f=3k/\Lam$}
\label{sss:x=1_qf3_by_Lam_qm}
%%%%%%%%%%%%%%%%%%%%%%%%%%%%%%%%%%%%%%%%%%%%%%%
%
\begin{figure}
    \centering
    \hspace{-2cm}
    \includegraphics[trim={2.5cm 0 2.5cm 0},clip,scale=0.7]{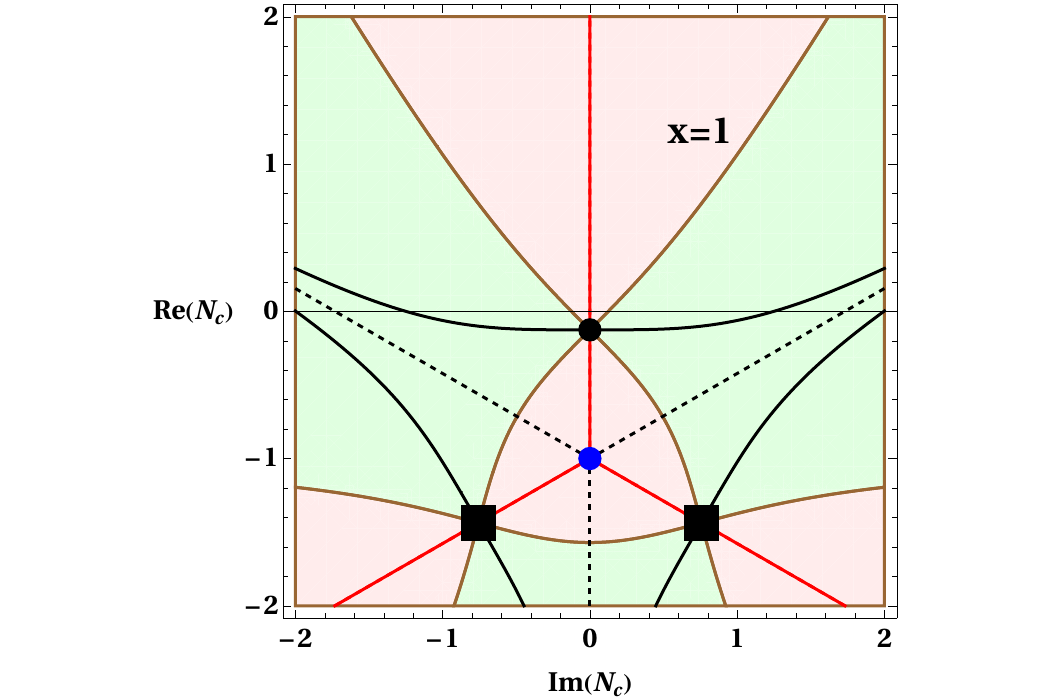}
    \caption{A plot similar to the Fig. (\ref{fig:PL_wwo_QM_qfqc_x0.5}) is shown for $x=1$ illustrating the doubly degenerate situation. The single, doubly degenerate no-boundary saddle indicated by blue dot has split into three non-degenerate saddles black markers upon including quantum corrections (corrections are amplified for visual purposes). Moreover, the black dot represents the relevant quantum corrected saddle point, where black squares are the irrelevant ones.}
    \label{fig:PL_w_wo_QM_qfqc_x1_new}
\end{figure}

For the doubly degenerate case $x = 1$ and $q_{f} = 3k/\Lam$ case, it turns out that one can compute exact expressions for the quantum corrected saddles. The quantum corrected saddle point equation for the double degenerate case is given by
\beq
\label{eq:Sad_Q_DD_deg}
\frac{6\hbar \Lam - (3\sqrt{k} - i \Lam N_{s})^3}{(3\sqrt{k} - i \Lam N_{s})} = 0 \, .
\eeq
Thus, in the semiclassical limit $\hbar\to 0$, we have $N_{\sigma} = -3i\sqrt{k}/\Lam$ as the doubly degenerate point. However, with quantum corrections, we have the exact corrected saddles as
\beq
\label{eq:Sad_Q_DD_deg}
N_{\sigma}^{(n)} = - 3i\Lam/\sqrt{k} + i \bigl( 6\hbar/\Lam^2\bigr)^{1/3} e^{2i\pi n/3} \, , \quad n = \{0,1,2\} .
\eeq
Note that in the double-degenerate case, the correction appears at the order $\hbar^{1/3}$, unlike the earlier two-degenerate cases. In this case, a single degenerate saddle splits into three non-degenerate saddles, see fig. (\ref{fig:PL_w_wo_QM_qfqc_x1_new}). Also note that the corrected saddles as mentioned in Eq. (\ref{eq:Sad_Q_DD_deg}) are {\it exact}, unlike the previous degenerate situations where saddles get corrected to all orders in $\hbar$. The on-shell $\hbar$-corrected $\mathfrak{h}$ and $\mathcal{H}$ at the quantum corrected saddles are given in table \ref{tab:hH_QC_qf=3/Lam_x=1_hbar}. From the table, it is clearly visible that the quantum fluctuations resolve this degeneracy as not only the doubly-degenerate saddle splits into three, but also each of the corrected saddles has a different $\mathcal{H}$ as shown in the table \ref{tab:hH_QC_qf=3/Lam_x=1_hbar}. 
\begin{table}[h]
\centering
\begin{tabular}{|c|c|c|}
\hline
$\hbar$ corrected saddle ($n$) & $\mathfrak{h}$ & $\mathcal{H}$ \\ \hline
$N_{\sigma}^{(0)}$ 
& $2 k^{3/2} \bl(3+4\al \Lam\br)/\Lam + \hbar [ 1- \log{(2\hbar\Lam/9k^{3/2})}]/6 $  
& $0$ \\
\hline
$N_{\sigma}^{(1)}$ 
& $2 k^{3/2} \bl(3+4\al \Lam\br)/\Lam + \hbar [ 1- \log{(2\hbar\Lam/9k^{3/2})}]/6$  
& $   - \hbar \pi/3 $ \\
\hline
$N_{\sigma}^{(2)}$ 
& $2 k^{3/2} \bl(3+4\al \Lam\br)/\Lam + \hbar [ 1- \log{(2\hbar\Lam/9k^{3/2})}]/6$  
& $   \hbar \pi/3 $ \\
\hline
\end{tabular}
\caption{On-shell values of one loop $\mathfrak{h}$ and $\mathcal{H}$ at the quantum corrected saddles for the double degenerate case $q_{f} = 3k/\Lam$ and $x=1$.}
\label{tab:hH_QC_qf=3/Lam_x=1_hbar}
\end{table}
%

%%%%%%%%%%%%%%%%%%%%%%%%%%%%%%%%%%%%%%%%%%%%%%
\section{
$G\hbar$ complexification
and Type-1 degeneracies
}
\label{complex_Ghbar}
%%%%%%%%%%%%%%%%%%%%%%s%%%%%%%%%%%%%%%%%%%%%%%

In the previous sections, we observed degeneracies that occur while studying the $N_c$- integral in Eq. (\ref{eq:sch_form}) can't be resolved by quantum effects alone. Although the inclusion of artificial {\it defect} helps in resolving these issues, it comes with ambiguities in the sense that it is not clear what kind of {\it defect} should be considered for lifting degeneracies and what should be avoided? It is better if a remedy can be found to address these issues without relying on the imposition of artificial {\it defects}. 

In the present minisuperspace model, it is noticed that type-1 degeneracies, which can't be fully resolved due to the presence of quantum effects, arise for $q_f\leq 3/\Lam$. In such situations, the overlap of flowlines occur along the imaginary axis in the complex $N_c$-plane. Such degeneracies, although they get resolved completely in the presence of artificial {\it defect}, quantum effects alone are not sufficient to lift them up. An interesting proposal taking care of such situation has been considered in \cite{Dorigoni:2014hea,Honda:2024aro, Kanazawa:2014qma}, where type-1 degeneracies are resolved by introducing a small complex deformation of $\hbar$. In the presence of gravity, these ideas can be further extended to consider introducing a small complex deformation of either Newton’s constant ($G$) or Planck’s constant ($\hbar$). The former has also been recently utilized in the study involving phases of the Euclidean gravitational path integral \cite{Maldacena:2024spf, Ivo:2025yek}. While a complex deformation of $G$ affects only the gravitational sector, the complex deformation of $\hbar$ instead influences the entire path integral and affects both the gravitational and matter sectors. In either case, one will encounter a complexified Planck's length $l_p^2 = G\hbar$ and Planck's mass $M_p^2 = \bl(G\hbar \br)^{-1}$. This clever procedure of lifting type-1 degeneracies by introducing a small complex deformation of couplings or $G\hbar$ is noticed to work well, while it is unclear at this stage the deeper reason responsible for it. It could be that the symmetry responsible for degeneracies gets broken in the presence of such complex deformations. This we will address in the next section \ref{sec:sym_type1}. 

A tiny complex deformation of $G\hbar$ leads to a loss of convergence in the lapse-$N_c$ integral given in Eq. (\ref{eq:sch_form}), when the contour of integration lies along the real line. In the presence of deformation $l_p^2=\lvert l_p^2 \rvert \exp\bl(i\theta\br)$, with $\lvert \theta\rvert \ll 1$, the modified region of convergence is given by
\beq
\label{eq:theta_convg_reg}
\Bl(\frac{\theta}{3},\frac{\pi+\theta}{3} \Br)
\cup
\Bl(\frac{2\pi+\theta}{3},\pi+\frac{\theta}{3} \Br)
\cup 
\Bl(\frac{4\pi+\theta}{3}, \frac{5\pi+\theta}{3} \Br) \, ,
\eeq
where the original region of convergence has been rotated by $\theta$. In such a situation, the original contour of integration (the real line) no longer lies entirely within the region of convergence. For $\theta>0$, the integral  diverges as $N_c\to +\infty$, while it diverges as $N_c\to -\infty$ for $\theta<0$. Therefore, in order to make the integral convergent and well-defined, one needs to simultaneously rotate the original (real) contour of integration infinitesimally into the complex plane so that it lies within the new regions of convergence.

Once the contour of integration is appropriately adjusted to take into account the effects of $\theta$, we proceed to analyse the $\theta$-rotated lapse-$N_c$ integral. This is achieved by systematically analysing the integral using Picard-Lefschetz methods. In the saddle-point approximation, it is seen that complex deformations like $l_p^2=\lvert l_p^2 \rvert \exp\bl(i\theta\br)$ don't alter the saddles following from the action $ \bar{S}_{\rm tot}^{\rm on-shell}[N_c] $. However, if quantum corrections are taken into account, the saddles get $\theta$-dependence. For cases where $q_f\neq 3k/\Lam$ and $x\neq 1$, the $\theta$-induced correction to the saddles is given by
\beq
\label{eq:QC_cor_sadslprorotlp1}
\mathcal{N}_{\pm}^{\rm nb,\theta} 
= \mp \frac{2 \pi x \lvert G \rvert e^{i\theta}}{V_3(1-x)\sqrt{3\Lam q_f k - 9k^2}} \, ,
\hspace{5mm}
\mathcal{N}_{\pm}^{\cancel{\rm nb},\,\theta} 
= \pm  \frac{2 \pi x \lvert G \rvert e^{i\theta}}{V_3(1-x)\sqrt{3\Lam q_f k - 9k^2}}\, ,
\eeq
where we reinstate $V_3$ and $8\pi G$. In the limit $\theta\to0$, these agree with the expressions mentioned in Eq. (\ref{eq:QC_cor_sads}) after taking into consideration that $V_3 = 8 \pi \lvert G \rvert$. In this section, we will reinstate the factor of $G$ everywhere and won't follow the $V_3=8\pi G$ convention followed in earlier sections.

Similarly for \textbf{type-2} degeneracies, when either $q_f=3/\Lam$ and $x\neq 1$ or $x=1$ and $q_f\neq 3/\Lam$, the $\theta$-modifications to the saddles are given by
\bea
\label{eq:quantum_cor_x_deg2rotlp2}
\mathcal{N}^{\rm nb,\,\theta}_\pm = \pm e^{i\theta/2}\Bigl(\frac{ 4\pi |G|x}{V_3\sqrt{k}\Lam(x-1)}\Bigr)^{1/2}; \,\,
\mathcal{N}^{\cancel{\rm nb},\,\theta}_\pm  = \pm e^{i\theta/2} \Bigl(\frac{4\pi |G| x}{V_3\sqrt{k}\Lam(1-x)}\Bigr)^{1/2}\, ,
\eea
or
\bea
\label{eq:quantum_cor_x_deg2rotlp3}
\mathcal{N}^{\rm nb,\,\theta}_{+} && = \pm e^{i\theta/2} e^{- i\pi/4}\sqrt{4\pi |G|}\bigl(\Lam q_{f}/3 -k\bigr)^{-1/4}/\sqrt{\Lam V_3}\, ,\\
\mathcal{N}^{\rm nb,\,\theta}_{-} && = \pm e^{i\theta/2} e^{+ i\pi/4}\sqrt{4\pi |G|}\bigl(\Lam q_{f}/3 -k\bigr)^{-1/4}/\sqrt{\Lam V_3},\notag
\eea
respectively. The appropriate expressions with $\theta=0$ are mentioned in Eq. (\ref{eq:quantum_cor_x_deg}) and (\ref{eq:quantum_cor_x_deg1}) respectively. It is worth stressing, that the $\theta$-dependence in the corrections for type-1 and type-2 are not same, in the former it goes as $e^{i\theta}$ while in later it goes as $e^{i\theta/2}$. 

Complex deformations like $l_p^2=\lvert l_p^2 \rvert \exp\bl(i\theta\br)$ also affect the quantum corrected lapse-$N_c$ action mentioned in Eq. (\ref{eq:qmAact}). Such deformations eventually modify Eq. (\ref{eq:qmAact}) into following
\begin{equation}
\label{eq:qccorA(Nc)}
\frac{i}{\hbar}\mathcal{A}(N_c)
=\frac{i V_3}{8\pi}\frac{e^{-i\theta}}{\lvert l_p^2\rvert}\bar{S}_{\rm tot}^{\rm on-shell}[N_c]
-\frac{1}{2}\ln\Delta(N_c),
\hspace{5mm}
\lvert l_p^2 \rvert= \lvert G\hbar \rvert \, ,
\end{equation}
where $\bar{S}_{\rm tot}^{\rm on-shell}[N_c]$ is mentioned in Eq. (\ref{eq:stot_onsh_rbc}) and $\Delta(N_c)$ is mentioned in Eq. (\ref{eq:DelNc_rbc}). For the corrected saddles this can be computed to first order in perturbation giving 
\begin{equation}
\label{eq:commGN}
\begin{split}
&\frac{i}{\hbar}\mathcal{A}(N_c=N_\sigma^{(q)}+|\hbar|\mathcal{N}_\sigma^{(q),\,\theta})
=\frac{iV_3}{8\pi}\frac{e^{-i\theta}}{|G\hbar|}S_{\rm tot}^{\rm on-shell}[N_\sigma^{(q)}]-\frac{1}{2}\ln\Delta(N_\sigma^{(q)})+\mathcal{O}(\hbar),\\
&\frac{i}{\hbar}\mathcal{A}(N_c=N_\sigma^{(q)}+|\sqrt{\hbar}|\mathcal{N}_\sigma^{(q),\,\theta})
=\frac{iV_3}{8\pi}\frac{e^{-i\theta}}{|G\hbar|}S_{\rm tot}^{\rm on-shell}[N_\sigma^{(q)}]-\frac{1}{2}\ln\Delta(N_\sigma^{(q)})\\
&\hspace{10mm}-|\hbar^{1/2}|\underbrace{\left[\frac{\Delta'(N_\sigma^{(q)})}{2\Delta(N_\sigma^{(q)})}\mathcal{N}_\sigma^{(q),\,\theta}-\frac{iV_3}{48\pi}\frac{e^{-i\theta}}{|G|}S_{\rm tot}^{{\rm on-shell}'''}[N_\sigma^{(q)}](\mathcal{N}_\sigma^{(q),\,\theta})^3\right]}_{\substack{
\text{Opposite $\theta$-rotation ($e^{i\theta/2}$)} \\
\text{relative to leading term}
}}+\mathcal{O}(\hbar),
\end{split}
\end{equation}
for the \textbf{type-1} and \textbf{type-2} cases, respectively. $S_{\rm tot}^{{\rm on-shell}'''}[N_\sigma^{(q)}]$ is $d^3S_{\rm tot}^{{\rm on-shell}}[N_c]/dN_c^3$ evaluated at the background saddles $N_\sigma^{(q)}$. Here, $N_\sigma^{(q)}$ is background saddle given in Eq. (\ref{eq:nb_sads}) and $\mathcal{N}_\sigma^{(q),\,\theta}$ is given in Eq. (\ref{eq:QC_cor_sadslprorotlp1}),(\ref{eq:quantum_cor_x_deg2rotlp2}), and (\ref{eq:quantum_cor_x_deg2rotlp3}). Although the corrections to the saddles and the action depend on the specific sign of $\theta$, either choice is sufficient to resolve the degeneracies. However, it has been noticed in the recent studies involving gravitational phases \cite{Maldacena:2024spf, Ivo:2025yek}, a particular choice is favorable. This is given by
\begin{equation}
\label{eq:maldacena}
\frac{1}{G}=\frac{1}{|G|}(1-i\epsilon)\, ,
\hspace{6mm} 0<\ep \ll 1
\end{equation}
with $\hbar$ being real. In order to be consistent with these studies, we will choose $\theta$ appropriately thereby giving $\theta=\epsilon$. Once the choice of $\theta$ is fixed, the degeneracy of overlapping flowlines gets broken accordingly. This can be explicitly seen by considering the quantity $iS_{\rm tot}^{\rm on-shell}(N_\sigma^{(q)})=h(N_\sigma^{(q)})+iH(N_\sigma^{(q)})$ at the saddles. When this is combined with the expression of quantum corrected action stated in eq. (\ref{eq:commGN}), and writing $i {\cal A}(N_c) = h^\ep(N_c) + i H^\ep(N_c)$, then it is noticed that at the saddle point with the convention of $G$-rotation taken as in eq. (\ref{eq:maldacena}), we have 
\begin{equation}
\label{eq:type1_grot_qm}
\begin{split}
\frac{i}{\hbar}\mathcal{A}(N_c=N_\sigma^{(q)}+\hbar\mathcal{N}_\sigma^{(q),\,\epsilon})
= &\,\frac{V_3}{8\pi\hbar}\frac{1}{|G|}(1-i\epsilon)\left[h(N_\sigma^{(q)})+iH(N_\sigma^{(q)})\right]\\
&-\frac{1}{2}\ln\Delta(N_\sigma^{(q)})+\mathcal{O}(\hbar) \, ,
\end{split}
\end{equation}
for the case of type-1 degeneracies. From Eq. (\ref{eq:type1_grot_qm}), we observe that the presence of non-zero $\epsilon$ introduces an extra phase given by
\begin{equation}
\label{eq:phasetype1}
H^\epsilon (N_\sigma^{(q)})=-\frac{V_3}{8\pi\hbar}\frac{\epsilon}{|G|}h(N_\sigma^{(q)}).
\end{equation}
For any two background saddles ($N_\sigma^{(q)}$) which are connected via overlapping flow-lines have same $H(N_\sigma^{(q)})$ (type-1 degeneracy); however, $h(N_\sigma^{(q)})$ is always different at these saddles by the definition. This implies, in the presence of complex deformation, $H^\epsilon (N_\sigma^{(q)})$ which is given in eq. (\ref{eq:phasetype1}) is then different for these saddles, thereby leading to type-1 degeneracy resolution due to complex deformations. 

For the case of type-2 degeneracies, where the background saddles merge for specific choices of boundary parameters, and have overlapping flow-lines due to merging of saddles, quantum effects and complex deformations are together needed for successful resolution of degeneracies. This is because quantum corrections help in splitting the saddles but not enough to prevent the overlapping flow-lines connecting the two split-saddles. This emergence of \textbf{type-1} degeneracy happens at a higher order in $\hbar$ than the previous cases. It is so because the Morse function ($h(N_\sigma^{(q)})$) begins to differ at the corrected saddles at order $\mathcal{O}(\hbar^{1/2})$ . In contrast, in the previous case the Morse function $(h(N_\sigma^{(q)}))$ was different even at the leading order $\mathcal{O}(1/\hbar)$. Substituting $iS_{\rm tot}^{\rm on-shell}(N_\sigma^{(q)})=h(N_\sigma^{(q)})+iH(N_\sigma^{(q)})$ in Eq. (\ref{eq:commGN}) with the complexified $G$ as mentioned in Eq. (\ref{eq:maldacena}), we have
\bea
\label{eq:db_type2_cd}
&&\frac{i}{\hbar}\mathcal{A}(N_c=N_\sigma^{(q)}+\sqrt{\hbar}\,\mathcal{N}_\sigma^{(q),\,\epsilon})
=\frac{V_3}{8\pi\hbar}\frac{1}{|G|}(1-i\epsilon)\left[h(N_\sigma^{(q)})+iH(N_\sigma^{(q)})\right]-\frac{1}{2}\ln\Delta(N_\sigma^{(q)})
\notag \\
&&
-\hbar^{1/2}|G|^{1/2}\left(\frac{8\pi}{V_3}\right)^{1/2}\underbrace{\biggl[\frac{\Delta'(N_\sigma^{(q)})}{2\Delta(N_\sigma^{(q)})}\mathcal{N}_\sigma^{(q)}-\frac{i}{6}S_{\rm tot}^{{\rm on-shell}'''}[N_\sigma^{(q)}](\mathcal{N}_\sigma^{(q)})^3\biggr]}_{\text{``X"}}\times \biggl(1+\frac{i\epsilon}{2}\biggr)+\mathcal{O}(\hbar)
\, ,
\notag \\
&&
= \frac{V_3 (1-i\ep)}{8 \pi \hbar |G|}
\left[h + iH \right]\Bl\rvert_{N_\sigma^{(q)}}
-\frac{\ln\Delta(N_\sigma^{(q)})}{2}
+\left(\epsilon-2i\right)
\left(\frac{8\pi \hbar |G| \Lam^4}{81V_3}\right)^{1/2}
%S_{\rm tot}^{{\rm on-shell}'''}[N_\sigma^{(q)}]
(\mathcal{N}_\sigma^{(q)})^3 +\mathcal{O}(\hbar),
\eea
where $\mathcal{N}_\sigma^{(q)}$ is the correction to the saddles due to the quantum correction with real $G$, given in Eq. (\ref{eq:quantum_cor_x_deg}) and (\ref{eq:quantum_cor_x_deg1}). The quantity defined by ``X" has been simplified using the relation $\Delta'(N_\sigma^{(q)})/\Delta(N_\sigma^{(q)}) = i S_{\rm tot}^{{\rm on-shell}'''}[N_\sigma^{(q)}](\mathcal{N}_\sigma^{(q)})^2$, which at the saddle point further simplifies by utilizing $S_{\rm tot}^{{\rm on-shell}'''}(N_\sigma^{(q)}) = 2\Lam^2/3$. The existence of type-1 degeneracy relies on the requirement of the quantity defined by ``X" in Eq. (\ref{eq:db_type2_cd}) to be real. From the second line of eq. (\ref{eq:db_type2_cd}) we notice that this happens when the corrections $\mathcal{N}_\sigma^{(q)}$ is purely imaginary. This is indeed the case as can be seen from the expression of corrected saddles mentioned in Eqs. (\ref{eq:quantum_cor_x_deg2rotlp2}) and (\ref{eq:quantum_cor_x_deg2rotlp3}) for suitable ranges of $x$ and $q_f$ with $\theta=0$. The explicit algebraic form for ``X" after substituting $\mathcal{N}_\sigma^{(q)}$ from Eqs. (\ref{eq:quantum_cor_x_deg2rotlp2}) and (\ref{eq:quantum_cor_x_deg2rotlp3}), can be read off from tables \ref{tab:hH_QC_qf=3/Lam_x} and \ref{tab:hH_QC_qfx=1_ep} where $X$ is the coefficient of $\hbar^{3/2}$ in the Morse function ($h$). From these, the emergence of type-1 degeneracy becomes transparent. 

\begin{figure}%[H]
    \centering
\includegraphics[trim={2cm 0 0  0},clip,
scale=0.8]{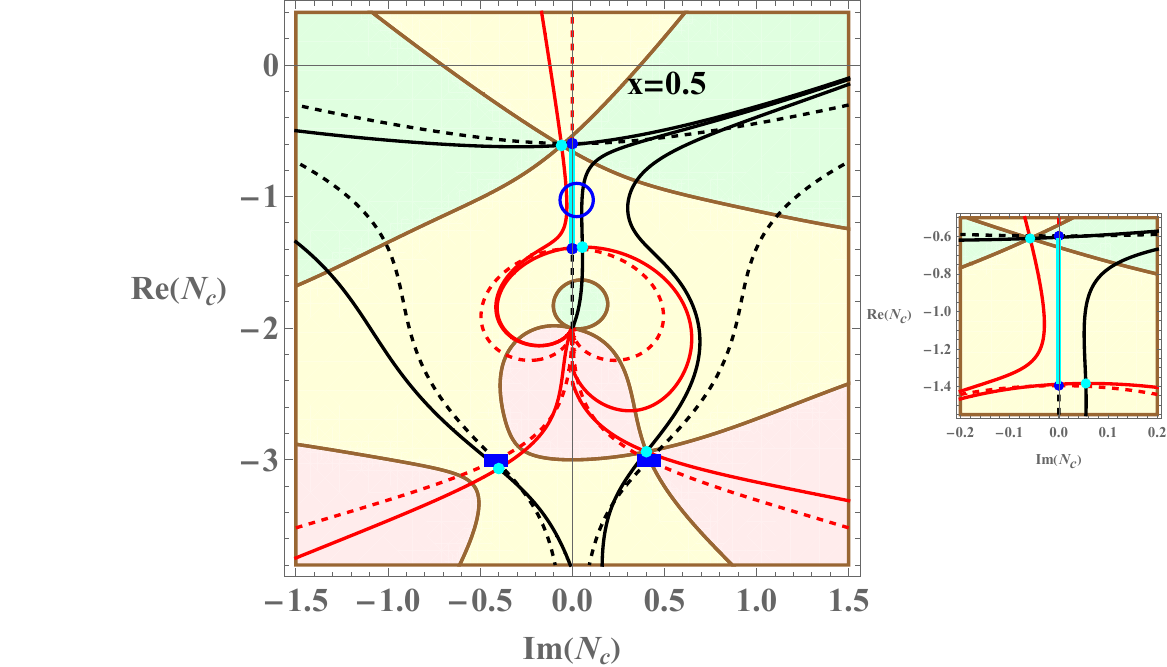}
    \caption{Plot illustrating the breaking of the residual \textbf{type-1} degeneracy (depicted as a cyan vertical line) induced by the rotation of $ G$. We take $G^{-1}=|G|^{-1}(1-i\epsilon),\epsilon=\pi/20,\hbar=1,V_3=8\pi|G|$. $P_{i} = -3i,\al = 2, \Lam=3, x = 1/2, q_{f} = 1$. Red lines (solid,dashed) are the steepest ascent lines, while black lines (solid,dashed) are the steepest descent lines with and without $G$-rotation. The saddles with $G$-rotation are depicted cyan dots. Note that the allowed and forbidden region also gets corrected due to the $G$-rotation. Quantum correction is always taken into account with and without $G$-rotation.}
    \label{fig:degbreak1}
\end{figure}

In the presence of complex deformations like the one mentioned in eq. (\ref{eq:maldacena}), an additional phase $-X\ep/2$ gets introduced. Note that this extra phase is also proportional to $\bl(\mathcal{N}_\sigma^{(q)}\br)^3$, which is always different at the two different saddles. Consequently, this phase is different at these quantum corrected saddles, and as a result, the degeneracy always gets resolved. Hence, we conclude that when considering the complex deformation in $G$, all the \textbf{type-1} degeneracies get resolved. This is illustrated in fig. \ref{fig:degbreak1}, where the residual Stokes ray (type-2 case), which is lying on the imaginary line, gets broken by the complexification of $G$.
%

%%%%%%%%%%%%%%%%%%%%%%%%%%%%%%%%%%%%%%%%%%%%%%%%%
\section{
Symmetries and Type-1 degeneracies
} 
\label{sec:sym_type1}
%%%%%%%%%%%%%%%%%%%%%%%%%%%%%%%%%%%%%%%%%%%%%%%%%

In the preceding sections, we introduced artificial linear {\it defects}, incorporated quantum corrections, and considered deformations in $G\hbar$ as mechanisms to resolve \textbf{type-1} degeneracies. Nonetheless, it remains entirely plausible that other alternative modifications capable of lifting the degeneracies exist beyond those explicitly considered. This naturally leads to the question: what class of modifications to the theory should one consider, and is sufficient to achieve such resolution? A sufficient criterion will be simply that the modification should induce a difference in the imaginary parts of the action evaluated at the two saddles which are connected via a Stokes ray. In this section, we aim to address this question from a different perspective by analyzing the underlying symmetries present in the theory. In particular, we shall investigate the question: what are the underlying symmetries responsible for the emergence of \textbf{type-1} degeneracies, whose breaking will resolve the overlapping of flow lines? 

To figure out the symmetry present in the theory, we start by observing that when $P_i$ and $\beta$ are purely imaginary, the exact transition amplitude in Eq. (\ref{eq:transAmp_rbc_exact}) becomes purely real. It turns out that this fact reflects the presence of the underlying {\it anti-linear} symmetry in the theory. It is simply the reflection symmetry around the imaginary axis in the complex-$N_c$ plane. Writing $N_c = N_x + i N_y$, the anti-linear (AL)-transformation of $N_c$ is given by
\begin{equation}
\label{eq:antilin}
{\rm AL} 
\hspace{5mm}
\Rightarrow
\hspace{5mm}
N_c\rightarrow-N_c^*\, , 
\hspace{3mm}
N_x\to-N_x\, ,
\hspace{3mm}
N_y\to N_y \, ,
\end{equation}
where a point on the imaginary axis is mapped to itself and lies on the axis of symmetry. Under this symmetry, the on shell action $\bar{S}_{\rm tot}^{\rm on-shell}[N_c]$ mentioned in Eq. (\ref{eq:stot_onsh_rbc}) and the one-loop fluctuation prefactor $\Delta(N_c)$ mentioned in Eq. (\ref{eq:DelNc_rbc}) transform as, 
\begin{equation}
\label{eq:anti-linear1}
\left\{i\bar{S}_{\rm tot}^{\rm on-shell}[ N_c]\right\}^*=i\bar{S}_{\rm tot}^{\rm on-shell}[-N_c^*] 
\, ,
\hspace{5mm}
[\Delta(N_c)]^*=\Delta(-N_c^*) \, ,
\end{equation}
where $(\,.)^*$ implies complex conjugation. We emphasize that this symmetry exists only when both the parameters $P_i$ and $\beta$ are purely imaginary. The symmetry mentioned in eq. (\ref{eq:antilin}) have also been observed in the finite-temperature and finite-density QCD, where the symmetry is known as $C\mathcal{K}$-symmetry (charge and complex conjugation) \cite{Nishimura:2014rxa, Nishimura:2014kla, Nishimura:2015loa, Tanizaki:2015pua, Tanizaki:2015gcv}. In those studies, after analytic continuation of the gauge field (analogous to $N_c$) to the complex plane, the reality of free energy is argued by choosing relevant complex saddle points that respect $C\mathcal{K}$-symmetry (a.k.a anti-linearity). Moreover, one finds that the fermionic determinant satisfies an analogous condition that obeyed by $\Delta(N_{c})$ in eq. (\ref{eq:anti-linear1}). 

From Eq. (\ref{eq:anti-linear1}), it is easy to see that when we add contributions from both the positive and negative real values of $N_c$ in Eq. (\ref{eq:TA_lps_int}), we get a real transition amplitude. Explicitly, when we add the integrand in Eq. (\ref{eq:TA_lps_int}) for $N_c$ and $-N_c$ and take the complex conjugate, we get
\begin{equation}
\label{eq:realamp}
\begin{split}
    & \left\{\bl[ \D(N_c) \br]^{-1/2}
\, \exp \bl[i \bar{S}_{\rm tot}^{\rm on-shell}[N_c]/\hbar \br] +\bl[ \D(-N_c) \br]^{-1/2}
\, \exp \bl[i \bar{S}_{\rm tot}^{\rm on-shell}[-N_c]/\hbar \br] \right\}^*\\
&=\bl[ \D(-N_c) \br]^{-1/2}
\, \exp \bl[i \bar{S}_{\rm tot}^{\rm on-shell}[-N_c]/\hbar \br] +\bl[ \D(N_c) \br]^{-1/2}
\, \exp \bl[i \bar{S}_{\rm tot}^{\rm on-shell}[N_c]/\hbar \br]
\end{split}
\end{equation}
which makes the amplitude real. While Eq. (\ref{eq:anti-linear1}) is straightforward to analyze for real $N_c$, extending the analysis to arbitrary complex-$N_c$ is more subtle. This is primarily due to the multi-valuedness of complex functions arising from branch cuts of the one-loop fluctuation ($\Delta(N_c)$). To preserve anti-linearity, one must consistently stick to a particular branch (else one needs to add $\pm 2\pi i$ to Eq. (\ref{eq:anti-linear})). The absolute value $\lvert \exp(i \bar{S}_{\rm tot}^{\rm on-shell}[N_c] )\rvert$ is same for the no-boundary saddles ($N_\pm^{\rm nb}$) and for the non no-boundary saddles ($N_\pm^{\cancel{\rm nb}}$), as these saddles are the anti-linear pairs satisfying 
\begin{equation}
\label{eq:antiLin_pairs_sad_qf>3k/lam}
\begin{split}
    N_+^{\rm nb}=-(N_-^{\rm nb})^*,\hspace{5mm} N_+^{\cancel{\rm nb}}=-(N_-^{\cancel{\rm nb}})^*, \,\, \hspace{3mm} \text{for \,} q_f>3/\Lam\,. 
\end{split}
\end{equation}
For $q_f\leq 3/\Lam$, saddles are invariant under anti-linearity ($C\mathcal{K}$ transformation in the case of QCD \cite{Nishimura:2014rxa, Nishimura:2014kla, Nishimura:2015loa, Tanizaki:2015pua, Tanizaki:2015gcv}) and lie on the imaginary axis. The presence of anti-linearity in the theory ensures that the difference in $\lvert \exp(i \bar{S}_{\rm tot}^{\rm on-shell}[N_c]) \rvert$ at any two anti-linear points will always be zero. Any non-zero value of this quantity would indicate the absence of anti-linearity, thereby breaking the symmetry. 

For the case of No-boundary Universe, the existence of this symmetry has been observed for the case of Dirichlet or the Neumann boundary conditions \cite{DiTucci:2019bui, Lehners:2023yrj}, and also pointed out earlier in \cite{Witten:2010cx, Kanazawa:2014qma} in the context of Airy's integral. Anti-linear symmetry has been studied recently in \cite{Witten:2025ayw,Harlow:2023hjb} in the context of the Euclidean path integral. Such symmetry is also noticed in finite density QCD/effective field theories, where it is known as $C\mathcal{K}$-symmetry \cite{Nishimura:2014rxa, Nishimura:2014kla, Nishimura:2015loa, Tanizaki:2015pua, Tanizaki:2015gcv} and also found in non-hermitian statistical mechanics in the context of $\mathcal{PT}$- symmetry \cite{Bender:1998ke,Meisinger:2012va}. In the case of generalised Robin boundary choices in gravity, this symmetry has not been explored previously. In the case of Robin, it is seen that the system has an additional symmetry when $\beta$ is purely imaginary ($\beta=-i|\beta|$), which hasn't been reported earlier in literature. This is clearly evident once we rewrite the on-shell Robin action mentioned in eq. (\ref{eq:stot_onsh_rbc}) using the transformed variable $\bar{N}_c$ stated in eq. (\ref{eq:lap_redef}). In terms of $\bar{N}_c$, the on-shell Robin-action becomes the action stated in eq. (\ref{eq:on_shell_Act_nb}). Then, under the transformation 
\begin{equation}
\label{eq:trans2}
\overline{\rm AL} 
\hspace{5mm}
\Rightarrow
\hspace{5mm}
\bar{N}_c\rightarrow-\bar{N}_c^* \, ,
\hspace{5mm} 
\Rightarrow
\hspace{5mm}
N_x+ i N_y \to
N_x - i \Bl(N_y+\frac{3}{|\beta|} \Br) \, ,
\end{equation}
the action mentioned in Eq. (\ref{eq:on_shell_Act_nb}) has the following transformation 
\begin{equation}
\label{eq:S_on_NcBar_tr_anti}
\{i\mathcal{\bar{S}}^{\rm on-shell}_{\rm tot}[\bar{N}_c]\}^*=-i\mathcal{\bar{S}}^{\rm on-shell}_{\rm tot}[-\bar{N}_c^*]
+ \underbrace{
\frac{2 |\beta|^2 \left(9 k+y^2\right)+3 \Lambda ^2-6 |\beta| \Lambda  y}{2 |\beta|^3}
+ 8\al\biggl(k y - \frac{y^3}{27}\biggr)}_{\mathbb{A}}
\, ,
\end{equation}
where we have substituted $P_i=-iy$ and $\beta=-i|\beta|$ in eq. (\ref{eq:on_shell_Act_nb}). The quantity defined by $\mathbb{A}$ is real. This is a version of anti-linearity present in Robin case, we refer to it as $\overline{\rm anti-linearity}$ or $\overline{\rm AL}$, where `bar' is used to represent the antilinear symmetry in the $\bar{N}_c$ plane. The transformation stated in Eq. (\ref{eq:trans2}) depends explicitly on the initial parameter $\beta$, and it does not exist for Neumann boundary conditions ($|\beta|=0$). However, Eq. (\ref{eq:trans2}) is present in the case of Dirichlet boundary conditions, where $|\beta|=\infty$. Thus, while anti-linearity is more fundamental and appears across all mini-superspace models, Eq. (\ref{eq:trans2}) arises only in specific situations, and its form varies depending on the model/boundary conditions under consideration. 

%%%%%%%%%%%%%%%%%%%%%%%%%%%%%%%%%%%%%%%%%%%%%%%%%
\subsection{$q_f > 3k/\Lam$}
\label{subs:Sym_qf>3/lam}
%%%%%%%%%%%%%%%%%%%%%%%%%%%%%%%%%%%%%%%%%%%%%%%%%

Once we have found out the symmetries present in the on-shell lapse $N_c$-action, it is worth asking their consequences, in particular its relation to the \textbf{type-1} degeneracies. To proceed, we first analyze the behavior of the flow-lines under $\overline{\rm AL}$ symmetry mentioned in eq. (\ref{eq:trans2}). We proceed as before by writing $ N_c=N_x+iN_y$ and $i\bar{S}_{\rm tot}^{\rm on-shell}[N_x+iN_y]=h(N_x,N_y)+iH(N_x,N_y)$. Plugging these in Eq. (\ref{eq:S_on_NcBar_tr_anti}), we get
\beq
\label{eq:hHtrans2}
\begin{split}
h(N_x,N_y) & = -h(N_x,-N_y-3/|\beta|) + \mathbb{A} \, ,
\\
H(N_x,N_y) & = H(N_x,-N_y-3/|\beta|)\, .
\end{split}
\eeq
The second line of above equation, that follows from the underlying symmetry transformation in Eq. (\ref{eq:S_on_NcBar_tr_anti}), defines a map between two points in the complex lapse plane that preserves the phase factor $H$, but with different Morse function $h$. Interestingly, for $q_f > 3k/\Lam$ the pairs of background saddle points $(N_{+}^{\rm nb}, N_{+}^{\cancel{\rm nb}})$ and $(N_{-}^{\rm nb}, N_{-}^{\cancel{\rm nb}})$ are related by the above symmetry. Specifically, we have:
\beq
\label{eq:sad_trans_rels_qf>3k_Lam}
N_{\pm}^{\cancel{\rm nb}} = (N_{\pm}^{\rm nb} + 3/\beta)^{*} \implies \bar{S}_{\rm tot}^{\rm on-shell}[N_{\pm}^{\cancel{\rm nb}}]^{*} = \bar{S}_{\rm tot}^{\rm on-shell}[N_{\pm}^{\rm nb}] - i \mathbb{A}\, ,
\eeq
this implies
\bea
\label{eq:sad_H_rels_qf>3k_Lam}
&&
H({\rm Re}(N_{\pm}^{\cancel{\rm nb}}), {\rm Im}(N_{\pm}^{\cancel{\rm nb}})) 
= 
H({\rm Re}(N_{\pm}^{\rm nb}), {\rm Im}(N_{\pm}^{\rm nb}))\, ,   
\notag \\
&&
h({\rm Re}(N_{\pm}^{\cancel{\rm nb}}), {\rm Im}(N_{\pm}^{\cancel{\rm nb}})) 
= -h({\rm Re}(N_{\pm}^{\rm nb}), {\rm Im}(N_{\pm}^{\rm nb}) ) + \mathbb{A} \, .
\eea
Indeed, the above relation mentioned in Eqs. (\ref{eq:sad_H_rels_qf>3k_Lam}) define the conditions for the emergence of \textbf{type-1} degeneracies between the pairs of background saddle points $(N_{+}^{\rm nb}, N_{+}^{\cancel{\rm nb}})$ and $(N_{-}^{\rm nb}, N_{-}^{\cancel{\rm nb}})$ for $q_f > 3k/\Lam$. Having now fully identified the underlying symmetry $\overline{\rm AL}$ is responsible for the \textbf{type-1} degeneracies for $q_f>3k/\Lam$ - it is sufficient to resolve the associated \textbf{type-1} degeneracies by modifying the action by adding any appropriate defect that explicitly breaks this symmetry. Interestingly, when corrections from quantum fluctuations are incorporated, this symmetry gets broken. In particular, the action corresponding to the quantum corrections mentioned in Eq. (\ref{eq:qmAact})
transforms under the symmetry as 
\beq
\label{eq:QC_act_trans}
{\cal A}_1[(N_c + 3/\beta)^{*}]^{*} 
= - {\cal A}_1(N_c) + \hbar\pi/2  \, .
\eeq

This clearly demonstrates that quantum correction does not respect the underlying symmetry that leads to \textbf{type-1} degeneracies. Consequently, it offers a resolution to the \textbf{type-1} degeneracy for $q_f>3k/\Lam$. This was also observed numerically in Sec. \ref{sec:qc_as_def}.

%%%%%%%%%%%%%%%%%%%%%%%%%%%%%%%%%%%%%%%%%%%%%%%%%
\subsection{$q_f \leq 3k/\Lam$}
\label{subs:Sym_qf<3/lam}
%%%%%%%%%%%%%%%%%%%%%%%%%%%%%%%%%%%%%%%%%%%%%%%%%

For $q_f \leq 3k/\Lam$, the pairs $(N_{+}^{\rm nb}, N_{-}^{\cancel{\rm nb}})$ and $(N_{-}^{\rm nb}, N_{+}^{\cancel{\rm nb}})$ are related by the symmetry $\overline{\rm AL}$ stated in eq. (\ref{eq:trans2}). This implies that for $q_{f} < 3k/\Lam$, we have: 
\beq
\label{eq:sad_trans_rels_qf<3k_Lam}
N_{\pm}^{\cancel{\rm nb}} = (N_{\mp}^{\rm nb} + 3/\beta)^{*} 
\implies 
\bar{S}_{\rm tot}^{\rm on-shell}[N_{\pm}^{\cancel{\rm nb}}]^{*} = \bar{S}_{\rm tot}^{\rm on-shell}[N_{\mp}^{\rm nb}] - i \mathbb{A} \, .
\eeq
From this one can extract that at the various saddles, the phase $H$ and the Morse-function $h$ satisfies the following relations 
\bea
\label{eq:sad_H_rels_qf<3k_Lam}
&&
H({\rm Re}(N_{\pm}^{\cancel{\rm nb}}), {\rm Im}(N_{\pm}^{\cancel{\rm nb}})) 
= H({\rm Re}(N_{\mp}^{\rm nb}), {\rm Im}(N_{\mp}^{\rm nb})) = 0\, , 
\notag \\
&&
h({\rm Re}(N_{\pm}^{\cancel{\rm nb}}), {\rm Im}(N_{\pm}^{\cancel{\rm nb}})) 
= -h({\rm Re}(N_{\mp}^{\rm nb}), {\rm Im}(N_{\mp}^{\rm nb})) 
+ \mathbb{A} \, .
\eea
The set of relations mentioned in eq. (\ref{eq:sad_H_rels_qf<3k_Lam}) give the condition for the emergence of \textbf{type-1} degeneracies between the pairs of background saddle points $(N_{+}^{\rm nb}, N_{-}^{\cancel{\rm nb}})$ and $(N_{-}^{\rm nb}, N_{+}^{\cancel{\rm nb}})$. 

However, unlike in the case of $q_f>3k/\Lam$, quantum corrections are not sufficient to break all \textbf{type-1} degeneracies for $q_f \leq 3k/\Lam$. While, quantum corrections indeed break the \textbf{type-1} degeneracy between the saddle pairs $(N_{+}^{\rm nb}, N_{-}^{\cancel{\rm nb}})$ and $(N_{-}^{\rm nb}, N_{+}^{\cancel{\rm nb}})$ — which are related by the symmetry $\overline{\rm AL}$ mentioned in eq. (\ref{eq:trans2}) - we still observe residual \textbf{type-1} degeneracies between other pairs of saddles for $q_f<3k/\Lam$: for instance between the pair $(N_{+}^{\rm nb}, N_{-}^{\rm nb})$ and the pair $(N_{+}^{\cancel{\rm nb}}, N_{-}^{\cancel{\rm nb}})$ for the case when $0<x<(1-\mu)^{-1}$ (see Sec. \ref{sss:qf<3_by_lam_qm} for detailed discussion). Moreover, it is worth noting that saddle point pairs for $q_f \leq 3k/\Lam$ that exhibit the residual \textbf{type-1} degeneracies are not related by the $\overline{\rm AL}$ transformation mentioned in Eq. (\ref{eq:trans2}). This hints that perhaps some other symmetry is probably responsible for this residual \textbf{type-1} degeneracy. 

To address these residual degeneracies present in the thimbles, we focus our attention on exploring the consequences of anti-linearity stated in Eq. (\ref{eq:antilin}). When $P_i$ and $\beta$ are purely imaginary, the one-loop quantum action ${\cal A}(N_c)$ transform as, 
\begin{equation}
\label{eq:anti-linear}
{\cal A}(-N_c^{*})^{*} = - {\cal A}(N_c)\, .
\end{equation}
To understand how anti-linearity might be related to the degeneracies, we first analyze how the flow-lines transform under this symmetry. Writing $N_c=N_x+i N_y$, the above anti-linear Eq. (\ref{eq:anti-linear}) becomes
\begin{equation}
\label{eq:anti-line2}
\{{\cal A}( -N_x+i N_y)\}^{*} = - {\cal A}(N_x+i N_y)\, .
\end{equation}
Expressing $i{\cal A}[ N_x+i N_y]=\mathfrak{h}(N_x,N_y)+i\mathcal{H}(N_x,N_y)$ and substituting it in the above equation, we get the following relations
\begin{equation}
\label{eq:handH}
\mathfrak{h}(N_x,N_y)=\mathfrak{h}(-N_x,N_y) \, ,
\hspace{5mm}
\mathcal{H}(N_x,N_y)=-\mathcal{H}(-N_x,N_y) \, .
\end{equation}
These are valid for all values of $q_f$, where we restrict ourselves to the principal Riemann sheet. Utilizing Eq. (\ref{eq:handH}), one can conclude that the steepest descent/ascent flow-lines defined by ($-$ sign for steepest descent and $+$ sign for steepest ascent)
\begin{equation}
\label{eq:floweq}
\frac{\partial N_x}{\partial \lambda}=\pm\frac{\partial \mathfrak{h}(N_x,N_y)}{\partial N_x}
\hspace{3mm}
{\rm and}
\hspace{3mm}
\frac{\partial N_y}{\partial \lambda}=\pm\frac{\partial \mathfrak{h}(N_x,N_y)}{\partial N_y} ,
\end{equation}
 are invariant under the symmetry mentioned in Eq. (\ref{eq:antilin}). Eq. (\ref{eq:handH}) also implies that the forbidden and allowed region as dictated by Picard-Lefschetz methods are also invariant under this symmetry. This underlying feature of flowlines has also been noticed in studies of fermion sign problem \cite{Tanizaki:2015pua, Tanizaki:2015gcv}.

It is worth noticing that flow-lines originating from two saddles related to each other by anti-linear transformation ($C\mathcal{K}$) can never overlap unless they have $\mathcal{H}(N_x,N_y)=\mathcal{H}(-N_x,N_y)=0$. However, such a situation doesn't correspond to a Stokes ray, since the Morse function $\mathfrak{h}(N_x,N_y)$ takes the same value at both points. Note, our interest lies in resolving Stokes-rays as their resolution implies unambiguous application of Picard-Lefschetz methods. Presence of anti-linear symmetry excludes the possibility of having \textbf{type-1} degeneracies between the pair of saddle points that are related via anti-linear transformation. 

Let us now consider specifically the points lying on the imaginary axis ($N_x=0$), which is the axis of {\rm AL}-symmetry on which the background saddles lie for $q_f\leq 3k/\Lam$. As previously noticed, these points remain invariant under the anti-linear transformation Eq. (\ref{eq:antilin}). Substituting $N_x=0$ in Eq. (\ref{eq:anti-line2}), we get
\begin{equation}
\label{eq:Str_AL_xy}
{\cal A}( i N_y)^{*} = - {\cal A}(i N_y) \, ,
\end{equation}
which implies that 
\begin{equation}
\label{eq:H_anti}
\begin{aligned}
    \mathcal{H}(0,N_y) &= 0                        && \text{for } \Delta(0,N_y) > 0 \,, \\
    \mathcal{H}(0,N_y) &= -\hbar\pi/2              && \text{for } \Delta(0,N_y) < 0 \,,    
\end{aligned}
\end{equation}
and,
\beq
\label{eq:h_anti}
\begin{split}
    \mathfrak{h}(0,N_y) = \frac{1}{36(3\sqrt{k}+N_y x \Lam)}&\Bigl[ x \Lam^3 N_y^4 + 12\sqrt{k}\Lambda^2 N_y^3 + 18\Lam(x q_f \Lam + 2\sqrt{k}y-6kx)N_y^2 \\
    &+ 36\sqrt{k}(3\Lam q_f+ y^2 -9k)N_y 
    +108 \sqrt{k}y q_f- 27 x q_f^2 \Lam \Bigr] \\
    & + 4\al\biggl( ky-\frac{y^3}{27}\biggr)- \hbar\ln(|\Delta(0,N_y)|)/2\, ,
\end{split}
\eeq
where we have $\Delta(0,N_y) = 1 + \Lam x N_y/(3\sqrt{k})$. With one-loop corrections, a constant phase ($-\hbar \pi/2$) arises due to the branch cuts; otherwise, it is zero. Since the Morse function mentioned in Eq. (\ref{eq:h_anti}) is a single-valued, upon plugging the saddle points into the above expression, we obtain
\begin{equation}
\label{eq:stokes_con_qf<3k/Lam}
\begin{aligned}
    \mathcal{H}_{\sg} &= \mathcal{H}_{\sg^{\prime}} \quad \& \quad \mathfrak{h}_{\sg} \neq \mathfrak{h}_{\sg^{\prime}}                        && \text{with} \quad  \sg \neq \sg^{\prime} \,. \\
\end{aligned}
\end{equation}
where $\mathcal{H}_{\sg}$ can take the value either zero or $-\hbar \pi/2$ depending on whether the $\Delta(0,-iN_{\sg})$ is either positive or negative, respectively. The Eq. (\ref{eq:stokes_con_qf<3k/Lam}) indeed corresponds to the conditions that leads to the \textbf{type-1} degeneracy between different pairs of saddles for $q_f \leq 3k/\Lam$, specifically between: $(N_+^{\rm nb}$ and $N_-^{\rm nb})$ and $(N_+^{\rm\cancel{\rm nb}}$ and $N_-^{\rm\cancel{\rm nb}})$ depending on $x$. Therefore, we find that the residual \textbf{type-1} degeneracy observed for $q_f\leq 3k/\Lam$, is a consequence of the anti-linear symmetry of the theory mentioned in Eq. (\ref{eq:antilin}). 

As discussed in Sec. \ref{complex_Ghbar}, this residual degeneracy can be lifted by introducing a small complex deformation in $G\hbar$. It is possible as this modification breaks anti-linearity
\begin{equation}
\label{eq:G_Nrotanti}
\begin{split}
&|G|^{-1}\{(1-i\epsilon){\cal A}(-N_c^{*})\}^{*} \neq - |G|^{-1}(1-i\epsilon) {\cal A}(N_c)\,,\,\,\,\\
&\text{where,\,\, $\frac{1}{G}=\frac{1}{|G|} (1-i\epsilon)$},\hspace{6mm} 1\gg\epsilon>0,
\end{split}
\end{equation}
with $\hbar$ taken to be real. When $\epsilon=0$, we recover Eq. (\ref{eq:anti-linear}). The degeneracy can also be resolved through the artificial {\it defect} as it also break anti-linearity, i.e., $\left\{iS^{\ep}[N_c]\right\}^*\neq iS^{\ep}[-N_c^*]$ where $S^{\ep}[N_c]$ is defined in Eq. (\ref{eq:Sact_arti_defect}). This has been explicitly shown in Sec. (\ref{lab:SadRel}) through various examples.

%%%%%%%%%%%%%%%%%%%%%%%%%%%%%%%%%%%%%%%%%%%%%%%%%
\subsection{Anti-linearity and its breaking}
\label{subs:anti-br}
%%%%%%%%%%%%%%%%%%%%%%%%%%%%%%%%%%%%%%%%%%%%%%%%%

We have just shown that for $q_f\leq 3/\Lam$, it is necessary to break anti-linearity in some situations to lift the \textbf{type-1} degeneracy. However, breaking of this symmetry isn't necessary for $q_f>3/\Lam$, as quantum corrections which respects anti-linearity are sufficient for resolving overlapping flow-lines as they breaks $\overline{\rm AL}$.  We therefore arrive at the important conclusion that, to fully resolve all the \textbf{type-1} degeneracies across all parameter regimes, the theory must be modified to break anti-linearity. 

Although not central to our present discussion, it is worth highlighting a few remarks when anti-linearity is broken in the theory. The immediate consequence is that the transition amplitude in Eq. (\ref{eq:TA_exp}) is no longer real: $G_{\rm RBC} - G^*_{\rm RBC} \neq0$. This was also noticed previously in the case of linear {\it defect}, (see for example Eq. (\ref{eq:Grbc_dd_degenlindef}). Moreover, for a generic deformation which breaks anti-linearity, the difference in $|\exp(iG^{-1}\bar{S}_{\rm tot}^{\rm on-shell}[N_c])|$ at the two anti-linear points will be ``nonzero", thereby breaking the symmetry between them. To be specific, consider the case when $G$ is complexified. In this specific case the difference in $|\exp(iG^{-1}\bar{S}_{\rm tot}^{\rm on-shell}[N_c])|$ at the anti-linear points $N_+^{\rm nb}$ and $N_-^{\rm nb}$ is explicitly given by
\begin{equation}
\label{eq:antilinGrot}
\begin{split}
|\exp(iG^{-1}\bar{S}_{\rm tot}^{\rm on-shell}[N_c])|_{N_+^{\rm nb}}- |\exp(iG^{-1}\bar{S}_{\rm tot}^{\rm on-shell}[N_c])|_{N_-^{\rm nb}}\\
=\frac{V_3\epsilon}{8\pi\hbar|G|} \exp\biggl[\frac{V_3h(N_+^{\rm nb})}{8\pi\hbar |G|}\biggr]\left\{ H(N_+^{\rm nb})-H(N_-^{\rm nb})\right\}+\mathcal{O}(\epsilon^2) \, .
\end{split}
\end{equation}
The above asymmetry in Eq. (\ref{eq:antilinGrot}) will lead to an asymmetry in the structure of Lefschetz thimbles and the corresponding regions of convergence between $\rm Re[N_c]>0$ and $\rm Re[N_c]<0$ in the complex plane. Moreover, since, $N_+^{\rm nb}$ and $N_-^{\rm nb}$ are the relevant saddles which enter in the computation of transition amplitude in Eq. (\ref{eq:TA_exp}), any small non-zero value of $\epsilon$ will also make the transition amplitude complex: $G_{\rm RBC}^{(\ep)}-G_{\rm RBC}^{(\ep) *}\neq 0$. This non-zero difference is explicitly given by
\begin{equation}
\label{eq:complextransamp}
\begin{split}
&\frac{G_{\rm RBC}^{(\ep)}[\rm Bd_f,\rm Bd_i]-G_{\rm RBC}^{(\ep)}[\rm Bd_f,\rm Bd_i]^*}{G_{\rm RBC}[\rm Bd_f,\rm Bd_i]} 
\\
&=i\epsilon\left[1-\frac{V_3}{4\pi\hbar|G|}\left\{ h(N_+^{\rm nb})-H(N_+^{\rm nb})\tan\left[\frac{V_3 H(N_+^{\rm nb})}{8\pi\hbar|G|}+\frac{\pi}{4}\right]\right\}\right]+\mathcal{O}(\epsilon^2).
\end{split}
\end{equation}
Although such asymmetries have already been observed numerically in Figs. (\ref{fig:PL_w_wo_LD_x0.5_cr}), (\ref{fig:PL_w_wo_LD_x1}), and (\ref{fig:degbreak1}), here we offer an alternative understanding based on symmetry considerations.

Let us consider a modification to the theory of the form
\begin{equation}
\label{eq:newdefect}
    \bar{S}_{\rm tot}^{\rm on-shell}[N_c]+\underbrace{\epsilon N_c}_{\text{defect}},
\end{equation}
and examine whether it can resolve the {\bf type-1} degeneracies. It is straightforward to verify that this modified action preserves both anti-linearity (Eq. (\ref{eq:antilin})) and $\overline{\rm AL}$ transformation stated in Eq. (\ref{eq:trans2}). Consequently, this modification cannot lift the \textbf{type-1} degeneracies. This is along with the observation that the extra modification does not introduce any difference in the imaginary parts of the action evaluated at the saddle points. On the other hand, artificial {\it defect} introduced in eq. (\ref{eq:Sact_arti_defect}) breaks both anti-linearity and $\overline{\rm AL}$ stated in Eq. (\ref{eq:trans2}), and hence helps in resolving all {\bf type-1} degeneracies.

%%%%%%%%%%%%%%%%%%%%%%%%%%%%%%%%%%%%%%%%%%%%%%%%%%
\section{
Degeneracies and KSW
}
\label{sec:KS}
%%%%%%%%%%%%%%%%%%%%%%%%%%%%%%%%%%%%%%%%%%%%%%%%%%

Complex metrics naturally appear in the gravitational path-integral which sums over all possible geometries. A particular example is a spacetime where the Universe starts with a smooth regular geometry at initial time instead of the cosmic singularity \cite{Gibbons:1990ns}. While, such complex metrics do offer benefits, they may not all correspond to physically meaningful scenarios, such as wormholes with vanishing action or certain bouncing geometries. Is there a way to sort out the physical ones from the unphysical complex metrics? 

Recently, Witten proposed a criterion to distinguish physically admissible complex metrics from the unphysical ones \cite{Witten:2021nzp}. This criterion is motivated from the seminal works of Louko-Sorkin (LS) and Kontsevich-Segal (KS). Louko and Sorkin proposed that a complex metric (like the regularized one at initial times) is `allowable' if the path integral for a real scalar field is formally convergent \cite{Louko:1995jw}. Independently, Kontsevich-Segal and Witten proposed (and generalized this) that the allowable metrics are those on which one can formulate well-defined quantum field theories (QFTs) \cite{Kontsevich:2021dmb}, in the sense that path-integral of the field defined on the metric should be convergent and should not blowup. While, this seem like a simple requirement, in practice the KSW (Louko-Sorkin-Kontsevich-Segal-Witten) puts strong constraints on the metrics which are physically ``allowable''. In the following, we will refer to it in its short form, KSW-criterion. 

Consider the background D($\geq 2$) spacetimes $g_{\mu\nu}$ on a manifold $\mathcal{M}$ on which we define the Euclidean path integral of a $p$-form real gauge field with field strength being a $q=p+1$-form denoted by $F$. According to the KSW-criterion, the path-integral for the $p$-form field is convergent on the given background metric if and only if the metric satisfies the following condition
\bea
\label{KSW}
\begin{split}
\mathcal{I}_q[A]& =\frac{1}{2q!}\int_M d^Dx\sqrt{det\,g}\,\,g^{\mu_1 \nu_1}\cdots g^{\mu_q \nu_q} F_{\mu_1 \mu_2...\mu_q} F_{\nu_1 \nu_2...\nu_q} \, ,\\
g_{\mu\nu} & \, \text{is allowable iff} \, \,\, {\rm Re}\left(\sqrt{det\,g}\,\, g^{\mu_1 \nu_1}\cdots g^{\mu_q \nu_q}F_{\mu_1 \mu_2...\mu_q} F_{\nu_1 \nu_2...\nu_q}\right) > 0 \, ,
\end{split}
\eea
for all $q \in \{0, ... , D\}$, where $\mathcal{I}_q[A]$ is the corresponding Euclidean action. For the metrics that are diagonal in a real basis, with diagonal elements $\lambda_{\mu}$ at a spacetime point $x \in \mathcal{M}$, i.e., $g_{\mu\nu} = \lambda_{\mu}(x)\delta_{\mu\nu}$, the criterion simplifies to the point-wise requirement of
\beq
\label{eq:KSW_criterion}
\Sigma \equiv \sum_{\mu=0}^{D-1} |\arg \, \lam_{\mu}(x)| < \pi \quad \forall x \in \mathcal{M}\, ,
\eeq
where $\arg (z) \in (-\pi,\pi]$ with $z$ being any complex variable. Although Eq. (\ref{eq:KSW_criterion}) is derived starting from an Euclidean path integral, one can obtain an exactly similar criterion by demanding the convergence of the Lorentzian path integral, see \cite{Jonas:2022uqb}, also the appendix (\ref{appen:KSEeuclor}) for more details. This form of KSW criterion has undergone extensive investigation, yielding several important physical implications. For instance, in the context of quantum cosmology, the KSW criterion supports the allowability of no-boundary solutions while eliminating quantum bounces. Recent studies have shown that imposing the KSW criterion on the no-boundary universe places strong constraints on inflationary models and anisotropies \cite{Hertog:2023vot, Hertog:2024nbh}.

In this section, we are interested in understanding the implications/compatibility of the KSW-criterion with degeneracies that arise in the quantum cosmology. In particular, we focus on the question of whether the geometries at the quantum-corrected saddles are allowable in the presence of various types of defects. Specifically, consider the 4D homogeneous and isotropic model of form (\ref{eq:frwmet_changed}), with scale factor $q(t)$ as the only degree of freedom. For a time contour $t(u)$ that runs in the complex plane, with the real parameter $u \in [0,1]$ and fixed boundary points: $t(0) = 0$ and $t(1) = 1$, the metric reads as:
\beq
\label{eq:frwmet_changed_comp_t}
{\rm d}s^2 = - \frac{N_c^2}{q(t(u))} t'(u)^2 {\rm d} u^2 
+ q(t(u)) \left[
\frac{{\rm d}r^2}{1-kr^2} + r^2 {\rm d} \OM_2^2
\right] \, .
\eeq
Here ($^{\prime}$) denotes derivative with respect to $u$-parameter. For later convenience, let us split the KSW function corresponding to the diagonalized isotropic and homogeneous metric into temporal and spatial parts as follows 
\beq
\label{eq:KSW_def}
\Sigma(N_c,t(u)) = \Sigma_{\rm temporal}(N_c,t(u))+ \Sigma_{\rm spatial}(N_c,t(u))\, ,
\eeq
where
\begin{equation}
    \label{KSW_temp_spa_defs}
    \begin{split}
         \Sigma_{\rm temporal}(N_c,t(u)) = \left|\arg\left(-\frac{N_c^2}{q(t(u))}t'(u)^2\right)\right|,\;
         \Sigma_{\rm spatial}(N_c,t(u)) = 3\left|\arg q(t(u))\right|\, .
    \end{split}
\end{equation}
Its worth noting that the spatial KSW function depends on the position $t(u)$ of the time paths in the complex plane, while the temporal part additionally depends on the tangent $t'(u)$ along the paths. We define a generic complex geometry $N_c \in \mathbb{C}$ as allowable, if and only if there exists a connected path $t(u)$ in the complex plane from $t(0)=0$ to $t(1)=1$, such that the corresponding on-shell solution $\bar{q}(N_c,t(u))$ satisfies the KSW criterion, namely:
\beq
\label{KSW_all_crt}
\Sigma(N_c,t(u)) < \pi \, .
\eeq
In the following we examine the allowability of a generic point $N_c$ in the complex plane in two steps: Ridge criterion and Extremal curve text, which we have detailed below.

%%%%%%%%%%%%%%%%%%%%%%%%%%%%%%%%%%%%%%%%%%
\subsection{Ridge Criterion}
\label{Rid_Cri}
%%%%%%%%%%%%%%%%%%%%%%%%%%%%%%%%%%%%%%%%%%

Ridge criterion technique uses the advantage of position $t(u)$ dependence of time paths and offers a quick and simple verification to determine whether a given geometry corresponding to $N_c$ violates the KSW bound or not. Due to the absolute sign in eq. (\ref{KSW_temp_spa_defs}), one finds that the temporal and spatial components of the KSW function satisfy the inequalities $\Sigma_{\rm temporal}(N_c,t(u)) \geq 0$ and $\Sigma_{\rm spatial}(N_c,t(u)) \geq 0$, respectively. Moreover, the full KSW function obeys $\Sigma(N_c,t(u)) \geq \Sigma_{\rm temporal}(N_c,t(u))$ and $\Sigma(N_c,t(u)) \geq \Sigma_{\rm spatial}(N_c,t(u))$.
Now, the above inequalities imply that the existence of a connected path from $t(0) = 0$ to $t(1) = 1$, along which $\Sigma_{\rm spatial}(N_c,t(u)) < \pi \, \forall \, u \in [0,1]$ is a \textit{necessary} condition, but not \textit{sufficient} condition to satisfy the KSW bound. Consequently, if no such connected path exists — i.e., if all paths violate this necessary condition — it is a sufficient criterion to conclude that the corresponding geometry $N_c$ or the associated metric violates the KSW criterion and becomes a disallowed point. 

\begin{figure}
    \centering
    \includegraphics[trim={2cm 0 2cm 0}, clip, scale = 0.53]{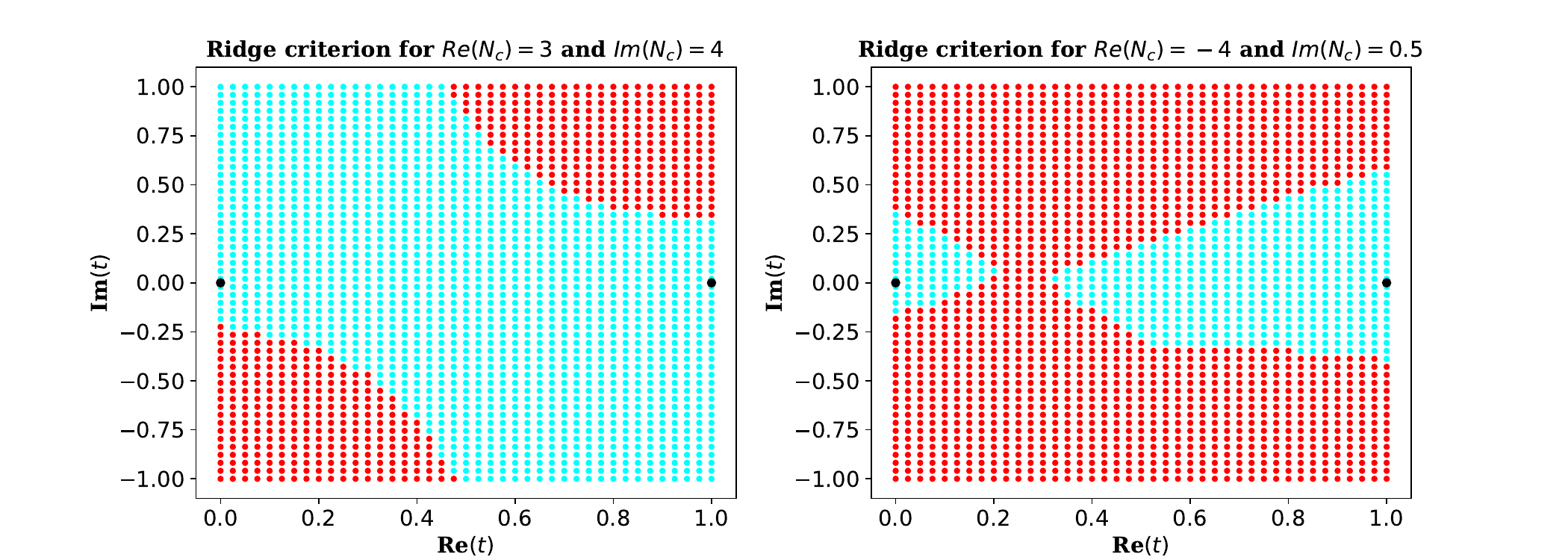}
    \caption{ Illustration of the Ridge criterion implemented in the complexified $t$ plane is shown here. The red dots indicate the points where $\Sigma_{\rm spatial} \geq \pi$ and cyan dots indicate the points where $\Sigma_{\rm spatial} < \pi$. The black dots corresponds to $t=0$ and $t=1$ points. For this numerical demonstration, we set the parameters as follows: $k=1, q_{f} = 10, \Lam =3, P_{i}=-3i$ and $x=3/4$. \textbf{Left:} This plot shows that there exists a simple path where one can reach $t=1$ point starting from $t=0$, while maintaining $\Sigma_{\rm spatial} < \pi$. However, this alone does not guarantee that the corresponding $N_c$ point satisfies the full KSW bound and must be decided based on the Extremal curve method. \textbf{Right:} This plot clearly shows an obstruction - resembling a “ridge” - between the way from $t=0$ to $t=1$, blocking all paths by the region with $\Sigma_{\rm spatial}\geq \pi$. This necessarily implies the violation of KSW bound and the corresponding geometry is concluded as disallowed. }
    \label{Fig:Ridge_plot}
\end{figure}

\begin{figure}
    \centering
    \includegraphics[trim={2cm 0 2cm 0}, clip, scale=0.55]{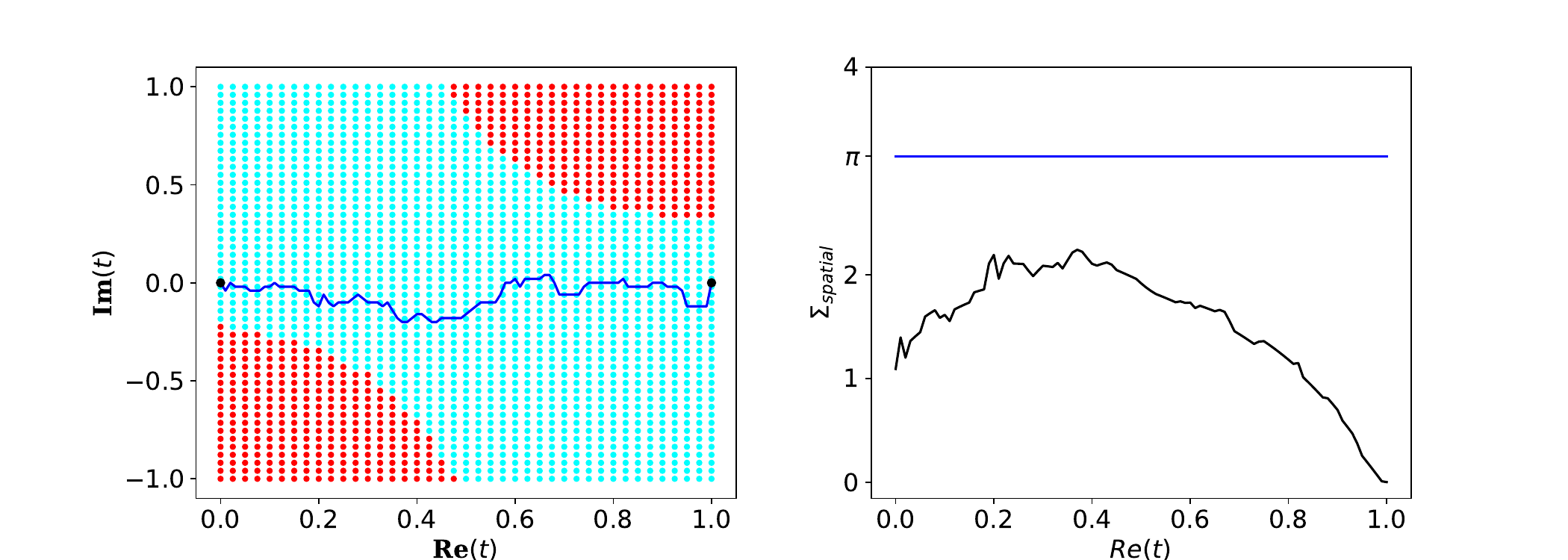}
    \caption{This plots shows a path that connects the end points $t=0$ and $t=1$ in the discretized complex $t$-plane. Along this path, the $\Sigma_{\rm spatial}< \pi $ is satisfied throughout the length(see right). Here, for numerical illustration, we set the parameters as follows: $k=1, q_{f} = 10, \Lam =3, P_{i}=-3i$ and $x=3/4$ for the lapse point ${\rm Re} (N_{c}) = 3$ and ${\rm Im}(N_c) = 4$.}
    \label{Fig:Ridge_all_plot}
\end{figure}

To assess this criterion, we implement a `Random walking algorithm' in which we move in the discretized complex time plane. We look for a path starting from $t=0$ and move around the plane by trying to go forward in ${\rm Re}(t)$ in small increments and reach $t=1$, while ensuring that the condition $\Sigma_{\rm spatial}(N_c,t(u)) < \pi$ is maintained throughout. If no such path can be found, the corresponding geometry is classified as KSW-disallowed. Meanwhile, if any such path is found, then further analysis is required to classify the geometry. In practice, for numerical analysis, one has to restrict focus to the compact region of complex time plane. Examining the region with ${\rm Re}(t) \in [0,1]$ and ${\rm Im}(t) \in[-1,1]$ is typically sufficient/robust enough to conclude the presence of path or not. Implementation of this criterion is shown in the Figure \ref{Fig:Ridge_plot}. Pictorially, such an obstruction to have an allowable path satisfying necessary condition resembles as a `Ridge' in the complex $t$ plane, blocking the way between $t=0$ and $t=1$ through the sea of $\Sigma_{\rm spatial}(N_c,t(u)) < \pi$ points, hence the name \textbf{\textit{Ridge Criterion}} (see \cite{Jonas:2022uqb}). In the figure \ref{Fig:Ridge_all_plot}, we have explicitly shown the presence of one such path using the `Random walk' approach.

%%%%%%%%%%%%%%%%%%%%%%%%%%%%%%%%%%%%%%
\subsection{Extremal Curve Test}
\label{Ext_cur}
%%%%%%%%%%%%%%%%%%%%%%%%%%%%%%%%%%%%%%

\begin{figure}
    \centering
    \includegraphics[trim={1.5cm 0 0 0}, clip, scale = 0.54]{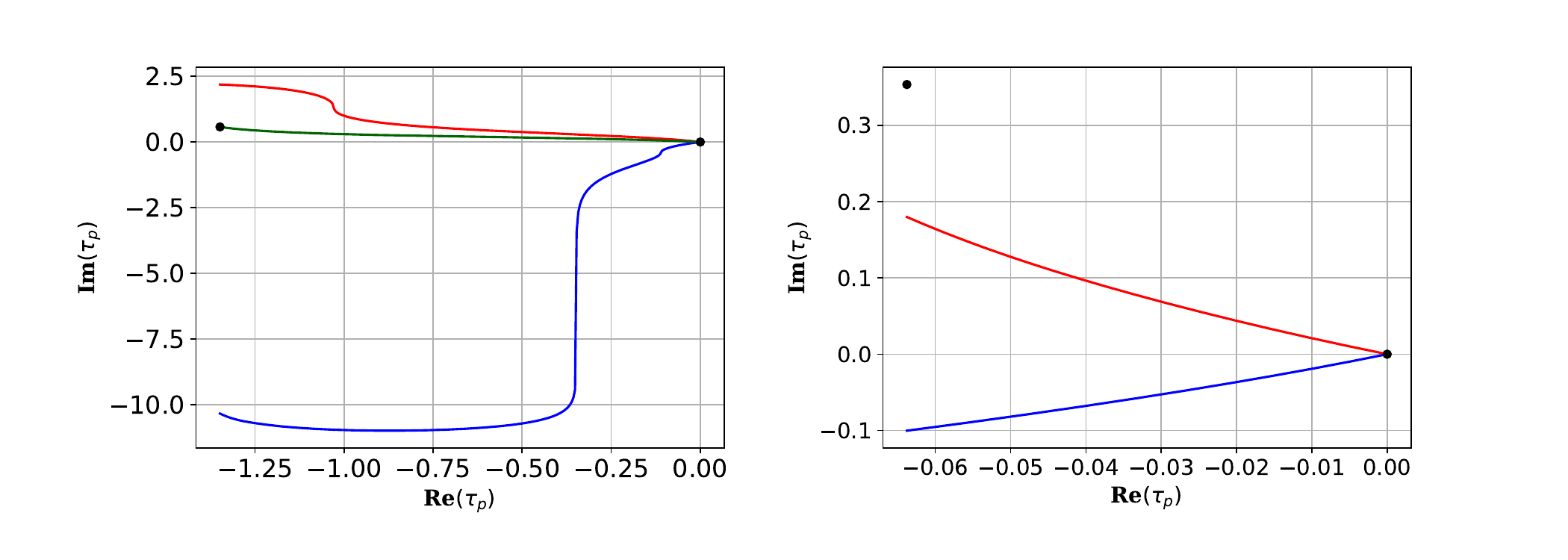}
    \caption{This figure provides an explicit demonstration of the Extremal Curve Test applied to the point that has passed the Ridge Criterion. The black dots denotes the end points: $\tau_{y}(\tau_x = 0) = 0 $ and $\nu$. The red and blue curves represent the two extremal solutions to  eq. (\ref{eq:ext_curve_1}) starting from $\tau_{y}(\tau_x=0) = 0$ point to $\tau_{x} = \rm Re(\nu)$. For numerical illustration, we choose the following parameter values:  $k=1, q_{f} = 10, \Lam =3, P_{i}=-3i$ and $x=3/4$. In the \textbf{left} panel, we display the extremal curve analysis for the lapse point $N_{c} = 3 + 4i$. The complexified Euclidean endpoint is found to be $\nu \approx -1.3479 + i 0.574282$. From the figure, it is observed that $|\tau_{y}|  > |\rm Im(\nu)| $ at $\tau_x = \rm Re(\nu)$ is satisfied. According to the extremal curve criterion, this confirms that the point is KSW allowable. The green curve illustrates one such allowable path, along which the KSW function remains bounded by $\pi/\ep < \pi$ where $\ep \approx 3.6582656$ throughout the length. In the \textbf{right} panel, we show the extremal curve analysis for the lapse point  $N_c = 1 +  0.25 i$, where the corresponding endpoint is $\nu \approx -0.06378 + i 0.353431$. In this case, $|\tau_{y}|  > |\rm Im(\nu)| $ at $\tau_x = \rm Re(\nu)$ is not achieved, indicating there exists no curve that can reach $\nu$ from $\tau_{y}(\tau_x =0) =0$ without violating KSW bound. Hence, this point is classified as KSW disallowed. }
    \label{Fig:Ext_curve}
\end{figure}

Although, Ridge criterion is robust enough for identifying KSW-disallowed points, it is not a sufficient condition to classify all lapse points definitively. Specifically, while the absence of a path with $\Sigma_{\text{spatial}} < \pi$ provides a sufficient indication of violation of the KSW bound, the presence of such a path does not guarantee satisfaction of the bound. Therefore, for those lapse points that pass the Ridge criterion, we apply the Extremal Curve Test as a follow up and decisive check. In contrast to the Ridge criterion, the Extremal Curve method offers a strong and conclusive classification of allowable and disallowable geometries under the KSW criterion. Before detailing this conclusive verification step, it is convenient to define the Euclidean physical time as follows (see Appendix (\ref{Eucl_Phy_Time_Cal}) for calculation of the on-shell relation between the two time co-ordinates):
\beq
\label{eq:Euc_phy_time}
{\rm d} \tau_{p} = i \frac{N_{c}}{\sqrt{q(t)}} {\rm d}t \, .
\eeq
The boundary points of Euclidean physical time coordinate are: $\tau_{p} (0) = 0$ at the initial hypersurface and $\tau_{p}(1) = \nu$ corresponds to the final hypersurface. The metric in this co-ordinate becomes
\beq
\label{eq:frwmet_tau_p}
{\rm d}s^2 =  {\rm d} \tau_{p}^2 
+ q(\tau_{p}) \left[
\frac{{\rm d}r^2}{1-kr^2} + r^2 {\rm d} \OM_2^2
\right] \, , \quad \because q(t(\tau_{p})) \equiv q(\tau_{p}) \, .
\eeq
For the time contour $\tau_{p}(N_c, u)$
that runs in the complex plane, with the real parameter $u$, the metric reads as
\beq
\label{eq:frwmet_changed_comp_t}
{\rm d}s^2 = \biggl(\frac{{\rm d}\tau_{p}}{{\rm d}u}\biggr)^2 {\rm d} u^2 
+ q(\tau_{p}(u)) \left[
\frac{{\rm d}r^2}{1-kr^2} + r^2 {\rm d} \OM_2^2
\right] \, .
\eeq
The KSW function reads as
\beq
\label{eq:KSW_def}
\Sigma(N_c, \tau_{p}(u)) = \Sigma_{\rm temporal}(N_c, \tau_{p}(u))+ \Sigma_{\rm spatial}(N_c, \tau_{p}(u))\, ,
\eeq
where
\begin{equation}
    \label{KSW_temp_spa_defs}
    \begin{split}
         \Sigma_{\rm temporal}(N_c,\tau_p(u)) = \left|\arg\left(\tau_{p}'(N_c, u)^2\right)\right|\, , \; \Sigma_{\rm spatial}(N_c,\tau_{p}(u)) = 3\left|\arg q(N_c, \tau_{p})\right|\, ,
    \end{split}
\end{equation}
where $(^\prime)$ denotes derivative w.r.t u. We define a generic geometry $N_c \in \mathbb{C}$ that has passed the Ridge test as allowable, if and only if there exists a curve $\tau_{p}(u): 0 \rightarrow \nu$ in the complex plane, such that the corresponding on-shell solution $\bar{q}(N_c,\tau_{p})$ satisfies the KSW criterion along the curve, namely:
\beq
\label{eq:KSW_Ext_curv_cond}
\left|\arg\left(\tau_{p}'(N_c, u)^2\right)\right| + 3\left|\arg \bar{q}(N_c, \tau_{p})\right| < \pi \, .
\eeq
Our strategy to verify the KSW criterion is based on the construction of an ``extremal curve" $\tau_{e}$ that saturates the inequality in eq. (\ref{eq:KSW_Ext_curv_cond}): 
\beq
\label{eq:ext_curve_1}
\left|\arg(\tau_{e}'(N_c, u))^2\right| + 3 \left|\arg \bar{q}(N_c, \tau_{e}(u))\right| = \pi \, .
\eeq
Upon complexifying the Euclidean time as $\tau_{p} = \tau_{x} + i \tau_{y}$, the extremal equation becomes
\beq
\label{eq:ext_curve_2}
2 \left|\tan^{-1}\biggl(\frac{d\tau_{y}}{d\tau_{x}}\biggr)\right| + 3 \left|\arg \bar{q}(N_c, \tau_{x}, \tau_{y})\right| = \pi \, ,
\eeq
which simplifies to
\beq
\label{eq:ext_curve_3}
\frac{d\tau_{y}}{d\tau_{x}}  = \pm \tan{\biggl(\frac{\pi}{2} - \frac{3}{2} \left|\arg \bar{q}(N_c, \tau_{x}, \tau_{y})\right|\biggr)} \, ,
\eeq
with the initial conditions as $\tau_{y}(\tau_{x} = 0) = 0$
\footnote{For numerical implementation, we start at $\tau_{y} = l \sin{\theta_{0}}$ when $\tau_{x} = l \cos{\theta_{0}}$ with $l \rightarrow 0$, where the initial angle $\theta_{0} = n\pi \pm (\pi - 3 |\arg \bar{q}(N_c,0,0)|)/2$ with $n \in \{0,1\}$.}. 
One obtains two extremal solutions in the complex physical time plane to the eq. (\ref{eq:ext_curve_3}), with the stated initial conditions. Then, it follows from the Petrovitsch's theorem \cite{Jonas:2022uqb, michel1901maniere}, the curves everywhere satisfying eq. (\ref{eq:KSW_Ext_curv_cond}) and starting from $\tau_{y}=\tau_x=0$ are constrained to remain between these two extremal curves. This suggests that an allowable curve — one that respects the KSW criterion — may exist between these extremal boundaries. Such a curve can be constructed by continuously deforming the right-hand side of eq. (\ref{eq:ext_curve_1}) and solving it. Following this, we classify a generic point $N_c$ as allowable, if the corresponding $\nu$ obeys \cite{Hertog:2023vot}: $|\tau_{y}|> |\rm Im(\nu)|$ at $\tau_x = \rm Re(\nu)$ or $|\tau_x| < |\rm Re(\nu)|$ at $\tau_{y}= \rm Im(\nu)$. The pictorial illustration of the extremal curve test is shown in the Figure \ref{Fig:Ext_curve}.

In the following, we are interested in studying the KSW criterion for the Robin boundary condition on the initial hypersurface and Dirichlet boundary condition on the final hypersurface for the closed geometries. Specifically, our primary interest lies in analyzing the response of no-boundary saddles in the presence of quantum correction. Furthermore, we aim to investigate the modification of KSW bound due to complex deformations of $G\hbar$. 

To begin the analysis of whether an off-shell lapse geometry is allowable or not, we first examine the spatial KSW function at $t=0$. In particular, we identify the non-allowable regions in the complex lapse plane at large $|N_{c}|$ at $t=0$. The on-shell solution for initial Robin and final Dirichlet boundary conditions is given by eq. (\ref{eq:qsol_RBC}). Recall that for any generic complex time path, the endpoints are fixed at $t=0$ and $t=1$. Since, KSW criterion is a point-wise criterion, it must be satisfied at every point along the complex time path. Thereby, the points for which $\Sigma_{spatial}(N_c,t=0) = 3|\arg(\bar{q}(N_c,t=0)) | > \pi$ are necessarily violate the KSW bound.  For the on-shell solution given in eq. (\ref{eq:qsol_RBC}), the initial scalar factor at the arbitrary $N_c$ is given by:
\beq
\label{eq:qsol_RBC_t_0}
 \bar{q}(N_c, t=0) = \frac{3 q_{f} + 2 N_c \beta - \Lam N_c^2}{3 + 2 N_c \beta} \, .
\eeq
Therefore, the complex $N_c$ points that satisfies
\beq
\label{eq:Weak_KSW_t_0}
\Bigl|\arg\Bigl(\frac{3 q_{f} + 2 N_c \beta - \Lam N_c^2}{3 + 2 N_c \beta} \Bigr) \Bigr| > \pi/3 \, ,
\eeq
necessarily violate the KSW bound and are classified as disallowed off-shell geometries. Writing $N_c=|N_c|e^{i\arg(N_c)}$, for large $|N_c|$, the above expression becomes:
\beq
\label{eq:qsol_RBC_t_0_lrg_Nc}
|\arg(- i N_{c} )| > \pi/3 \, ,
\eeq
where we used $\beta = - i \Lam x/(2\sqrt{k})$, $k=1$ with $x$ being positive real quantity. The asymptotic disallowed region of the complex lapse plane is given by $\arg(N_c) \in  \bigl( -\pi,\pi/6\bigr)\cup \bigl(5\pi/6,\pi\bigr)$. It is worth noticing that asymptotically, the negative imaginary axis ($\arg(N_c)=-\pi/2$) and the entire real line ($\arg(N_c)=0$) in the lapse plane are disallowed. 

\begin{figure}
    \centering
    \includegraphics[trim={2cm 0 2cm 0}, clip, scale = 0.4]{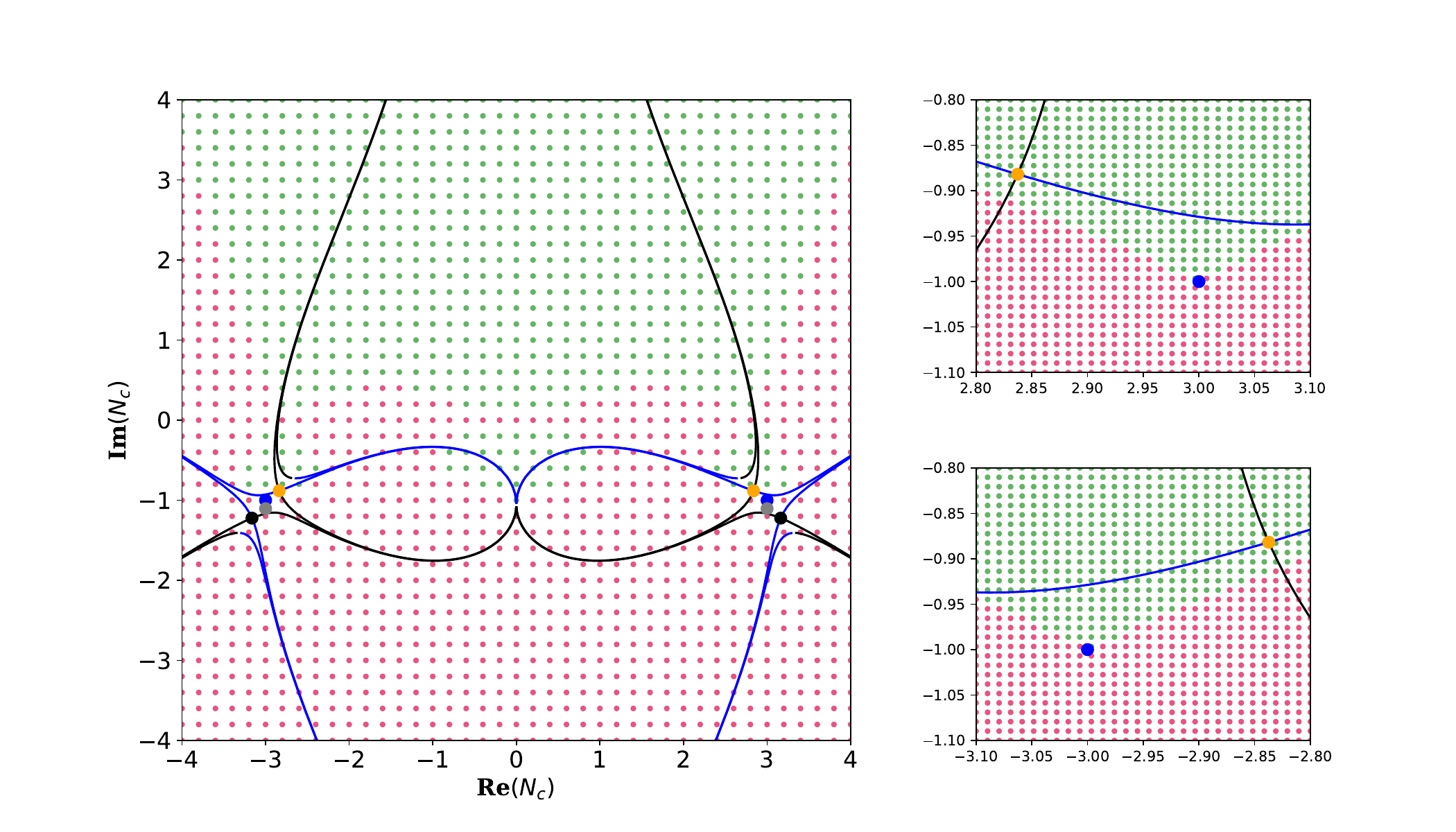}
    \caption{ Illustration of KSW numerical analysis results. The area covered with  pink dots represent the disallowed region, where no complex time path exists that connects the endpoints while satisfying the KSW bound. In contrast, the light green dots indicate the allowed region, where such a path is found or exists. The blue and grey dots denotes the semiclassical no-boundary and non-no-boundary saddles, respectively. The orange and black dots denotes the Quantum corrected no-boundary and non-no boundary saddles, respectively. The black lines represents the steepest ascents and the blue lines corresponds to the steepest descents of one-loop exponent. For this numerical illustration we set the parameters: $\al = 2, k=1, q_{f} = 10, \Lam =3, P_{i}=-3i, \hbar = 1$ and $x=0.95$. }
    \label{Fig:KSW_x_0.95}
\end{figure}

\begin{figure}
    \centering
    \includegraphics[trim={2cm 0 2cm 0}, clip, scale = 0.4]{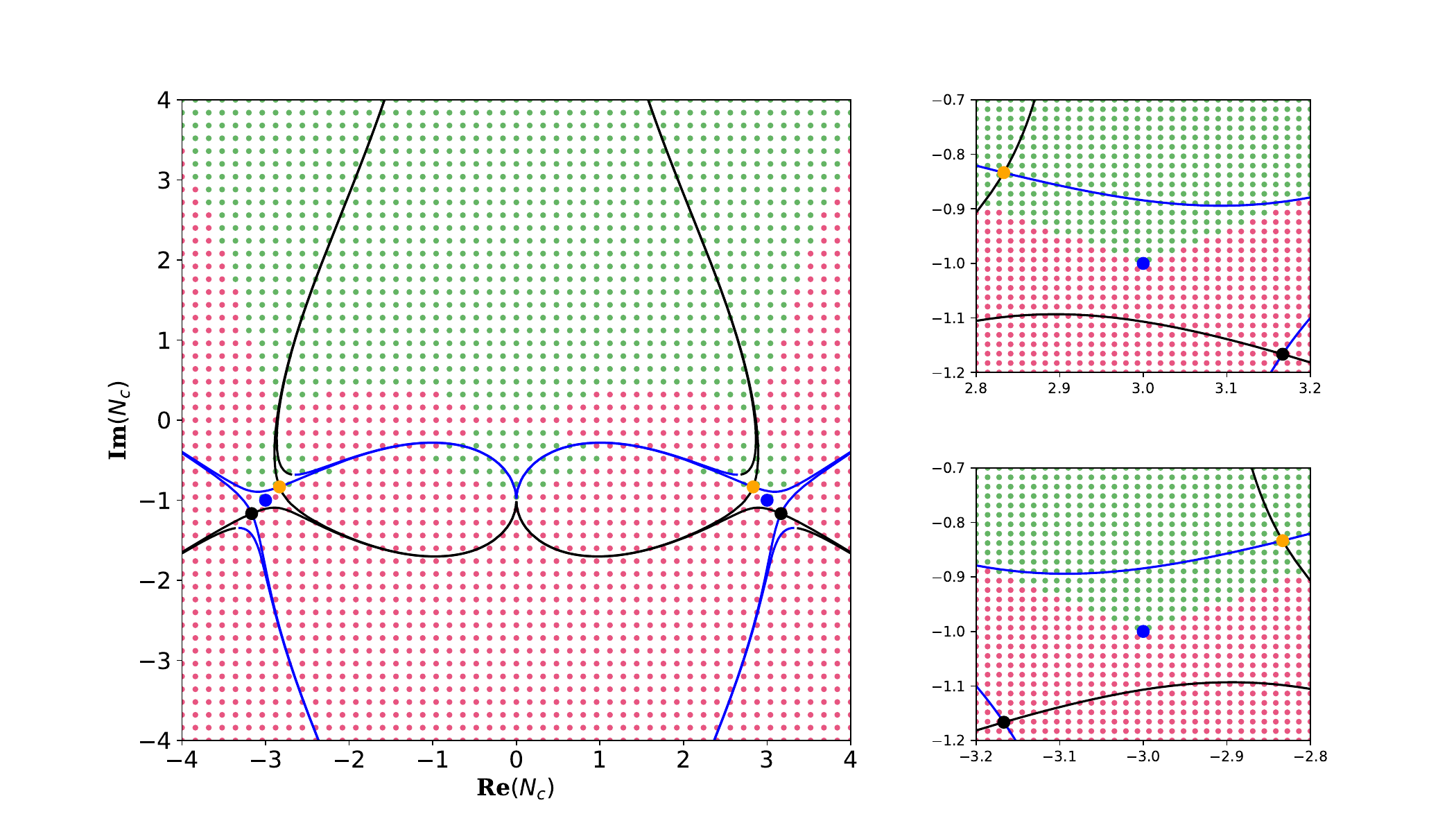}
    \caption{Similar to Fig. \ref{Fig:KSW_x_0.95}, this figure corresponds to the degenerate case with $x=1$. In this scenario, we find that two of the quantum-corrected saddles are pushed into the allowed region, while the other two are pushed into the disallowed region.}
    \label{Fig:KSW_x_1}
\end{figure}

To assess the allowability of metrics elsewhere in the complex lapse plane, we have to resort to the tests described in Section \ref{Rid_Cri} and Section \ref{Ext_cur}, which involve numerical analysis. The results of detailed numerical analysis of off-shell geometries have been illustrated in the figs. [\ref{Fig:KSW_x_0.95}] and [\ref{Fig:KSW_x_1}]. From the figs. [\ref{Fig:KSW_x_0.95}] and [\ref{Fig:KSW_x_1}], we observe that the semiclassical No-boundary saddles consistently lie along the boundary of the KSW allowable region (analytical details been briefed below), whereas the non–No-boundary saddles are located strictly within the disallowed region. Moreover, we find that for $x=0.95$, (see Figure \ref{Fig:KSW_x_0.95}), the quantum-corrected No-boundary saddle is pushed into the allowed region, whereas for $x=0.75$, it moves into the disallowed region. In contrast, the quantum-corrected non–No-boundary saddle remains in the disallowed region in both cases. In the degenerate case $x=1$, the two doubly degenerate saddles split into four non-degenerate saddles, with two residing in the allowed region and the other two in the disallowed region. A detailed analytical understanding of the conditions under which the quantum-corrected No-boundary saddles move into the allowable region requires a more careful investigation, which we leave for future work. A notable observation from all our numerical analysis is that the Lefschetz thimbles associated with the relevant saddle points passes through the disallowed region.
%

%%%%%%%%%%%%%%%%%%%%%%%%%%%%%%%%%%%%%%%%%%%%%%%%%
\subsection{Anti-linearity and KSW}
\label{subs:AL_ksw}
%%%%%%%%%%%%%%%%%%%%%%%%%%%%%%%%%%%%%%%%%%%%%%%%%

In Sec. \ref{sec:sym_type1}, we figured out that the present minisuperspace model admits the antilinear symmetry. It will be worth analyzing this symmetry in the context of KSW criterion too. To proceed, we first note that the on-shell solution of the scale factor mentioned in Eq. (\ref{eq:qsol_RBC}) satisfies
\beq
\label{eq:on-shell_anti_pro}
    \bar{q}(N_c, t)^{*}=\bar{q}(-N_{c}^{*}, t),\,\,\forall t\in [0,1],
\eeq
under the antilinear transformation mentioned in Eq. (\ref{eq:antilin}),
where $\bar{q}(t)\equiv \bar{q}(t,N_c)$.
Using the identity $|\arg(\bar{z})|=|\arg(z)|,\,\,\arg\in (-\pi,\pi]$,
we find that 
\begin{equation}
\label{eq:antispatial}
|\arg(q(N_c,t))|=|\arg(q(N_c,t)^*)|=|\arg(q(-N_c^*,t))|.
\end{equation}
From Eq. (\ref{eq:antispatial}) it is easy to see that the KSW function mentioned in Eq. (\ref{eq:KSW_criterion}) is the same at the anti-linear points $N_c$ and $-N_c^{*}$, i.e., 
\beq
\label{eq:antilinearlsksw}
    \Sigma(N_c, t) = \Sigma(-N_c^{*}, t).
\eeq
The above equation implies that, if a point ($N_c$) is KSW allowed (or disallowed), then the anti-linear point ($-N_c^*$) is also allowed (or disallowed). This shows that the KSW criterion also respects the antilinear symmetry of the theory. This characteristic of the KSW criterion has been reflected in the plots. Moreover, this property holds even at the level of complex time path $t(N_c,u)$ satisfying 
\beq
\label{eq:anti_sym_time_path}
     t(N_c, u)^{*}=t(-N_c^{*}, u),\,\,\forall u\in[0,1],
\eeq
where $t(u)\equiv t(N_c,u)$. 
%

%%%%%%%%%%%%%%%%%%%%%%%%%%%%%%%%%%%%%%%%%%%%%%%%%
\subsection{KSW near No-boundary saddles}
\label{subs:KSW_nb_sad}
%%%%%%%%%%%%%%%%%%%%%%%%%%%%%%%%%%%%%%%%%%%%%%%%%

\begin{figure}
    \centering
    \includegraphics[trim={0 0 2cm 2cm}, clip,scale = 0.45]{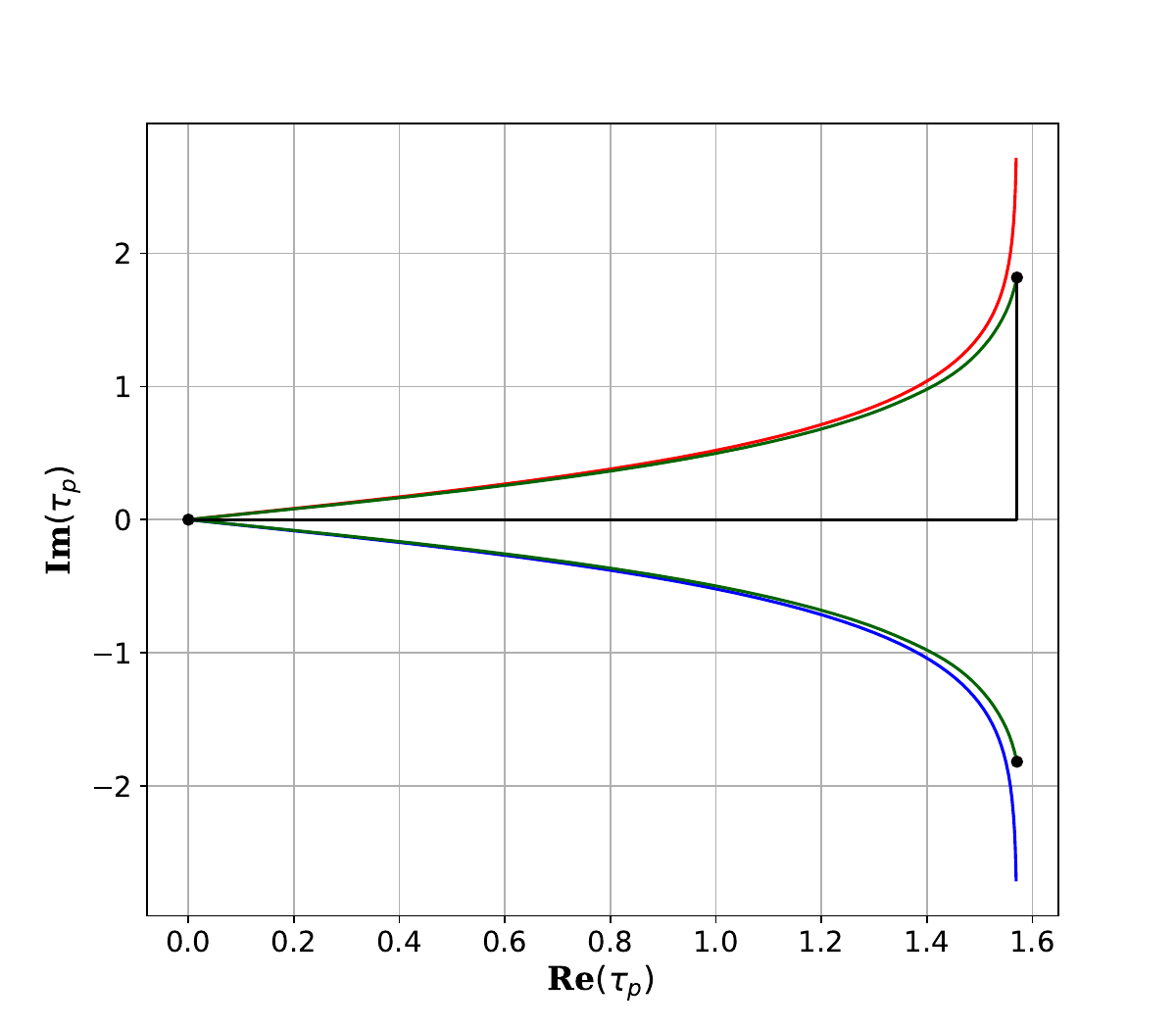}
    \caption{This plot illustrates the allowability of background no-boundary saddles under the Extremal Curve Test.  The parameters are set to $k = 1, q_f = 10, \Lam = 3$, and $P_{i} = -3i$. The resultant no-boundary saddles are found to be $N_{\pm}^{\rm nb} = \pm 3 - i $, with the corresponding $\nu$ at the saddles given as $\nu_{\pm} = \pi/2 \pm i \arcsinh{(3)}$, respectively. Here, the red and blue lines represent the extremal curves for the no-boundary saddles. The black dots denotes the $\tau_y(\tau_x =0) =0$ and $\nu_{\pm}$. The black line connecting the initial point to $\nu_{+}$ represents the Hartle-Hawking no-boundary contour. A similar contour exists for the  $N_{-}^{\rm nb}$ saddle, connecting to $\nu_{-}$. An explicit allowable path connecting the two endpoints for the no-boundary saddles is shown by the green curve. Along this curve, the KSW function remains bounded by $\pi/\ep < \pi$, where $\ep \approx 1.036938$, thereby confirming the allowability of no-boundary saddles.  }
    \label{Fig:Nb_ext_all_curves}
\end{figure}

We now move our focus to study the KSW criterion near the No-boundary saddle points. For the closed universes, at the semiclassical no-boundary saddles, the on-shell solution is given by (see Appendix \ref{Eucl_Phy_Time_Cal})
\beq
\label{eq:on_sol_nb}
\bar{q}(N_{\pm}^{\rm nb},\tau_{p}) =  (3/\Lam)\sin^2({\sqrt{\Lam/3} \tau_p}) \, ,
\eeq
with the $\nu = \sqrt{3/\Lam}[\pi/2 \pm i \arcsinh{(\sqrt{\Lam q_f/3 - k}})]$ at the final hypersurface. For $q_{f}<3k/\Lam$, the $\nu$ is purely real, describing a purely Euclidean geometry. At $q_f = 3/\Lam$, $\nu$ reaches $\pi/2$, corresponding to the waist of the de Sitter space. For $q_{f} > 3k/\Lam$, $\nu$ becomes complex with a fixed real part and an increasing imaginary part as universe evolves. This time contour is shown in the Figure \ref{Fig:Nb_ext_all_curves}, and is known as the Hartle-Hawking No-boundary contour. To make calculations look cleaner, we work in rescaled Euclidean physical time plane i.e $\tilde{\tau_p} =\sqrt{\Lam/3}\tau_p$ plane. In the re-scaled plane, the final point becomes as $\tilde{\nu} = \pi/2 \pm i \sinh^{-1}{(\sqrt{\Lam q_f/3 - k})}$. For the semiclassical No-boundary saddles, the extremal curve equation in the $\tilde{\tau_p}$ plane becomes as follows:
\beq
\label{eq:ext_curve_nb_sad}
|\arg(d\tilde{\tau}_{p}/du)| + 3 |\arg \sin{(\tilde{\tau}_{p})}| = \pi/2\, .
\eeq
Using the triangle inequality: $|z_{1}+z_{2}|\leq |z_{1}| + |z_{2}|$, where $z_{1}$ and $z_{2}$ being any generic complex variables. The extremal curve of the semiclassical No-boundary saddles must satisfy
\beq
\label{eq:red_ext_curve_cond_nb_sad}
|\arg ( \sin^3{(\tilde{\tau}_{p})} \, d\tilde{\tau}_{p}/du )| \leq \pi/2\, .
\eeq
Again, according to the Petrovitsch theorem, we find that the extremal curve of semiclassical No-boundary saddles, that satisfies eq. (\ref{eq:ext_curve_nb_sad}) must remain bounded between the reduced extremal curves, which satisfy the following simplified equation: 
\beq
\label{eq:red_ext_curve_eq_nb_sad}
|\arg( \sin^3{(\tilde{\tau}_{p})} \, d\tilde{\tau}_{p}/du )| = \pi/2\, .
\eeq
For this case, the analytical expression for the reduced extremal curve with the initial condition $\tilde{\tau}_{p}(0) = 0$ can be obtained explicitly. Writing $\tilde{\tau}_{p} = \tilde{\tau}_x + i \tilde{\tau}_y$, the expression for the reduced extremal curve is given by:\footnote{This equation leads to multiple curves, however one should consider the curve that satisfies $\tau_{y} = l \sin{\pi/8}$ when $\tau_{y} = l \cos{\pi/8}$, in limit $l \to 0,$ that are the no-boundary conditions. }
\beq
\label{eq:red_ext_curve_sol}
\cos{(3\tilde{\tau}_x)} \cosh{(3\tilde{\tau}_y)} - 9 \cos{(\tilde{\tau}_x)}  \cosh{(\tilde{\tau}_y)} + 8 = 0 \, .
\eeq
The reduced extremal curve starts from $\tilde{\tau}_y(\tilde{\tau}_x =0) = 0$ and asymptotes the vertical line $\tilde{\tau}_x = \pi/2$, indicating that the curve never crosses this boundary. Interestingly, in the case of No-boundary saddles, the extremal curve coincides with this reduced extremal curve.  As a result, the allowable condition for the extremal curve test $|\tilde{\tau}_x| < |\rm Re(\tilde{\nu})|$ at $\tilde{\tau}_y = \rm Im(\tilde{\nu})$ is achieved for all values of $q_f$. Thereby ensuring the KSW allowability of the background Hartle-Hawking No-boundary saddle points. One such allowable contour for the No-boundary saddle has been shown in the Fig \ref{Fig:Nb_ext_all_curves}.
% %%%%%%%%%%%%%%%%%%%%%%%%%%%%%%%%%%%%%%%%
% %%%%%%%%%%%%%%%%%%%%%%%%%%%%%%%%%%%%%%%%
Let us now briefly comment on the allowability of the region near the background No-boundary saddles, expressed as $N_{\pm}^{\delta,\rm nb} = N_{\pm}^{\rm nb} + \delta  $, where $\delta$ represents the perturbation around the saddle point. At $t = 0$, the spatial KSW function near the saddles, to leading order in $\delta$, reduces to
\beq
\label{eq:spa_KSW_cor_sads}
    \Sigma_{\rm spatial}(N_{\pm}^{\rm  nb,\delta} = N_{\pm}^{\rm nb} + \delta, t=0) = 3\left|\arg\left(-\frac{2 i \delta \sqrt{\Lambda  q_f-3}}{x \sqrt{\Lambda  q_f-3}\mp i \sqrt{3} (x-1)}\right)\right|
\eeq
If this quantity exceeds $\pi$, it necessarily implies that the point near to the saddle is disallowed. However, if it is less than $\pi$ doesn't guarantee the allowability. Notice, that the allowability of near saddle points depends upon $x$ and $q_f$ factors. For instance, the $x=1,q_f\neq 3/\Lam$ degenerate case, the spatial KSW function at  $t=0$ becomes
\beq
\label{eq:allox=1}
   \Sigma_{\rm spatial}( N_{\pm}^{\delta,\rm nb} = N_{\pm}^{\rm nb} + \delta, t=0) = 3|\arg(-i\delta)|\, .
\eeq
Complex lapse points that are satisfying $3|\arg(-i\delta)| \geq \pi$ are necessarily disallowed. However, those satisfying  $3|\arg(-i\delta)| < \pi$ are not necessarily implying allowability. Analyzing the quantum corrected saddles for this degenerate case, where 
\bea
\label{eq:quantum_cor_x_deg12}
\delta^{\pm}_{+} && = \pm e^{-i\pi/4}\bigl(\Lam q_{f}/3 -k\bigr)^{-1/4}/\sqrt{2\Lam}\, , \notag \\
\delta^{\pm}_{-} && = \pm e^{i\pi/4}\bigl(\Lam q_{f}/3 -k\bigr)^{-1/4}/\sqrt{2\Lam}\, ,
\eea
we find that both $\delta_+^+$ and $\delta_-^-$ violate the inequality, and hence non-allowable. This remark is illustrated in the Figure. \ref{Fig:KSW_x_1}. Complete analytic understanding of a generic point in the vicinity of the saddle requires a more detailed and systematic study, particularly involving the behavior of the extremal curves. Such an investigation is beyond the scope of the present work and will be addressed in future studies.

%%%%%%%%%%%%%%%%%%%%%%%%%%%%%%%%%%%%%%%%%%%%%%%%%%%%
\section{$\theta$-KSW criterion
}
\label{Mod_KSW_bound}
%%%%%%%%%%%%%%%%%%%%%%%%%%%%%%%%%%%%%%%%%%%%%%%%%%%
%
\begin{figure}
\subfigure[]{\includegraphics[trim={0.6cm 0 1cm 1.5cm}, clip, width=8cm,height=7cm]{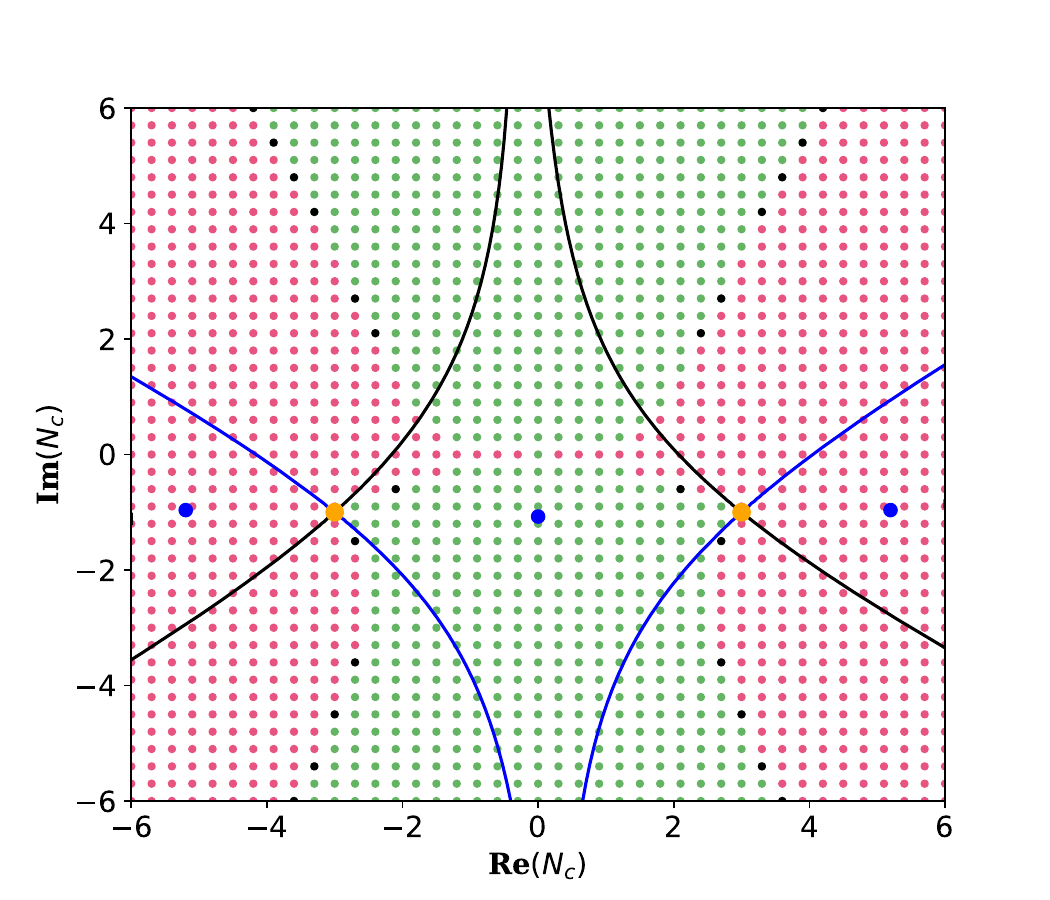}}
\subfigure[]{\includegraphics[trim={0.8cm 2.5cm 7cm 2cm}, clip, scale = 0.45]{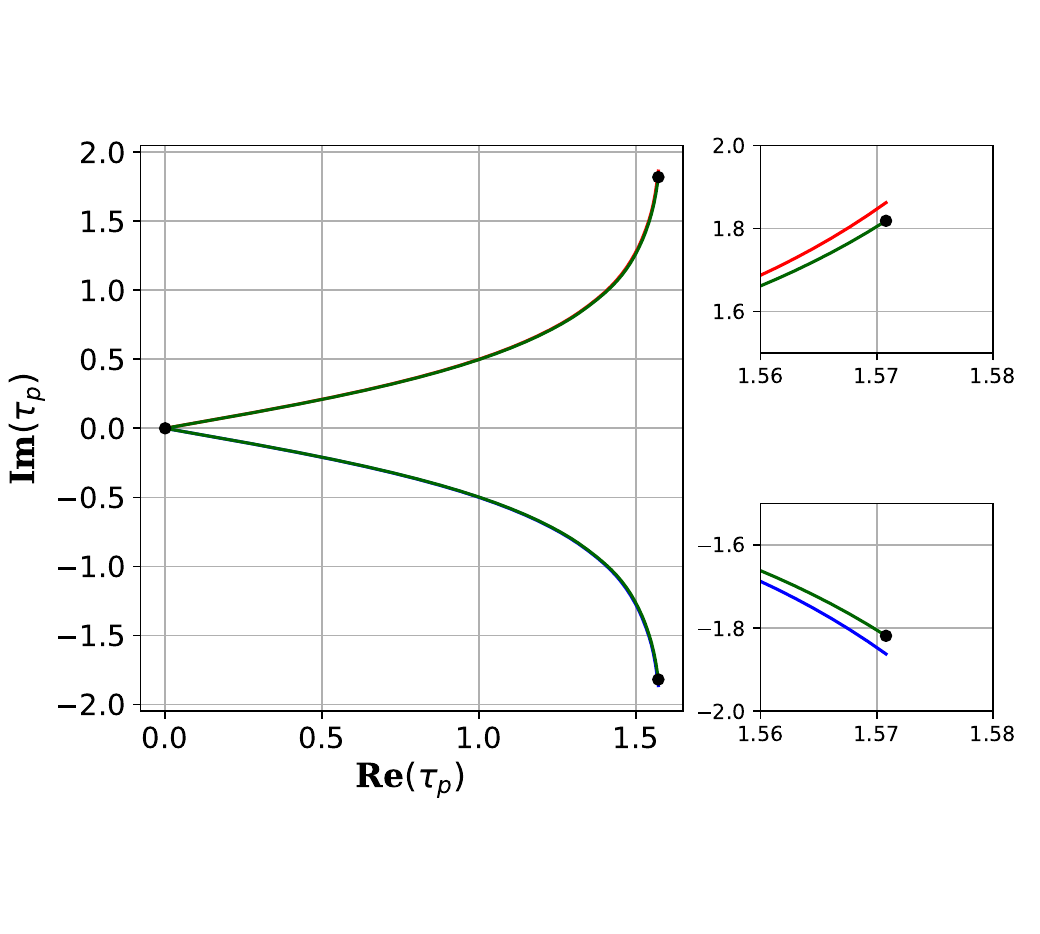}}
\caption{This figure shows the $\theta$-KSW constraint implementation for Neumann and Dirichlet boundary conditions imposed on the initial and final hypersurfaces, respectively. The parameters are set to : $\al = 2, x=0,  k=1, q_{f} = 10, \Lam =3, P_{i}=-3i$, and $\theta = 0.05$. a) The red dots and green dots represent the disallowed and allowed geometries under $\theta$-KSW constraint, respectively. The black dots correspond to geometries that become disallowed under the $\theta$-KSW constraint, despite being allowed when $\theta = 0$. The orange dots denote no-boundary saddles, namely, $N_{\pm}^{\rm nb} = \pm 3 - i$. The blue dots correspond to the points at which the on-shell action vanishes: $N_{1} \approx 6.15624\times10^{-16} - 1.07406 i, N_{2} \approx 5.19655 - 0.96297 i$ and $N_{3} \approx -5.19655 - 0.96297 i$, (see appendix \ref{Zer_Act} for the discussion on zeros of on-shell action). b) This plot shows the allowability of no-boundary saddles, even after complex deformation of plank length. The green curve is an allowable path with KSW function being $\pi/\ep < \pi$, where $\ep \approx 1.003806$.} 
\label{Fig:KSW_theta_0.05}
\end{figure}
\begin{figure}
\subfigure[]{\includegraphics[trim={0.6cm 0 1cm 1.5cm}, clip, width=8cm,height=7cm]{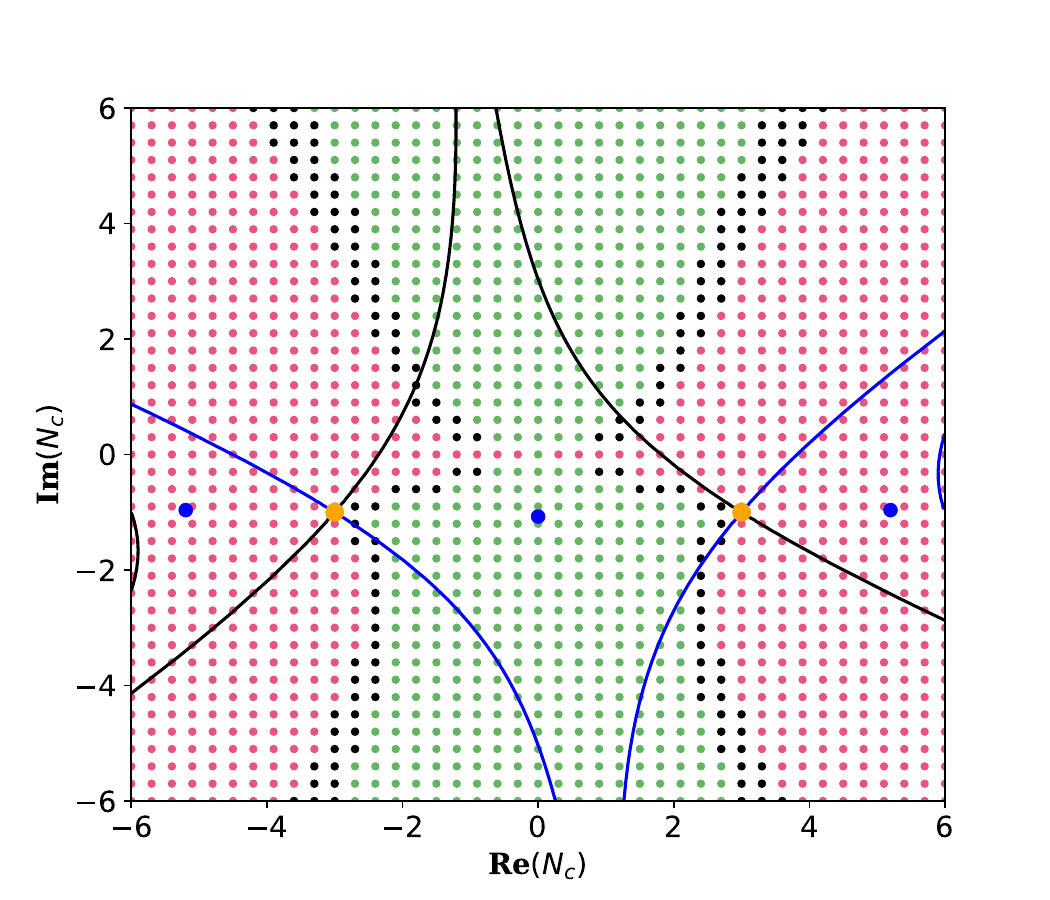}}
\subfigure[]{\includegraphics[trim={0 0 1cm 2cm}, clip, scale = 0.5]{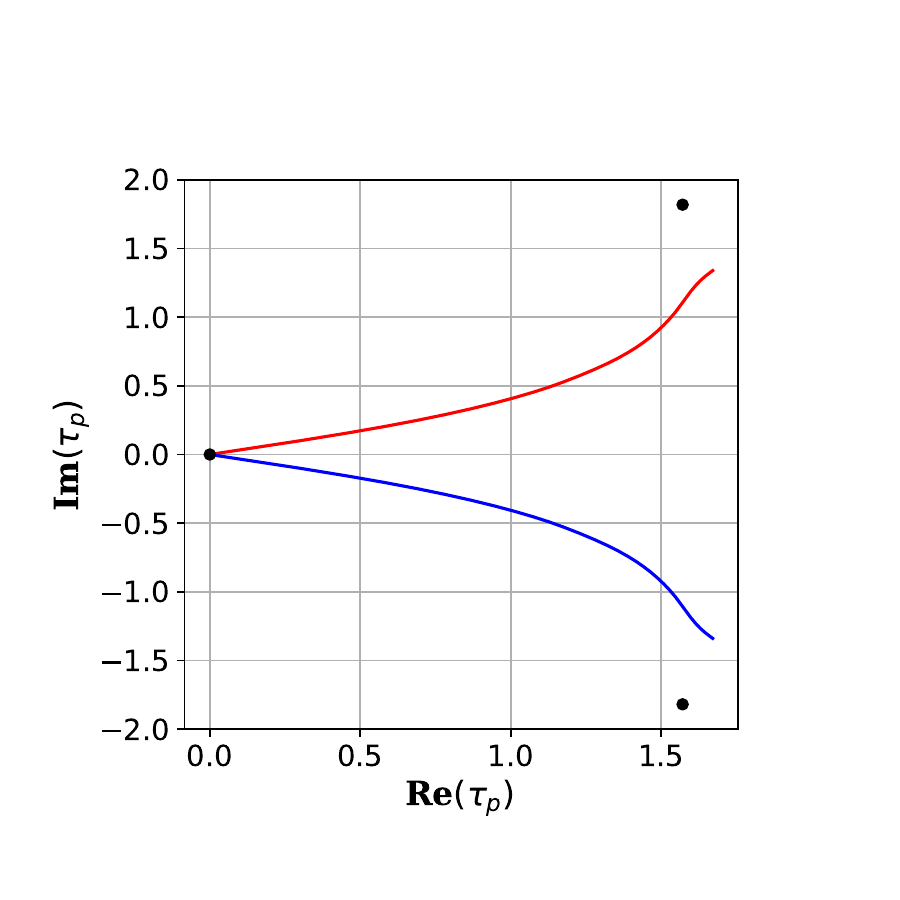}}
\caption{ A similar plot to Fig. \ref{Fig:KSW_theta_0.05}, with $\theta = 0.3$ is shown. This value of $\theta$ violates the bound given by Eq. (\ref{eq:cons_theta_qf>3_Lam}), thereby implying that no-boundary saddles are disallowed. The right panel figure clearly demonstrates the disallowablity of the no-boundary saddle, as there exists no curve starting from $\tau_p = 0$ can reach $\nu$ without crossing the extremal curve.}
\label{Fig:KSW_theta_0.3}
\end{figure}

Previously we saw that {\bf type-1} degeneracies are caused by the symmetries present in the system. These symmetries are anti-linearity (mentioned in eq. (\ref{eq:antilin})) which exists for all boundary choices, and symmetry specific to Robin boundary condition mentioned in eq. (\ref{eq:trans2}). These symmetries together lead to overlapping of flowlines. Loop corrections coming from quantum fluctuations of fields when present, lead to breaking of symmetry in eq. (\ref{eq:trans2}). This helps in resolving some of the {\bf type-1} degeneracies, leaving behind a system which still respects anti-linearity with some residual {\bf type-1} degeneracies. These residual degeneracies can be lifted by breaking anti-linearity, which is possible to achieve by introducing a tiny complex deformation of $G\hbar$. Lifting the degeneracies is crucial as otherwise it is not possible to implement Picard-Lefschetz methods unambiguously. 

However, the presence of an additional tiny phase like this can cause issues with convergence, where it becomes important to re-access the KSW-criterion to find the set of allowed metrics on which a suitable QFT can be defined. Hence, it is instructive to examine how the KSW criterion is modified when such a phase is present. Similar to the criterion in Eq. (\ref{KSW}), the geometry $g_{\mu\nu}$ will be allowable iff  
\bea
\label{eq:KSW_modified}
 {\rm Re}\left(e^{-i\theta} \sqrt{det\,g} \,\, g^{\mu_1\nu_1}\cdots g^{\mu_q\nu_q}F_{\mu_1\mu_2...\mu_q}F_{\nu_1\nu_2...\nu_q}\right) > 0 \, .
\eea
Analyzing the above constraint in the diagonal basis where the metric $g_{\mu\nu}(x)=\lambda_{\mu}(x) \delta_{\mu\nu}$ for all $0\leq q\leq D$, we get
\begin{equation}
\label{eq:KSW_mod_eg_conds}
    \begin{split}
        & {\rm Re}\biggl( e^{-i\theta}\prod_{\mu=1}^D\sqrt{\lambda_\mu}\biggr)>0 \,\,\text{and} \\
        & {\rm Re}\biggl( e^{-i\theta} \prod_{\mu=1}^D\sqrt{\lambda_\mu}\prod_{\nu\in S}\lambda_\nu^{-1}\biggr)>0\,\,\forall S,
    \end{split}
\end{equation}
for $q=0$ and $0<q\leq D$ respectively, where $S$ is all possible subsets of $(1, \, 2, \cdots, \,D)$. In obtaining Eq. (\ref{eq:KSW_mod_eg_conds}), we assume that the fields ($F$) are real. For $q=0$, we get
\begin{equation}
    \label{eq:q=0ksw}
    -\pi<-2\theta+\arg(\lambda_1)+\arg(\lam_2)+\cdots +\arg(\lam_D)<\pi.
\end{equation}
For $0<q\leq D$, one needs to flip any ``$q$" number of `$+$' signs to `$-$' in front of ``$\arg$" in Eq. (\ref{eq:q=0ksw}).
These constitute a total of $1+\sum_{q=1}^{D}{}^DC_q=2^D$ independent constraints, which are concisely expressed by the following
\beq
\label{eq:KSW_theta_conds}
-\pi < \pm a_1 \pm a_2 \pm \cdots \pm a_D \pm 2 \theta < \pi, \quad \text{where} \quad a_\mu=\arg(\lambda_\mu).
\eeq
Interestingly, changing the deformation angle $\theta$ to $-\theta$ does not affect the inequalities, since the constraints arising from a $q$-form field with $\theta > 0$ are identical to those from a $(D-q)$-form field with $\theta < 0$, due to the identity ${}^D C_q = {}^D C_{D - q}$. Taking all $0 \leq q \leq D$ forms into account, one gets the $\theta$-modified KSW-criterion:
\beq
\label{eq:mod_KSW_1}
   |2\theta|+ |a_1|+|a_2|+\cdots+|a_D|<\pi.
\eeq
or equivalently,
\beq
\label{eq:mod_KSW_bound}
\boxed{ \sum_{\mu=1}^D|\arg(\lambda_\mu)|<\pi-2|\theta| }
\eeq
Clearly, this $\theta$-modified KSW bound is tighter than the original bound corresponding to $\theta = 0$. Some geometries that were permissible under the original criterion might become disallowed under the rotated bound, particularly those satisfying the inequality $\pi-2|\theta| < \sum_{i=1}^D |\arg(\lambda_i)| < \pi$. Since $\theta$ is usually taken to be tiny, geometries that lie on the verge of allowability (near $\pi$) become non-allowable. This behavior is demonstrated for the case with Neumann boundary condition on the initial hypersurface and Dirichlet boundary condition on the final hypersurface, in a simplified setup with no quantum corrections, see figures \ref{Fig:KSW_theta_0.05}, and \ref{Fig:KSW_theta_0.3}. Our observation - that by choosing an appropriate $\theta$, one can always violate the original KSW bound - was previously noted by Witten in \cite{Witten:2021nzp}, where it was argued that a similar phase might come in the conformal rescaling of the metric. In the same work, a possible resolution was suggested that involves rotating the integration contour for perturbative fluctuations of the fields, $A\rightarrow Ae^{i\theta/2}$, which compensates for the extra phase, thereby keeping the bound unchanged. This might be a sensible thing to do, as we have already noticed that in order to define the path integral for gravity, one also needs to suitably rotate the real $N_c$-contour of integration infinitesimally into the complex plane. However, we will not pursue this direction in this paper and stick to the real fields and analyze the consequences. In particular, we will be interested in the question of whether a small-$\theta$ required for resolving the degeneracy makes the No-boundary saddles non-allowable.
Also note that since the LHS of Eq. (\ref{eq:mod_KSW_bound}) is positive semi-definite, we must have $|\theta|\leq \pi/2$. 

We now proceed to determine the bound on $\theta$ by requiring that the background No-boundary saddles remain within the KSW-allowed region under the complex phase. Following the earlier analysis of semiclassical No-boundary saddles, under a complex deformation of the Planck length by an angle $\theta$, we obtain the equation for the reduced extremal curve as follows:
\beq
\label{eq:mod_red_ext_curve}
\frac{\cos{(3\tilde{\tau}_x)} \cosh{(3\tilde{\tau}_y)} - 9 \cos{(\tilde{\tau}_x)} \cosh{(\tilde{\tau}_y)} + 8}{ \sin{(3\tilde{\tau}_x)} \sinh{(3\tilde{\tau}_y)} - 9 \sin{(\tilde{\tau}_x)} \sinh{(\tilde{\tau}_y)} } = \pm \tan{|\theta|} \, .
\eeq
As in the case of the No-boundary saddles without deformation, we find that even under complex deformation, the reduced extremal curve coincides with the extremal curve. Requiring the No-boundary saddles to remain allowable imposes the condition: $|\tilde{\tau}_y| > |\rm Im(\tilde{\nu})|$ at $|\tilde{\tau}_x| = |\rm Re(\tilde{\nu})|$ for all values of $q_f$. This translates into having a bound on the permissible amount of deformation, which is given by
\beq
\label{eq:cons_theta_nb}
\biggl| \frac{ (4\cot{|\theta|} + \sqrt{1+ 16 \cot^2{|\theta|} })^{2/3} - 1}{ (4\cot{|\theta|} + \sqrt{1+ 16 \cot^2{|\theta|}})^{1/3} }\biggr| > |\sinh{\rm Im(\tilde{\nu})}| \, .
\eeq
It is observed that for $q_{f} \leq 3/\Lambda$, where the No-boundary geometry remains in the Euclidean phase, with $\rm Im(\tilde{\nu}) = 0 $, the modified KSW bound is satisfied for any value of $|\theta| < \pi/2$. However, for $q_{f}>3/\Lam$, the above condition becomes
\beq
\label{eq:cons_theta_qf>3_Lam}
\biggl| \frac{ (4\cot{|\theta|} + \sqrt{1+ 16 \cot^2{|\theta|} })^{2/3} - 1}{ (4\cot{|\theta|} + \sqrt{1+ 16 \cot^2{|\theta|}})^{1/3} }\biggr| > \sqrt{\frac{\Lam q_f}{3} - 1} \, .
\eeq
Therefore, we confine the deformation angle $\theta$, such that above condition is satisfied. Furthermore, for $|\theta| \ll 1$, we have
\beq
\label{eq:cons_small_theta_qf>3_Lam}
2  \biggl( \frac{\Lam q_f}{3} - 1 \biggr)^{-1/2} \gg |\theta|^{1/3} \, .
\eeq
Hence, an infinitesimal complex deformation of the Planck length is permissible as long as the above condition is satisfied, thereby ensuring that the No-boundary saddles remain allowable even after the deformation. The situations in which this constraint is satisfied or violated are depicted in Figs. (\ref{Fig:KSW_theta_0.05}) and (\ref{Fig:KSW_theta_0.3}), respectively, for the case with Neumann boundary condition on the initial hypersurface and Dirichlet boundary condition on the final hypersurface, where the two relevant saddle points correspond to No-boundary geometries. 
%%%%%%%%%%%%%%%%%%%%%%%%%%%%%%%%%%%%%%%%%%%%%%%%%%

%%%%%%%%%%%%%%%%%%%%%%%%%%%%%%%%%%%%%%%%%%%%%%%%%
\section{Conclusions and Outlook}
\label{sec:conc}
%%%%%%%%%%%%%%%%%%%%%%%%%%%%%%%%%%%%%%%%%%%%%%%%%

In this paper, we study the gravitational path-integral in the Lorentzian signature for Einstein-Hilbert (EH) Gravity including Gauss-Bonnet term, by restricting the geometries in the path-integral to $\mathbb{R} \times S^3$ topology, with the metric characterized by lapse $N_c$ and scale-factor $q(t)$. Such path-integrals are exactly doable in four spacetime dimensions for Robin boundary conditions (which includes Dirichlet and Neumann BC as special cases). However, when analyzing the integral using WKB and Picard-Lefscehtz methods, one notices the occurrence of degeneracies which prevent the unambiguous application of Picard-Lefschetz methods. This paper studies these degenerate situations and analyses them systematically. 

We start by first computing the gravitational path-integral with Robin boundary conditions. In the metric ansatz of the mini-superspace approximation within the ADM gauge, this can be computed exactly for the Gauss-Bonnet gravity. Although this has already been computed in past works \cite{Ailiga:2023wzl, Ailiga:2024mmt}, in the present paper we re-derive those results alternatively by using Hubbard–Stratonovich (HS) transformations, by directly computing the lapse $N_c$-integral given in Eq. (\ref{eq:TA_lps_int}). This method of computing the transition amplitude wasn't done previously. The benefit of this is that it can be directly compared with the saddle and Picard-Lefschetz analysis of the lapse $N_c$-integration, where a geometric understanding exists as each $N_c$ corresponds to a specific complex spacetime metric. 

We then proceed to analyse the lapse integral specified in (\ref{eq:TA_lps_int}) using WKB and Picard-Lefschetz methods. This process involves finding the set of saddles for the specific choice of boundary conditions, the set of steepest ascent/descent flow-lines corresponding to each saddle, and determining their {\it relevance} as dictated by the Picard-Lefschetz methods. This straight-forward procedure works fine as long as there are no degeneracies present as it allows us to decide the {\it relevance} of saddles unambiguously. However, in our studies, emergence of degeneracies is noticed which we classify in two types: \textbf{type-1} and \textbf{type-2}. \textbf{Type-1} degeneracies are those where the flow-lines from different saddles overlap with each other while \textbf{type-2} degeneracies occur when the saddles merge depending on the specific choice of boundary conditions. Existence of {\bf type-1} degeneracies leads to failure in deciding the relevance of saddles unambiguously, while existence of {\bf type-2} degeneracies leads to failure of WKB methods. Hence, it is crucial to resolve these degeneracies to compute the integral utilizing Picard-Lefschetz methods. This paper tries to understand the cause of these degeneracies and proposes various methods to resolve them. 

Type-1 degeneracies occur when ${\rm Im} \bl(i \bar{S}^{\rm on-shell}_{\rm tot} \br)$ is the same at two different saddles, while type-2 occurs for special boundary choices. Both these degeneracies gets resolved in the presence of artificial {\it defect} (introduced by hand) like the one considered in eq. (\ref{eq:Sact_arti_defect}). Presence of such tiny {\it defects} modifies the ${\rm Im} \bl( i \bar{S}^{\rm on-shell}_{\rm tot} \br)$ for each saddle slightly thereby lifting the type-1 degeneracies. For type-2, such defect leads to non-degenerate saddles by introducing a small correction. One wonders if such a process of resolving degeneracies can be achieved without introducing any artificial {\it defects}, and whether the theory itself can offer a resolution to this. 

It is seen that there indeed exists a way out, bypassing the need to introduce artificial {\it defects}. This is achieved partially by considering quantum corrections. In the gravitational system studied in this paper, it is noticed that the quantum fluctuations of the scale-factor give rise to corrections that, in turn, end up modifying the lapse-$N_c$ action. Such corrections are able to resolve some of the degenerate situations, but not all. For example, for $q_f>3k/\Lam$, all type-1 degeneracies are resolved. But for $q_f<3k/\Lam$, resolution is achieved only for some cases. Type-2 degeneracies, on the other hand, are always resolved due to quantum corrections. Furthermore, it is seen that complexifying either $G$ (\cite{Maldacena:2024spf, Ivo:2025yek}) or $\hbar$ (\cite{Dorigoni:2014hea,Honda:2024aro}) or both, which is equivalent to complex deformation of either Planck's length $l_p^2 = G \hbar$ or Planck's mass $M_p^2 = \bl( G \hbar \br)^{-1}$ by a small phase helps in resolving {\bf type-1} degeneracies entirely. This led us to propose that a combination of quantum corrections and complex deformation of $(G\hbar)$ by a small phase resolves both {\bf type-1} and {\bf type-2} degeneracies. 

These studies motivate one to ask if there are any symmetries responsible for the occurrence of degeneracies. In the process of studies of introducing artificial {\it defects}, or quantum corrections plus complex deformations of $G\hbar$, we have realized that the {\bf type-1} degeneracies occur when ${\rm Im} \bl( i \bar{S}^{\rm on-shell}_{\rm tot} \br)$ is same for two or more saddles. But is there some underlying symmetry which is causing this? In this study we have come across two symmetries which appear in the lapse-$N_c$ on-shell action $\bar{S}^{\rm on-shell}_{\rm tot}$: anti-linearity ($\rm AL$) (mentioned in eq.(\ref{eq:antilin})) and $\overline{\rm AL}$ (mentioned in eq. (\ref{eq:trans2})). This symmetry has been observed in some form or other in various earlier works involving QCD or gravity \cite{DiTucci:2019bui, Lehners:2023yrj, Witten:2010cx, Kanazawa:2014qma, Witten:2025ayw, Harlow:2023hjb, Nishimura:2014rxa, Nishimura:2014kla, Nishimura:2015loa, Tanizaki:2015pua, Tanizaki:2015gcv, Bender:1998ke, Meisinger:2012va}. 

Existence of $\overline{\rm AL}$ symmetry eq. (\ref{eq:trans2}) implies that for $q_f > 3k/\Lam$, points in the lapse-$N_c$ plane which are related by (\ref{eq:trans2}) have same $H(N_x, N_y)$ as seen from eq. (\ref{eq:hHtrans2}). This immediately gives a hint of the occurrence of degeneracies if the saddles are related to each other by the symmetry $\overline{\rm AL}$ stated in eq. (\ref{eq:trans2}). In this case flow-line overlap each other resulting in Stoke-rays. If this symmetry is broken, then {\bf type-1} degeneracies are lifted for the case $q_f > 3k/\Lam$. It turns out, that corrections coming from quantum fluctuations of the scale-factor break this symmetry and hence naturally offers a way to resolve {\bf type-1} degeneracy for $q_f > 3k/\Lam$. This already suggest that quantum corrected lapse-$N_c$ action should be utilized in the study of Picard-Lefschetz analysis, as it overcomes {\bf type-1} degeneracies for $q_f > 3k/\Lam$. 

For $q_f < 3k/\Lam$, things are a bit different. The quantum corrected lapse $N_c$ action mentioned in eq. (\ref{eq:qmAact}), although breaks the $\overline{\rm AL}$ symmetry in eq. (\ref{eq:trans2}), still retains the anti-linearity stated in eq. (\ref{eq:antilin}), as can be seen from eq. (\ref{eq:anti-linear}). Due to this, there is still some residual degeneracies for $q_f < 3k/\Lam$ which are left unresolved even after breaking of symmetry (\ref{eq:trans2}). These residual {\bf type-1} degeneracies occur along the imaginary axis in the complex $N_c$-plane. For points lying on the imaginary axis, as a consequence of anti-linearity, they have the same $H(0, N_y)$ as mentioned in eq. (\ref{eq:stokes_con_qf<3k/Lam}), indicating that anti-linearity is directly responsible for these residual degeneracies. Thus breaking anti-linearity is necessary to resolve them. Complex-deformation of $G\hbar$ appears as a suitable method of breaking anti-linearity, thereby offering a way to overcome {\bf type-1} degeneracies. This leads us to propose a way of resolving all the degeneracies by simultaneous incorporation of quantum corrections and complex-deformation of $G\hbar$. This procedure breaks both the symmetries which are causing the degeneracies. 

We then proceed to check the compatibility of the methods of lifting degeneracies (either via artificial {\it defects} or via quantum corrections plus complex deformations of $G\hbar$) with the KSW-criterion. It is seen that the KSW-criterion which is shown to respect anti-linearity has points corresponding to No-boundary geometries lying on the boundary of the allowed-disallowed regions. However, when quantum corrections are incorporated, then the No-boundary geometries can get pushed either to the allowed region or disallowed region depending of external parameters, while irrelevant saddles always gets pushed in the disallowed region. This is interesting in the sense as the quantum corrections pushes the {\it relevant} No-boundary saddles which were previously lying on the boundary, to the KSW-allowed regions (at least for some parameters), thereby making them allowable geometries on which sensible QFTs can be defined. 

Furthermore, the complex-deformation of $G\hbar$ leads to modification of the KSW-criterion, where the RHS of the bound gets adjusted to incorporate the deformations. We call this $\theta$-KSW criterion which is mentioned in the eq. (\ref{eq:mod_KSW_bound}). This follows directly from the requirements of convergence of the path-integral for the field theory to have a well-defined QFT on the complex spacetime. It is noticed that saddles which lie on the boundary can become disallowed for certain values of $\theta$-deformation angle. For example the No-boundary geometry can become disallowed if $\theta$ is above certain value. However, if we demand that No-boundary saddles to be allowed, then it puts a constraint on deformation angle-$\theta$. When the Universe expands and size becomes large, then the $\theta$-parameter is restricted to be very small, tending to zero as the size of the Universe goes to infinity. This indicates that for a large-sized Universe, only tiny complex deformations are allowed if the No-boundary geometries are required to be always KSW-allowed. 

In hindsight, we realize the presence of anti-linearity symmetry in various forms (AL and $\overline{\rm AL}$) is causing the occurrence of degeneracies, which when broken by either by suitable {\it defects} or by complex-deformations of $(G\hbar)$, resolves degeneracies systematically. Although, corrections coming from quantum fluctuations of the scale-factor break $\overline{\rm AL}$ but not the {\rm AL}. These however are sufficient to lift the {\bf type-2} degeneracies. Overall, it is seen to resolve {\bf type-1} degeneracies we need to break AL-symmetry, which is done via complex deformations of $(G\hbar)$, while {\bf type-2} degeneracies are taken care of by quantum corrections. This is the first step in understanding the symmetries responsible for the degeneracies in the system, whose resolution is mandatory for unambiguous application of Picard-Lefschetz and WKB methods. This work however focuses only on the simple gravitational system where we work within the mini-superspace ansatz. How these results will be affected if one goes beyond mini-superpsace is not clear and yet to be seen.

%%%%%%%%%%%%%%%%%%%%%%%%%%%%%%%%%%%%%%%%%%%%%%%
\bigskip
\centerline{\bf Acknowledgements} 
%%%%%%%%%%%%%%%%%%%%%%%%%%%%%%%%%%%%%%%%%%%%%%

We are thankful to Romesh Kaul for illuminating discussions on symmetries. 
We would also like to thank Oliver Janssen for useful discussion on KSW numerics. We are grateful to the anonymous referee for bringing to our attention some key references involving symmetries.

%%%%%%%%%%%%%%%%%%%%%%%%%%%%%%%%%%%%%%%%%%%%%%%%%
\appendix
\section{Euclidean Physical Time Calculations}
\label{Eucl_Phy_Time_Cal}
%%%%%%%%%%%%%%%%%%%%%%%%%%%%%%%%%%%%%%%%%%%%%%%%%

In this section, we will derive the expression of Euclidean physical time for the generic on-shell solution given by
\beq
\label{eq:on_shell_sol}
\bar{q}(t) = \frac{\Lam}{3}N_c^2 t^2 + c_{1} t + c_{2}
\eeq
which can be re-expressed as
\beq
\label{eq:on_shell_sol_1}
\bar{q}(t) = \frac{\Lam N_c^2}{3} \Biggl[ \biggl(t + \frac{3 c_1}{2 \Lam N_c^2}\biggr)^2 - \biggl(1 + \frac{3 c_1}{2 \Lam N_c^2}\biggr)^2 \Biggr] + q_f \, .
\eeq
Using eq. (\ref{eq:Euc_phy_time}) we have the Euclidean physical time as
\bea
\label{eq:Euc_time}
\tau_{p} && = i \sqrt{\frac{3}{\Lam}} \Biggl[ \arctanh{\Biggl( \frac{2\Lam N_c^{2} t + 3 c_{1}}{2 N_c \sqrt{3 \bar{q}(t) \Lam}}\Biggr)} - \arctanh{\Biggl( \frac{ 3 c_{1}}{2 N_c \sqrt{3 c_2 \Lam}}\Biggr)} \Biggr] \, ,
\eea
that gives value of $\tau_{p}$ at final hypersurface as
\bea
\label{eq:nu_value}
\nu && = i \sqrt{\frac{3}{\Lam}} \Biggl[ \arctanh{\Biggl( \frac{2\Lam N_c^{2}  + 3 c_{1}}{2 N_c \sqrt{3 \bar{q}(1) \Lam}}\Biggr)} - \arctanh{\Biggl( \frac{ 3 c_{1}}{2 N_c \sqrt{3 c_2 \Lam}}\Biggr)} \Biggr] \, .
\eea
The on-shell solution in terms of Euclidean physical time variable is given by
\beq
\label{eq:on_shell_tau_p}
\bar{q}(\tau_{p}) = \Bigl( c_{2} - \frac{3 c_{1}^2}{4\Lam N_{c}^2}\Bigr) \cosh^2{\biggl( \arctanh{\biggl(\frac{3c_{1}}{2 N_c \sqrt{3 c_{2} N_c}}\biggr)} - i \sqrt{\frac{\Lam}{3}} \tau_{p}\biggr)} \, .
\eeq
Either imposing Robin boundary condition or Neumann boundary condition at the initial hypersurface and Dirichlet boundary condition at the final hypersurface, we find the final value of $\nu$ at the no-boundary saddles as
\bea
\label{eq:nu_value_Nb}
\nu_{\rm nb} && = \sqrt{3/\Lam} \bigl[ \pi/2 + i  \arcsinh{ (i \sqrt{k}+ \Lam N_{\pm}^{\rm nb} /3)}  \bigr] \, .
\eea
Plugging the semi-classical no-boundary saddle expression given in eq. (\ref{eq:nb_sads}) in the above equation, one obtains
\bea
\label{eq:nu_value_Nb_qf}
\nu_{\rm nb} && = \sqrt{3/\Lam} \bigl[ \pi/2 \pm i  \arcsinh{ (\sqrt{\Lam q_{f}/3 - k})}  \bigr] \, .
\eea
At these saddles, for $q_{f} < 3k/\Lam$ the $\nu_{nb}$ is purely real, suggesting the final hypersurface is purely Euclidean in nature. For $q_{f} > 3k/\Lam$, $\nu_{nb}$ becomes complex indicating the transition phase from Euclidean to Lorentzian. Moreover, for large $q_{f}>3k/\Lam$, from above expression, it is evident that $\nu_{nb}$  becomes purely imaginary in the complex Euclidean physical time plane. This indicates that the corresponding solution evolves into a Lorentzian classical universe at late times. The corresponding on-shell solution at the semiclassical saddles in terms of Euclidean physical time is given by
\bea
\label{eq:on_shell_tau_p_nb}
\bar{q}(\tau_{p}) && = (3k/\Lam)[1 - \cos^2{(\sqrt{\Lam/3}\tau_{p})}/k] \, . \\
\eea
For the closed universe $k = 1$, the scale factor along the complexified Euclidean time contour at the semiclassical saddle is given by
\bea
\label{eq:on_shell_tau_p_nb_closed}
\bar{q}(\tau_{p}) && = (3/\Lam)\sin^2{(\sqrt{\Lam/3}\tau_{p})} \, .
\eea
This corresponds to the smooth gluing of a four-sphere $S^{4}$ to a de sitter spacetime $dS^{4}$ with the three sphere waist of radius $\sqrt{3/\Lam}$, realising the no-boundary geometry for the above stated complexified $\tau_{p}$ contour.

%%%%%%%%%%%%%%%%%%%%%%%%%%%%%%%%%%%%%%%%%%%%%%%%%%%%%%%
\section{KSW criterion from Lorentzian path integral}
\label{appen:KSEeuclor}
%%%%%%%%%%%%%%%%%%%%%%%%%%%%%%%%%%%%%%%%%%%%%%%%%%%%%%%

Originally, the KSW criterion was derived in \cite{Witten:2021nzp}, starting from an Euclidean path integral. Here, we would like to obtain the same criterion starting from the Lorentzian path integral defined by 
\begin{equation}
    \label{eq:lorenpath}
    \int \mathcal{D}A e^{i\mathcal{S}_q[A]} ,\hspace{3mm} \mathcal{S}_q[A] =\frac{1}{2q!}\int_\mathcal{M} d^Dy\sqrt{-det\,g}\,g^{\mu_1 \nu_1}\cdots g^{\mu_q \nu_q} F_{\mu_1 \mu_2...\mu_q} F_{\nu_1 \nu_2...\nu_q}.
\end{equation}
The convergence of the path integral requires the following to hold
\begin{equation}
\label{eq:loreconvercond}
    \text{Im}[\sqrt{-det\,g}\,g^{\mu_1 \nu_1}\cdots g^{\mu_q \nu_q} F_{\mu_1 \mu_2...\mu_q} F_{\nu_1 \nu_2...\nu_q}]>0.
\end{equation}
If Eq. (\ref{eq:loreconvercond}) holds for all $q\in (0,\cdots,D)$-forms with $q=p+1$, one can say that the complex matrix $g_{\mu\nu}$ is allowable, similar to the Euclidean case. However, in the Lorentzian signature, we immediately note the presence of an ambiguity. It is because of the ``minus" sign in $det \,g$, as $-1$ can be written as $e^{\pm i\pi}$. It is not a priori clear what sign one should take. For different choices, one obtains different criterion. As before going to a diagonal basis, where $g_{\mu\nu}(y)=\lambda_\mu(y)\delta_{\mu\nu}$, we get the pointwise criterion for all $y\in \mathcal{M}$ (taking the ``$-$" sign)
\begin{equation}\label{eq:plussign}
\begin{split}
    \pi<\sum_{\mu=1}^D\arg \lambda_{\mu} (y)<3\pi,\,\hspace{3mm}\text{and}\,\hspace{3mm}
     \pi<\sum_{\mu\in S_1}\arg \lambda_{\mu} (y)-\sum_{\nu\in S_2} \arg \lambda_{\nu} (y)<3\pi,
    \end{split}
\end{equation}
for $q=0$ and $0<q\leq D$ respectively. Where $S_1$ is any subset of $D-p$ elements of $S$ and $S_2$ is any subset of $q$ elements of $S$ such that $S_1\cap S_2= \text{null}$, $S=\{1,2,3,\cdots,D\}$. On the other hand, taking the ``$+$" sign, we get
\begin{equation}\label{eq:minussign}
\begin{split}
    -\pi<\sum_{\mu=1}^D\arg \lambda_{\mu} (y)<\pi,\,\hspace{3mm}\text{and}\,
     -\pi<\sum_{\mu\in S_1}\arg \lambda_{\mu} (y)-\sum_{\nu\in S_2} \arg \lambda_{\nu} (y)<\pi,
    \end{split}
\end{equation}
for $q=0$ and $0<q\leq D$ respectively. Now, we observe that Eq. (\ref{eq:plussign}) can never be satisfied when considering all $p$-forms. If it is satisfied for $q$-form, it wouldn't be satisfied for $D-q$ form and vice versa. So, we are left with Eq. (\ref{eq:minussign}), which is exactly the same as what we get from the Euclidean path integral, and it can be compactly written as
\begin{equation}
    \sum_{\mu=1}^D\left|\arg \lambda_\mu\right|<\pi,
\end{equation}
where $\arg(z)\in (-\pi,\pi]$.
This conclusion is not surprising if we think of ``naive" wick rotation (as in flat space QFT), which relates Lorentzian to Euclidean path integral. And the ``$+$" sign relates precisely to the direction of the wick rotation and $\sqrt{-1}=+i$. Hence, the KSW criterion is the same, regardless of whether one starts with a Euclidean or Lorentzian path integral.

%%%%%%%%%%%%%%%%%%%%%%%%%%%%%%%%%%%%%%%%%%%%%%%%%%%%%%%
\section{Comments on zeros of the Action}
\label{Zer_Act}
%%%%%%%%%%%%%%%%%%%%%%%%%%%%%%%%%%%%%%%%%%%%%%%%%%%%%%%

In Ref.\cite{Witten:2021nzp}, it was argued that instanton solutions corresponding to vanishing Einstein-Hilbert action are unphysical, and such pathological configurations must be eliminated by the KSW criterion. Following the same lines, we are interested in examining the allowability of ``off-shell" geometries in the lapse ($N_c$) plane yielding to the vanishing on-shell action,i.e., $\bar{S}_{\rm tot}^{\rm on-shell}[N_c]=0$. These geometries satisfy all the components of Einstein's equation except the ``time-time" component (which is the Hamiltonian constraint). Let us first consider the $\beta=0$ case, when the Neumann boundary condition is imposed on the initial hypersurface and the Dirichlet boundary condition on the final hypersurface. For the no-boundary choice, $P_{i} = -3i\sqrt{k}$ and we have
\bea
\label{eq:stot_onsh_nbc}
\bar{S}_{\rm tot}^{\rm on-shell}[ N_c] &=& \Lam^2 N_c^3/9 + i\Lam\sqrt{k}N_c^2 - \Lam q_f N_c - 3 i q_f \sqrt{k} - 8 i \al k^{3/2}.
\eea
The zeroes of the action are given by:
\bea
\label{eq:zeros_onsh_nbc}
N_{1} && = - \frac{3i\sqrt{k}}{\Lam} + 3 e^{-i\pi/6} \frac{\Lam q_f - 3k}{\Delta^{1/3}} +  e^{i\pi/6} \frac{\Delta^{1/3}}{\Lam^2} \, , \\
N_{2} && = - \frac{3i\sqrt{k}}{\Lam} - 3 e^{i\pi/6} \frac{\Lam q_f - 3k}{\Delta^{1/3}} -  e^{-i\pi/6} \frac{\Delta^{1/3}}{\Lam^2} \, , \\
N_{3} && = - \frac{3i\sqrt{k}}{\Lam} + 3 i \frac{\Lam q_f - 3k}{\Delta^{1/3}} - i \frac{\Delta^{1/3}}{\Lam^2} \, ,
\eea
where we define the $\Delta = 27 \Lam^3 k^{3/2} + \sqrt{27 q_f \Lam^{7} (\Lam^2 q_f^2 - 9k q_f \Lam + 27 k^2)} > 0$ which is positive, $\Delta > 0$, for all values of $q_f$. Here, $N_{3}$ is purely imaginary, while $N_{1}$ and $N_{2}$ are the complex lapse points related by an anti-linearity. Our numerical analysis reveals that the zero-action solution $N_3$ which corresponds to a Euclidean geometry, is classified as KSW-allowed while, the other two zero-action solutions, $N_1$ and $N_2$ are KSW disallowed, see Figure (\ref{Fig:KSW_theta_0.05}). The fact that both $N_1$ and $N_2$, which are related by anti-linearity, are disallowed further confirms that the KSW criterion respects anti-linear symmetry. From our analysis, in contrast to the \cite{Witten:2021nzp}, we found an instance where one of the zero-action geometries is KSW allowed. Similarly, for the RBC-DBC case, two of the zeroes of the action lie in the allowable region and the other two are confined to the disallowed region. This raises the question regarding sufficiency of KSW criterion. Indeed, in \cite{BenettiGenolini:2025jwe}, it is noted that KSW allowable criterion is necessary but not sufficient criterion for the complex metrics in the context of gravitational index for supergravity theories with asymptotically flat and AdS spaces\footnote{The authors of \cite{BenettiGenolini:2025jwe} have posted a corrected version \cite{BenettiGenolini:2026raa}, in which they clarify that the KSW condition is equivalent to the microscopic constraint, and that the KSW criterion is sufficient for the computation of the gravitational index.}. A similar situation is also encountered here. However, addressing this sufficiency completely is beyond the scope of this paper. We leave this question for future publication.

%%%%%%%%%%%%%%%%%%%%%%%%%%%%%%%%%%%%%%%%%%%%%%%%%%%%%%%%%

%%%%%%%%%%%%%%%%%%%%%%%%%%
\end{document}